\documentclass[12pt]{article}
\usepackage{amsmath}
\numberwithin{equation}{section}
\usepackage{amssymb}
\usepackage{graphicx,psfrag,epsf}
\usepackage{enumerate}
\usepackage {geometry}
\usepackage{natbib}
\usepackage{multirow}
\usepackage{apalike}
\usepackage{appendix}
\usepackage{bm}
\usepackage{mathtools}
\usepackage{booktabs}
\usepackage{url} % not crucial - just used below for the URL 
\usepackage{subfigure}
\usepackage{xcolor}
\usepackage{hyperref}
\newcommand{\rvs}[1]{{\color{black} #1}}
\newcommand{\addrvs}[1]{{\color{black} #1}}
\usepackage{titling}

\begin{document}

\def\spacingset#1{\renewcommand{\baselinestretch}%
{#1}\small\normalsize} \spacingset{1}

%%%%%%%%%%%%%%%%%%%%%%%%%%%%%%%%%%%%%%%%%%%%%%%%%%%%%%%%%%%%%%%%%%%%%%%%%%%%%%
\spacingset{1.45} % DON'T change the spacing!

  \title{\bf Variational Estimation for Multidimensional Generalized Partial Credit Model}

\author{Chengyu Cui$^1$, Chun Wang$^2$, and Gongjun Xu$^1$\\
    \textit{\small{$^1$ Department of Statistics, University of Michigan}}\\
    \textit {\small{$^2$ College of Education, University of Washington}}}
    \date{}
  \maketitle

%\bigskip
\begin{abstract}
Multidimensional item response theory (MIRT) models have generated increasing interest in the psychometrics literature. Efficient approaches for estimating MIRT models with dichotomous responses have been developed, but constructing an equally efficient and robust algorithm for polytomous models has received limited attention. To address this gap, this paper presents a novel Gaussian variational estimation algorithm for the multidimensional generalized partial credit model (MGPCM). The proposed algorithm demonstrates both fast and accurate performance, \rvs{as illustrated through a series of simulation studies and two real data analyses}.
%Many approaches are proved to be efficient in estimating the MIRT models with dichotomous responses. However, little attention has been paid to construct an efficient and robust algorithm for polytomous models. In this paper a Gaussian variational estimation algorithm is proposed for multidimensional generalized partial credit model (MGPCM). We demonstrate that this new variational algorithm is fast and accurate. An illustrative real data analysis is also presented based on the Trends in International Mathematics and Science Study database.
\end{abstract}

\noindent%
\vfill

\newpage
\section{Introduction}
\label{sec:intro}
There \addrvs{is} a wide range of psychometric models for analyzing data in educational and psychological surveys. Models including discrete and continuous latent factors have received great attention due to repeated empirical evidence of adequate model fit and success of interpretation that aligns with substantive theory. Among them, a variety of multidimensional item response theory (MIRT) models \citep{reckase2009multidimensional} have been proposed to account for various multidimensional structures of the latent constructs. Within the family of MIRT models, one of the most studied models is the \addrvs{multidimensional two-parameter logistic model} (M2PL) \citep{reckase2009multidimensional} for dichotomous response, as well as the Multidimensional Three-parameter Logistic Model (M3PL) and Multidimensional Four-parameter Logistic Model (M4PL), which are often used 
%when the lower and upper asymptote may exist. This is often the case 
when students can guess the answer to the test item correctly (resulting in a lower asymptote in educational measurement) or when the chance of answering an item correctly does not approach 1 (resulting in a higher asymptote in psychopathology measurement).  Moreover, the multidimensional graded response model \citep{cai2010high} and multidimensional (generalized) partial credit model \citep{yao2006multidimensional} have also been proposed to handle polytomous items. These models are regarded as extensions of IRT models for dichotomous response in various ways to characterize latent cognitive structures from more complicated datasets.

In the IRT literature, the \addrvs{marginal maximum likelihood estimator} (MMLE) is considered to be a consistent and efficient approach for parameter estimation by maximizing the marginal log-likelihood of item parameters \citep{bock1981marginal,bock1988full}. In particular, this estimator is efficient in the asymptotic regime where the test length is small with a rather large number of examinees, which is often the case in applications.  However, integrating out the latent ability for marginal likelihood evaluation is notoriously time-consuming, involving multidimensional integrals whose computational complexity grows exponentially with the number of latent dimensions. This results in great difficulties especially when the latent structure is complex or within a domain of large dimension. Several methods have been proposed previously to deal with this computational challenge, such as Gaussian quadrature methods \citep{bock1981marginal,schilling2005high}, Laplace approximation methods \citep{lindstrom1988newton,wolfinger1993generalized,andersson2021estimation}, Metropolis-Hastings Robbins-Monro algorithms \citep{cai2010high}, stochastic expectation maximization (EM) algorithms \citep{von2010stochastic,zhang2020improved}, and variational approximation methods \citep{rijmen2013fitting,cho2021gaussian,cho2022regularized}. Among these, Gauss-Hermite quadrature approximation does not scale well to high-dimensional scenarios, \rvs{and Laplace approximation, which is closely related to variational method \citep{opper2009variational}, may suffer from numerical inaccuracies when dimensions get high or when the likelihood function is in a skewed shape. Additionally, many variants of Laplace approximations, though overcoming some deficiencies, suffer from inflexibility or may be hard to implement \citep{ormerod2010explaining}}. The methods based on Monte Carlo simulations such as Metropolis-Hastings Robbins-Monro and stochastic EM algorithms, on the other hand, are more robust but may be computationally inefficient. \rvs{Notably, in addition to being numerically accurate and computationally efficient, the variational method also provide good interpretability and the variational distributions contain additional useful information \citep{blei2017variational}}.

Despite the extensive literature on estimation methods for models of dichotomous response \citep{cai2010high,chen2019joint,cho2021gaussian,feuerstahler2014estimation,meng2020marginalized}, little attention has been given to the efficient and robust estimation of MIRT models for polytomous responses \citep{bock1988full,kim2020polytomous}. In the paper, we propose a Gaussian variational expectation-maximization algorithm for the multidimensional Generalized Partial Credit Model (MGPCM), which is both computationally efficient and numerically stable. The variational method, first proposed in computer science and statistical learning, has since become an efficient approach for large-scale computation in fields such as pattern recognition and document retrieval \citep{titterington2004bayesian,blei2006variational}. The application of the variational method in analyzing statistical models has also received wide attention \citep{blei2017variational,ormerod2010explaining}. For instance, \cite{ormerod2012gaussian} adopted a Gaussian variational approximation approach for the estimation of generalized linear mixed effects models. In the field of psychometrics and educational measurement, variational methods have been used \addrvs{for efficient estimation of} multidimensional 1PL models \citep{rijmen2013fitting} and multi-dimensional 2/3PL models \citep{cho2021gaussian}, as well as analysis in cognitive diagnostic models \citep{yamaguchi2020variational, oka2023scalable}. Motivated by \cite{cho2021gaussian}, in this paper, we generalize their method to the estimation of MGPCM. This generalization is non-trivial because the MGPCM uses a generalized logit link function that requires a completely new derivation of the variational lower bound to fully enable closed-form parameter update in the EM algorithm.

The rest of the paper is organized as follows. Section 2 provides a brief introduction to the model. In Section 3, we introduce the polytomous Gaussian variational expectation-maximization algorithm (pGVEM) and give its derivation. Section 4 evaluates the performance of the pGVEM algorithm compared to the traditional EM algorithm \rvs{and other existing methods through a comprehensive simulation study}. In Section 5, we apply the pGVEM algorithm to analyze data from an international educational assessment database \rvs{and a Big-Five  personality assessment}. Finally, in Section 6, we conclude the paper and suggest potential future research directions.

\section{Model Setting}
\addrvs{Generalized partial credit Model} (GPCM) \citep{muraki1992generalized,embretson2013item}, also named as a compensatory multidimensional two-parameter partial credit model (M-2PPC) \citep{yao2006multidimensional}, is a popular IRT model for polytomous responses. It allows for the assessment of partial scores for constructed response items and intermediate steps that students have accomplished on the path toward solving the items completely. Suppose we have $N$ examinees and $J$ test items, with the random variable $Y_{ij}$ denoting the partial credit of person $i$'s response to item $j$. For each item $j$, $a_j$ is a discrimination parameter that implies the strength of association between latent trait and item responses. $\beta_{jk}$ is a threshold parameter that separates two adjacent response categories. The \addrvs{partial credit model} \citep{masters1982rasch} is a special case of GPCM where the discrimination parameter is fixed to be the same across different items. The item response function of GPCM that characterizes the probability of a specific response is given by
\begin{equation}
    \Pr(Y_{ij}=k|\theta_i,a_j,\beta_{jk})=\frac{\exp[\sum_{r=0}^ka_j(\theta_i-\beta_{jr})]}{\sum_{v=0}^{K_j}\exp[\sum_{r=0}^va_j(\theta_i-\beta_{jr})]},
    \label {GPCM}
\end{equation}
where $k=0,1,\cdots,K_j-1$ and $K_j$ is the number of differential partial credit scores for the $j^{th}$ item.

%In some paper \citep{yao2006multidimensional} the Generalized Partial Credit Model is expressed in a more delicate form with more interpretable parameters:
%\begin{equation}\Pr(Y_{ij}=k|\theta_i,\bm\beta)=\frac{\exp(k\beta_{2,j}\theta_i-\sum_{t=1}^k\beta_{\delta_t,j})}{\sum_{v=0}^{K_j-1}\exp(v\beta_{2,j}\theta_i-\sum_{t=1}^v\beta_{\delta_t,j})}\end{equation}
%Here $\beta_j=(\beta_{2,j},\beta_{\delta_1,j}=0,\beta_{\delta_2,j},\cdots,\beta_{\delta_{K_j},j})$ is the item parameters. Sometimes the difficulty will be scaled by the discrimination parameter $a_j$(or $\beta_{j,2}$) for a explicit comparison across different categories. 
The \addrvs{multidimensional generalized partial credit model} (MGPCM) is a natural multidimensional extension of GPCM. The main idea is replacing the uni-dimensional latent ability with a $D$-dimensional vector, and each dimension represents a facet of the multidimensional construct (i.e., science and literacy). Similarly, the discrimination parameters $\bm a_j$ also become \addrvs{$D$-dimensional vectors} to reflect the discrimination power of item $j$ with respect to each facet (i.e., dimension) of the multidimensional construct $\boldsymbol{\theta}$. The threshold parameters stay the same as in uni-dimensional models. Equation \eqref{GPCM} therefore is updated as follows:
\begin{equation}
    \Pr(Y_{ij}=k|\bm\theta_i,\bm a_j,\beta_{jk})=\frac{\exp\{\sum_{r=0}^k(\bm a_j^\prime\bm\theta_i-\beta_{jr})\}}{\sum_{v=0}^{K_j-1}\exp\{\sum_{r=0}^v(\bm a_j^\prime\bm\theta_i-\beta_{jr})\}}.
\label{mgpcm1}
\end{equation}
\rvs{Here $\bm a_j^\prime\bm \theta_i$ indicates the inner product of $\bm a_j$ and $\bm\theta_i$ as  $\bm a_j^\prime\bm \theta_i=\sum_{d=1}^Da_{jd}\theta_{id}$ where $a_{jd}$ and $\theta_{id}$ are the $d$th component of $\bm a_j$ and $\bm\theta_i$, respectively}. With a slight re-parameterization, we have the following item response function for MGPCM which we will use throughout the paper:
\begin{equation}\Pr(Y_{ij}=k|\bm \theta_i,\bm a_j,b_{jk})=\frac{\exp(k\bm a_j^{\prime}\bm \theta_i-b_{jk})}{\sum_{v=0}^{K_j-1}\exp(v\bm a_j^{\prime}\bm \theta_i-b_{jv})}.
\label{mgpcm2}
\end{equation}
In Equation \eqref{mgpcm2}, $b_{jk}$ replaces $\sum_{r=0}^k\beta_{jr}$ in Equation \eqref{mgpcm1} for each $k=0,1,\cdots,K_j-1$. Note that for model identification, we can only have $K_j-1$ estimable threshold parameters \addrvs{for each item $j$}, and hence we fix $b_{j0}=0$.

\section{Gaussian Variational Approximation}
\label{sec:meth}
\subsection{Derivation of algorithm}
In this section, we describe the derivation of the GVEM algorithm. In the following, we denote the collection of item parameters by $M_p$ being the total number of parameters to be estimated, i.e. $M_p=\{\boldsymbol{a}_j\in\mathbb{R}^D,b_{jk}\in\mathbb{R}:j=1,\cdots,J,k=1,\cdots,K_j-1\}$ for MGPCM. As we discussed above, the parameter $b_{j0}$ is fixed as 0 for all $j$. \rvs{To be consistent with the common convention, we assume the latent vector $\boldsymbol{\theta}$ follows a multivariate normal distribution of $\bm 0$ mean and covariance $\bm\Sigma_{\theta}$ with density function denoted by $\phi(\cdot)$.} 
%The diagonals of $\bm \Sigma_\theta$ are fixed at 1, whereas off-diagonals are free to estimate in confirmatory analysis or zeros in exploratory analysis (with  later proper rotations \citep{browne2001overview}).
The marginal probability of response vector $\bm Y_i=(Y_{i1},\cdots,Y_{iJ})^\prime$ for person $i$ is defined as follows
\begin{align*}
\Pr(\bm Y_{i}|M_p)=&\int_{\bm\theta_i}\prod_{j=1}^J\prod_{k=0}^{K_j-1}I_{(Y_{ij}=k)}\Pr(Y_{ij}=k|\bm \theta_i,M_p)\phi(\bm \theta_i)d\bm \theta_i\\=&\int_{\bm\theta_i}\prod_{j=1}^J\prod_{k=0}^{K_j-1}I_{(Y_{ij}=k)}\frac{\exp(k\bm a_j^\prime\bm \theta_i-b_{jk})}{\sum_{v=0}^{K_j-1}\exp(v\bm a_j^\prime\bm \theta_i-b_{jv})}\phi(\bm \theta)d\bm \theta_i,
\end{align*}
where $I_{(Y_{ij}=k)}$ is an indicator function equal to 1 if $Y_{ij}=k$ and zero otherwise, and therefore the marginal log-likelihood function for all responses from the examinees is given by
\begin{equation}
l(M_p|\bm Y)=\sum_{i=1}^N\log\Pr(\bm Y_i|M_p)=\sum_{i=1}^N\log\int_{\bm \theta_i}\prod_{j=1}^J\Pr(Y_{ij}|\bm \theta_i,M_p)\phi(\bm \theta_i)d\bm \theta_i,
\label{marginal}
\end{equation}
where \rvs{$\bm Y=(\bm Y_1,\cdots,\bm Y_{N})^\prime$ is the $N\times J$  matrix of realized categorical responses}.

%We follow a similar approach to \cite{cho2021gaussian} to 
Following the variational estimation literature \citep{blei2017variational,cho2021gaussian}, we first derive a variational lower bound for MGPCM. 
We define $KL\{p(\cdot)\|q(\cdot)\}$ the Kullback-Leibler divergence of probability distribution $p$ and $q$. For an arbitrary probability density function $q(\cdot)$, the marginal log-likelihood in Equation \eqref{marginal} has the following lower bound \citep{blei2017variational}:
\rvs{\begin{align}
& l(M_p|\bm Y)=\sum_{i=1}^N\int_{\bm \theta_i}q_i(\bm \theta_i)d\bm \theta_i\log\Pr(\bm Y_i|M_p)\nonumber\\
& =\sum_{i=1}^N\int_{\bm \theta_i}\log\Big[\frac{P(\bm Y_i,\bm \theta_i|M_p)q_i(\bm \theta_i)}{P(\bm \theta_i|\bm Y_i,M_p)q_i(\bm \theta_i)}\Big]q_i(\bm \theta_i)d\bm \theta_i\nonumber\\
& =\sum_{i=1}^N\bigg{[}\int_{\bm \theta_i}\big[\log P(\bm Y_i,\bm \theta_i|M_p)\big] q_i(\bm \theta_i)d\bm \theta_i-\int_{\bm \theta_i}\big[\log q(\bm \theta_i)\big] q_i(\bm \theta_i)d\bm \theta_i+KL\Big\{q_i(\bm \theta_i)\big\|P(\bm \theta_i|\bm Y_i,M_p)\Big\}\bigg{]}\nonumber\\
&\geq\sum_{i=1}^N\int_{\bm \theta_i}\big[\log P(\bm Y_i,\bm \theta_i|M_p)\big] q_i(\bm \theta_i)d\bm \theta_i-\sum_{i=1}^N\int_{\bm \theta_i}\big[\log q(\bm \theta_i)\big] q_i(\bm \theta_i)d\bm \theta_i.\label{lbo}
\end{align}}
The last inequality holds if and only if the KL divergence between the variational distribution $q_i(\cdot)$ and the posterior distribution $P(\cdot|\bm Y_i,M_p)$ is 0, which indicates $q_i(\bm \theta_i)=P(\bm \theta_i|\bm Y_i,M_p)$. In the literature of variational inference, the right-hand side of Equation \eqref{lbo} is defined to be the evidence lower bound \citep{blei2017variational}, which is equivalent, up to a constant with respect to $q(\cdot)$, to the KL divergence between the assumed variational distribution $q(\cdot)$ and the conditional density of the latent variables given the observations.

In the following, we construct an approximation for the \addrvs{marginal maximum likelihood estimator }from the evidence lower bound. The primary objective is to identify a suitable distribution that can approximate the posterior distribution $P(\bm \theta_i|\bm Y_i,M_p)$. Motivated by this insight, we propose to construct an EM-type algorithm to compute the \addrvs{marginal maximum likelihood estimator}. In the E-step, we evaluate the expectation of the complete data log-likelihood, and the expectation is taken with respect to the latent variables $\bm\theta_i$ under its variational probability density function $q_i(\cdot)$:
\begin{equation*}
\sum_{i=1}^N\int_{\bm \theta_i}\log P(\bm Y_i,\bm \theta_i|M_p)q_i(\bm \theta_i)d\bm \theta_i,
\end{equation*}
Here the density function $q_i(\bm \theta_i)$ is chosen to minimize the KL divergence $KL\{q_i(\bm \theta_i\|P(\bm \theta_i|\bm Y_i,$ $M_p)\}$ as the best approximation to the posterior distribution. The second term in the evidence lower bound is left out since it is irrelevant to item parameters. However, a problem with respect to minimizing the KL divergence is that it is hard to find an explicit formula for the posterior distribution of $\bm \theta_i$ with respect to the previous estimated item parameters $\hat M_p$, as it involves computing $D$-dimensional integrals. Numerical methods, such as the Gauss–Hermite approximation, Monte Carlo expectation-maximization, and stochastic expectation-maximization, are often used to provide fast approximation. Herein we adopt the Gaussian variational inference method. It is widely accepted that the posterior distribution of the latent ability $P(\theta_i|Y_i,M_p)$ can be approximated by a Gaussian distribution \citep{changstout1993, wang2015}, and hence we aim to find an optimal $q_i(\bm\theta_i)$ in the family of Gaussian distribution while minimizing the KL divergence between $q_i(\bm\theta_i)$ and $P(\bm \theta_i|\bm Y_i,M_p)$. \rvs{Since the posterior distribution can be expressed as 
\begin{equation*}
P(\bm \theta_i|\bm Y_i, M_p)=\frac{P(\bm Y_i,\bm \theta_i|M_p)}{P(\bm Y_i|M_p)},
\end{equation*}
we only need to evaluate $P(\bm Y_i,\bm \theta_i|M_p)$ to find a proper $q_i(\cdot)$ as \
\begin{equation*}KL\{q_i(\bm \theta_i)||P(\bm \theta_i|\bm Y_i,M_p)\}=KL\{q_i(\bm \theta_i)||P(\bm \theta_i,\bm Y_i|M_p)\}+C.\end{equation*}}
Under the setting of MGPCM, the logarithm of joint distribution function of $\bm \theta_i$ and $\bm Y_i$ is
\begin{align}
&\log P(\bm Y_i,\bm \theta_i|M_p)=\log P(\bm Y_i|\bm \theta_i,M_p)+\log\phi(\bm \theta_i)\nonumber\\
&=\sum_{j=1}^J\Bigg\{\sum_{k=0}^{K_j-1}I_{(Y_{ij}=k)}\Big[k\bm a_j^\prime\bm \theta_i-b_{jk}-\log(\sum_{v=0}^{K_j-1}\exp(v\bm a_j^\prime \bm \theta_i-b_{jv}))\Big]\Bigg\}+\log\phi(\bm \theta_i).\label{gvem1}\end{align}

The nonlinear softmax function, defined by $f_v(\bm x)=\exp(x_v)/[\sum_{k=1}^n\exp(x_k)]$ for an $n$-dimensional vector $\bm x$, is the main cause of the intractability of integral. To overcome this problem, a variational lower bound based on the approximation to the softmax function is proposed and by augmenting Equation \eqref{gvem1} with variational parameters, the evidence lower bound can be computed explicitly without resorting to numeric integration. Among the many approximations for the softmax function, we adopt a One-Versus-Each bound \citep{aueb2016one}, which well approximates the softmax function and captures the model features. We start with the following inequality: 
\begin{equation}
f_v(\bm x)=\frac{e^{x_v}}{\sum_{k=1}^ne^{x_k}}\geq\prod_{k=1,k\neq v}^n\frac{e^{x_v}}{e^{x_v}+e^{x_k}}=2\prod_{k=1}^n\frac{e^{x_v}}{e^{x_v}+e^{x_k}}.\label{logistic}
\end{equation}
Denote by $k_{ij}$ the realized partial credit score for the response of the $i^{th}$ person for the $j^{th}$ item. Then by applying the bound \eqref{logistic} to \eqref{gvem1}, we have
\begin{align*}
\log P(\bm Y_i,\bm\theta_i|M_p)=&\log\phi(\bm \theta_i)+\sum_{j=1}^J\Big[\sum_{k=0}^{K_j-1}I_{(Y_{ij}=k)}\log\frac{\exp(x_{ijk})}{\sum_{v=0}^{K_j-1}\exp(x_{ijv})}\Big]\nonumber\\\geq&\log\phi(\bm \theta_i)+\sum_{j=1}^J\Bigg\{\sum_{k=0}^{K_j-1}I_{(Y_{ij}=k)}\big[\log2+\sum_{v=0}^{K_j-1}\log\frac{\exp(x_{ijk})}{\exp(x_{ijv})+\exp(x_{ijk})}\Big]\Bigg\}\nonumber\\=&\log\phi(\bm\theta_i)+\sum_{j=1}^J\Bigg\{\log2-\sum_{k=0}^{K_j-1}\log\big[1+\exp(x_{ijk}-x_{ijk_{ij}})\big]\Bigg\}.
\end{align*}
Here we denote $x_{ijk}=k\bm a_j^\prime \bm \theta_i-b_{jk}$ for short. We wish to draw attention to our selection of the ``One-Versus-Each bound". It can be established for \eqref{logistic} that a strict inequivalence holds true in all cases, except for the exceptional circumstance ${x_v}/{x_k}\to\infty$ for all $k\neq v$ with at most one exception. Additionally,  the approximation is \addrvs{the closest} when $x_v$ is among the largest of all $x_k$. The idea of maximum likelihood estimation indicates that, when partial credit score $Y_{ij}$ is recorded as $k_{ij}$, $k_{ij}\bm a_j^\prime\bm\theta_i-b_{jk_{ij}}$ is the most likely to be the largest among all $v\bm a_j^\prime\bm\theta_i-b_{jv}$ for $v=0,1,\cdots,K_j-1$. Therefore the feature of the One-Versus-Each bound does fit well as an approximation to the \addrvs{marginal maximum likelihood estimator}.

% Moreover, the recorded response $Y_{ij}=k$ shall indicate an emphasis on the likelihood of the category k from a Bayesian prospect. Therefore the characteristic of the bound contributes a lot to an accurate and favourable estimation compared with other softmax bounds.

The logistic sigmoid function \eqref{logistic} can be further approximated by a local variational approach:
\begin{align*}
\log P(\bm Y_i,\bm \theta_i|M_p)\geq&\log \phi(\bm \theta_i)+\sum_{j=1}^J\Bigg\{\log2-\sum_{k=0}^{K_j-1}\eta(\xi_{ijk})\big[(x_{ijk}-x_{ijk_{ij}})^2-{\xi_{ijk}}^2\big] \\&-\sum_{k=0}^{K_j-1}\frac12(x_{ijk}-x_{ijk_{ij}}-\xi_{ijk})-\sum_{k=0}^{K_j-1}\log(1+e^{\xi_{ijk}}) \Bigg\} ,
\end{align*}
where $\bm\xi=\{\xi_{ijk}\}_{i,j,k}$ are called variational parameters, which will be iteratively updated together with item parameters in the M-step. Here the function $\eta(x)$ is defined as ${(e^x-1)}/{[4x(e^x+1)]}$ \citep{jaakkola2000bayesian}. We use this local variational approximation for a suitable expectation of $\log P(\bm Y_i,\bm\theta_i|M_p)$ (i.e., given below in Equation (3.5)) that can be written as a quadratic form with respect to $\bm\theta_i$. This will facilitate the selection of $q_i(\cdot)$ in the family of Gaussian distributions.

Next we substitute $x_{ijk}$ by $k\bm a_j^\prime\bm\theta_i-b_{jk}$ and write the above lower bound of joint distribution function as
\begin{align}\log P(\bm Y_i,\bm \theta_i|M_p)\geq&\sum_{j=1}^J\Bigg\{-\sum_{k=0}^{K_j-1}\Big[\eta(\xi_{ijk})(k-k_{ij})^2\bm \theta_i^\prime \bm a_j\bm a_j^\prime\bm \theta_i-2(k-k_{ij})\eta(\xi_{ijk})(b_{jk}-b_{jk_{ij}})\bm a_j^\prime
\bm \theta_i\nonumber\\
&+ \frac12(k-k_{ij})\bm a_j^\prime\bm \theta_i+\eta(\xi_{ijk})(b_{jk}-b_{jk_{ij}})^2-\frac12(b_{jk}-b_{jk_{ij}})\nonumber\\&-\eta(\xi_{ijk}){\xi_{ijk}}^2-\frac12\xi_{ijk}+\log(1+e^{\xi_{ijk}})\Big]+\log2\Bigg\}+\log \phi(\bm \theta_i)\nonumber,\end{align}
and therefore the expectation of the log-likelihood, which needs computing in the E-step, takes the following form:
\begin{align}E(M_p,\bm\xi):=&\int_{\bm \theta_i}\log P(\bm Y_i|\bm \theta_i,M_p)q_i(\bm \theta_i)d\bm \theta_i+\int_{\bm \theta_i}\log \phi(\bm \theta_i)q_i(\bm \theta_i)d\bm \theta_i\nonumber\\\geq&\int_{\bm \theta_i}\sum_{j=1}^J\Bigg\{\log2-\sum_{k=0}^{K_j-1}\Big[\eta(\xi_{ijk})(k-k_{ij})^2\bm \theta_i^\prime \bm a_j\bm a_j^\prime\bm \theta_i-2(k-k_{ij})\eta(\xi_{ijk})(b_{jk}-b_{jk_{ij}})\bm a_j^\prime\bm 
\theta_i\nonumber\\+&\frac12(k-k_{ij})\bm a_j^\prime\bm \theta_i+\eta(\xi_{ijk})(b_{jk}-b_{jk_{ij}})^2-\frac12(b_{jk}-b_{jk_{ij}})\nonumber\\-&\eta(\xi_{ijk}){\xi_{ijk}}^2-\frac12\xi_{ijk}+\log(1+e^{\xi_{ijk}})\Big]\Bigg\}q_i(\bm \theta_i)d\bm \theta_i+\int_{\bm \theta_i}\log \phi(\bm \theta_i)q_i(\bm \theta_i)d\bm \theta_i.\label{gvem2}\end{align}
For a minimized KL divergence, $q_i(\theta_i)$ is selected as follows:
$$
\log q_i(\bm \theta_i)\propto\sum_{j=1}^J\sum_{k=0}^{K_j-1}\Big\{(k-k_{ij})\big[2\eta(\xi_{ijk})(b_{jk}-b_{jk_{ij}})-0.5\big]\bm a_j^\prime\bm \theta_i-\eta(\xi_{ijk})(k-k_{ij})^2\bm \theta_i^\prime \bm a_j\bm a_j^\prime\bm \theta_i\Big\}-\frac{\bm \theta_i^\prime\bm \Sigma_\theta^{-1}\bm \theta_i}2.
$$
As the choice of $q_i(\cdot)$ has been confined in the Gaussian family, it suffices to give the update for the mean and covariance matrix:
\begin{equation}
\bm \mu_i=\bm\Sigma_i\times\sum_{j=1}^J\sum_{k=0}^{K_j-1}(k-k_{ij})\Big[2\eta(\xi_{ijk})(b_{jk}-b_{jk_{ij}})-\frac12\Big]\bm a_j^\prime;\label{updatemean}
\end{equation}
\begin{equation}
\bm \Sigma_i^{-1}={\bm \Sigma}_\theta^{-1}+2\sum_{j=1}^J\sum_{k=0}^{K_j-1}\eta(\xi_{ijk})(k-k_{ij})^2\bm a_j\bm a_j^\prime.\label{updatecov}
\end{equation}
In each iteration, the item parameters and variational parameters $\xi_{ijk},\bm a_j,b_{jk}$ are obtained from the previous M-step, and taken as the initial value if it is the first iteration.

For the M-step, the item parameters are chosen to maximize the above expectation of the lower bound obtained by plugging Equation \eqref{updatemean} and \eqref{updatecov} into Equation \eqref{gvem2}:
\begin{align}E(M_p,\bm \xi)\geq&\sum_{i=1}^N\sum_{j=1}^J\log 2+\sum_{i=1}^N\sum_{j=1}^J\sum_{k=0}^{K_j-1}\Bigg\{
-\eta(\xi_{ijk})(k-k_{ij})^2\bm a_{j}^\prime\Big[\bm \Sigma_i^{(t)}+(\bm \mu_i^{(t)})(\bm \mu_i^{(t)})^\prime\Big]\bm a_{j}\nonumber\\
&+(k-k_{ij})[2\eta(\xi_{ijk})(b_{jk}-b_{jk_{ij}})-\frac12]\bm a_j^\prime\bm \mu_i^{(t)}-\eta(\xi_{ijk})(b_{jk}-b_{jk_{ij}})^2+\frac12(b_{jk}-b_{jk_{ij}})\nonumber\\&+\eta(\xi_{ijk}){\xi_{ijk}}^2+\frac12\xi_{ijk}-\log(1+e^{\xi_{ijk}})
\Bigg\}-\frac N2\log|\bm \Sigma_{\theta}^{(t)}|\\&-\sum_{i=1}^N\frac12Tr\{(\bm \Sigma_{\theta}^{(t)})^{-1}[\bm \Sigma_i^{(t)}+(\bm \mu_i^{(t)})(\bm \mu_i^{(t)})^\prime\Big]\nonumber\\ \coloneqq &\, \underline{E}(M_p,\bm\xi).\label{gvem3}\end{align}

Through maximizing the lower bound $\underline{E}(M_p,\bm\xi)$, we can derive a new set of item parameters that could potentially maximize the left-hand side. Updating the variational parameters helps to prevent the iteration from leading to a smaller value of the target expectation by shrinking the inequality too much when the right-hand side is maximized. The efficiency of this \addrvs{majorization-maximization} approach depends on the goodness of fit of the adopted softmax bound.

To maximize the lower bound on the expectation concerning the item parameters $M_p$ and variational parameters $\bm \xi$, we employ a Gauss-Seidel scheme to handle the nonlinear terms regarding the parameters. Each iterative update uses the most recently updated copies of the parameters. The update are given as follows.

For each $j=1,\cdots, J$,
\begin{equation}
\bm a_j=\frac12\Big[\sum_{i=1}^N\sum_{k=0}^{K_j-1}\eta(\xi_{ijk})(k-k_{ij})^2(\bm \Sigma_i+\bm \mu_i\bm \mu_i^\prime )\Big]^{-1}\Big\{\sum_{i=1}^N\sum_{k=0}^{K_j-1}(k-k_{ij})\Big[2\eta(\xi_{ijk})(b_{jk}-b_{jk_{ij}})-\frac12\Big]\bm \mu_i\Big\},\label{updatedisc}
\end{equation}
with $\xi_{ijk}$, $b_{jk}$ from the last iteration or initialization and $\bm \mu_i$, $\bm\Sigma_i$ from E-step. Next for the threshold parameters, for each $j=1,\cdots,J,k=1,\cdots K_j-1$,

\begin{equation}b_{jk}=\frac{\sum_{i=1}^N[\mbox B_1(i,j,k)I_{(k\neq k_{ij})}+I_{(k=k_{ij})}\sum_{v=0,v\neq k}^{K_j-1}\mbox B_2(i,j,v,k)]}{2\sum_{i=1}^N(\eta(\xi_{ijk})I_{(k\neq k_{ij})}+I_{(k=k_{ij})}\sum_{v=0,v\neq k}^{K_j-1}\eta(\xi_{ijv}))},\label{updatethres}
\end{equation}
where
\begin{align*}
\mbox B_1(i,j,k)=2\eta(\xi_{ijk})(k-k_{ij})\bm a_j^\prime\bm \mu_i+0.5+2\eta(\xi_{ijk})b_{jk_{ij}};\\\mbox B_2(i,j,v,k)=-2\eta(\xi_{ijk})(v-k)\bm a_j^\prime\bm \mu_i-0.5+2\eta(\xi_{ijv})b_{jv}.
\end{align*}
Here $\bm a_j$ are from the previous step and $\bm \mu_i$, $\bm\Sigma_i$ are from the previous E-step. Finally for the variational parameters $\xi_{ijk}$, for each $i=1,\cdots,\addrvs{N},j=1,\cdots, J,k=0,\cdots K_j-1$, we have update as
\begin{equation}
\xi_{ijk}^2=[(k-k_{ij})\bm a_j^\prime \bm \mu_i-(b_{jk}-b_{jk_{ij}})]^2+(k-k_{ij})^2\bm a_j^\prime\bm \Sigma_i\bm a_j.\label{updatevarpara}
\end{equation}
with all other parameters obtained from the latest updates.

In the exploratory analysis where we do not have any prior information on the item factor loadings, \addrvs{the assumed covariance} $\Sigma_\theta$ is fixed as $I_D$ and later proper rotations \citep{browne2001overview} are imposed to allow the factors to be correlated and thus allow for analysis of latent structures. But for confirmatory factor analysis, we update the covariance as
\begin{equation}
\bm \Sigma_\theta=\frac1N\sum_{i=1}^N(\bm \Sigma_i+\bm \mu_i\bm \mu_i^\prime).\label{updatecovtheta}
\end{equation}
 and scale its diagonal entries to be 1.
 
\subsection{Standard Error Estimation}
\label{ssc_e}
\rvs{Computing standard errors (SEs) of the item parameter estimates is crucial for various applications, such as multidimensional computerized adaptive testing, item parameter calibration as well as differential item functioning. 
Challenges arise in estimating SEs when dealing with a high-dimensional latent domain and polytomous responses as in  MGPCM.
%The estimation of approximation method is challenging, especially when the latent domain is high-dimensional and one record of polytomous response is recorded for one pair of individual and item. 
The commonly used method for estimating SEs is based on the approximated Fisher's information matrix. However, taking the inverse of a prohibitively large information matrix (due to high dimensions and long test length) can be unstable when the sample size of the examinees is not large enough. An alternative numerical approximation using Gaussian quadrature in EM estimation has been proposed \citep{cagnone2013latent}, but it is computationally expensive and sensitive to dimensionality. The \addrvs{supplemented expectation maximization} (SEM) algorithm has also been developed in the IRT literature \citep[e.g.,][]{tian2013numerical}. %to be more efficient and accurate. 
\rvs{However, in pilot simulations we found that none of these methods is capable of providing stable estimations, especially when the dimension $D$ and the number of categories $K$ is large.} Therefore, to estimate the standard errors of item parameters under the pGVEM framework for MGPCM, we adopt a bootstrap approach that uses a resampling procedure. Bootstrap is an efficient alternative when the standard SEs estimation is mathematically intractable \citep{efron1986bootstrap}. The resampling procedure avoids the direct computation of SEs.
%when making inferences related to the accuracy of the statistic. 

The bootstrap procedure in the pGVEM framework is implemented as follows. First we simulated $B$ bootstrap datasets based on $\hat M_p=\{\hat{\bm a}_j,\hat b_j\}_j$ estimated from the pGVEM scheme. Then we apply the pGVEM method to estimate the item parameters for each of the \addrvs{bootstrap} datasets, denoted by $\hat M_p^{(1)},\cdots,\hat M_p^{(B)}$. The standard errors are estimated by
\begin{equation*}
    \widehat{SE}_{v}=\sqrt{\frac{1}{B-1}\sum_{i=1}^B(\hat v^{(i)}-\hat v)^2},
\end{equation*}
where $v$ denotes item parameter $a_{jr}$ or $b_{jk}$ and $\hat v^{(i)}$ is its $i$th bootstrap estimate. Given that our objective is to estimate SEs rather than the distributions of the estimators,  in our study, we take the number of bootstrap samples to be $50$, which generates stable results numerically.
}
\subsection{Determining Latent Dimension}
In this section, we discuss how to select the appropriate number of latent dimensions. We propose to use the information criterion such as AIC or BIC to compare the model fit with different dimensions. In the  \addrvs{MGPCM}, direct computation of the residual sum of squares is costly. So we adopt the modified version of the information criterion where the expectation are replaced by its lower bound given by \eqref{gvem3}:
\begin{align}AIC^{*}=&2(\|\hat{\bm A}\|_0+\|\hat{\bm B}\|_0+\|\mbox{nondiag}(\hat{\bm \Sigma}_\theta)\|_0/2)-2\underline{E}(\hat M_p,\hat{\bm \xi}),\label{modaic}
\\BIC^{*}=&\log (N)(\|\hat{\bm A}\|_0+\|\hat{\bm B}\|_0+\|\mbox{nondiag}(\hat{\bm \Sigma}_\theta)\|_0/2)-2\underline{E}(\hat M_p,\hat{\bm \xi}),
\label{modbic}\end{align}
where $\hat {\bm A}=(\hat{\bm a}_1,\cdots,\hat{\bm a}_J)$ and $\hat{\bm B}=(\hat {\bm b}_1,\cdots,\hat{\bm b}_J)$ are assembled matrices of the discrimination and threshold parameters, respectively,  $\mbox{nondiag}(\hat{\bm \Sigma}_\theta)$ denotes the nondiagonal entries of $\hat{\bm \Sigma}_\theta$,  and the zero norm $||\cdot||_0$ counts the number of nonzero entries of the assembled matrix. \rvs{Here note that the term $\|\hat{\bm B}\|_0$ does not increase with dimension $D$ and it denotes the number of all effective \addrvs{threshold} parameters. In addition, since the covariance matrix $\hat{\bm\Sigma}_{\theta}$ is symmetric with unit diagonal entries, we count the effective number of parameters in $\hat{\bm\Sigma}_{\theta}$ as $\|\mbox{nondiag}(\hat{\bm \Sigma}_\theta)\|_0/2$.  The major advantage of the proposed criteria is that the lower bound of expectation is readily obtained with the updated item parameters and variation parameters, with no extra computation cost.
}

\section{Simulation Studies}
\label{sec:simu}
\subsection{Study \MakeUppercase{\romannumeral1}}

We conducted simulation studies to compare the empirical performance of the proposed pGVEM algorithm with \rvs{the EM algorithm with fixed quadrature, Metropolis-Hastings Robbins-Monro (MHRM) algorithm \citep{cai2010high}, and the stochastic EM (StEM) for MGPCM, which are implemented in the R package `mirt' \citep{chalmers2012mirt}, in terms of mean squared error and bias of the estimation together with their \addrvs{computation} time.} The simulations were conducted in the exploratory factor analysis (EFA) scenario, where no constraints on the item factor loading structure were imposed during the analysis. EFA is generally more computationally challenging than confirmatory factor analysis. In our study, we assumed that the latent covariance matrix was $I_D$ to remove scale and rotational indeterminacy, and no further assumptions on the structure of the loading matrix $A$ were made during the analysis.

After estimating the parameters by our pGVEM algorithm, we performed proper oblique rotation to allow the factors to be correlated. Many methods, including varimax, direct oblimin, quartimax, equamax, and promax \citep{browne2001overview, hendrickson1964promax}, are available for factor rotation in the literature. In our simulation study, we applied promax rotation as it is one of the most computationally efficient oblique rotation methods in large-scale factor analysis. \rvs{For the estimation implemented in the `mirt' package, we use the built-in promax rotation to obtain the estimation, and for the pGVEM estimation, we use the function promax in R, with default $m=4$, to perform the rotation after the iteration ends.}

The manipulated conditions include: (1) sample size $N=200,500$; (2) test length $J=10,20$; (3) \rvs{number of categories $K=3,6$; (5) low and high correlation among the latent traits; (6) small and large scale loadings. For each condition, a total number of \rvs{100} replicated cases were simulated. In the context of partial credit models, it is noted that the scaling of loadings plays a pivotal role in shaping the likelihood function. Specifically, when loadings are high and there exist multiple categories for partial credit scoring, the following case usually occurs: the probability of attaining the highest or lowest scores becomes disproportionately large. Consequently, this dominance of extreme scores may result in insufficient records of intermediate scores, thereby making the estimation of threshold parameters problematic. Therefore in the simulation studies, we considered two cases: (1) low scale loading: parameter $a_{jr}$ was simulated from $Unif(0.5,1)$ for all $j=1,\cdots,J,\;r=1,\cdots, D$; (2) high scale loading: parameter $a_{jr}$ was simulated from $Unif(1,2)$ for all $j=1,\cdots,J,\;r=1,\cdots, D$. 
The threshold parameters $b_{jk}$ are simulated from $N(0,1)$ for all $j=1,\cdots, J$ and $k=1,\cdots, K-1$. 
For the latent variables, they were simulated from a multivariate normal distribution with 0 mean and covariance matrix $\bm \Sigma_\theta$. The diagonal entries were fixed as 1 and off-diagonal entries were generated from a uniform distribution $Unif(0.1,0.3)$ in the low correlation case and $Unif(0.5,0.7)$ in the high correlation case. For the responses generated from the simulated model parameters, we perform simulation only on cases where for all $j=1,\cdots,J$ and $k=0,\cdots K-1$,
\begin{equation*}
    \#\{i \mid Y_{ij}=k,i=1,\cdots,N\}>0.
\end{equation*}
Here \# denoting the set counting operator. Skipping any item with all responses being 0 is necessary since the threshold parameter linked to each category of this item cannot be identified within the finite sample context.
%when no corresponding score has been recorded.
For the convergence criterion, the algorithm was terminated when the change of all item parameters between two iterations dropped below a pre-specified threshold, i.e.,
\begin{equation}
\frac1{J\times D+J\times K}\sum_{j=1}^J\Big[\sum_{r=1}^D\big(a_{jr}^{(t)}- a_{jr}^{(t-1)}\big)^2+\sum_{k=1}^{K-1}\big(b_{jk}^{(t)}-b_{jk}^{(t-1)}\big)^2\Big]<10^{-5}.
\label{convergence}
\end{equation}

The estimation errors are presented separately in the form of mean squared error and bias for the discrimination and threshold parameters and the covariance matrix, averaged across the test items:
\begin{align*}
Bias_a=\frac1{JD}\sum_{j=1}^{J}\sum_{r=1}^{D} \hat { a}_{jr}-a_{jr},\quad &
MSE_a=\frac1{JD}\sum_{j=1}^{J}\|\bm a_j-\hat {\bm a}_j\|^2_2;\\
Bias_a=\frac1{J(K-1)}\sum_{j=1}^{J}\sum_{k=1}^{K-1} \hat { b}_{jr}-b_{jr},\quad&
MSE_b=\frac{1}{J(K-1)}\sum_{j=1}^J\sum_{k=1}^{K-1}(b_{jk}-\hat b_{jk})^2;\\
Bias_\Sigma=\frac2{D(D-1)}\sum_{l<h}\hat\Sigma_{hl}-\Sigma_{hl},\quad&
MSE_\Sigma=\frac2{D(D-1)}\|\bm \Sigma_\theta-\hat{\bm \Sigma}_\theta\|_F^2.
\end{align*}
Note that %the parameters here are all regularized: 
here both the true and estimated discrimination parameters have been rotated by the promax rotations. The number of Markov chain samples drawn in the MHRM algorithm was by default $5,000$ in the R package ‘mirt’. The convergence criterion and optimizer were all set as default in the `mirt' package. } 

%aj sim Uni(0.5,1)
\begin{figure}
\begin{center}
\subfigure{\includegraphics[width=6.8in,height=3in]{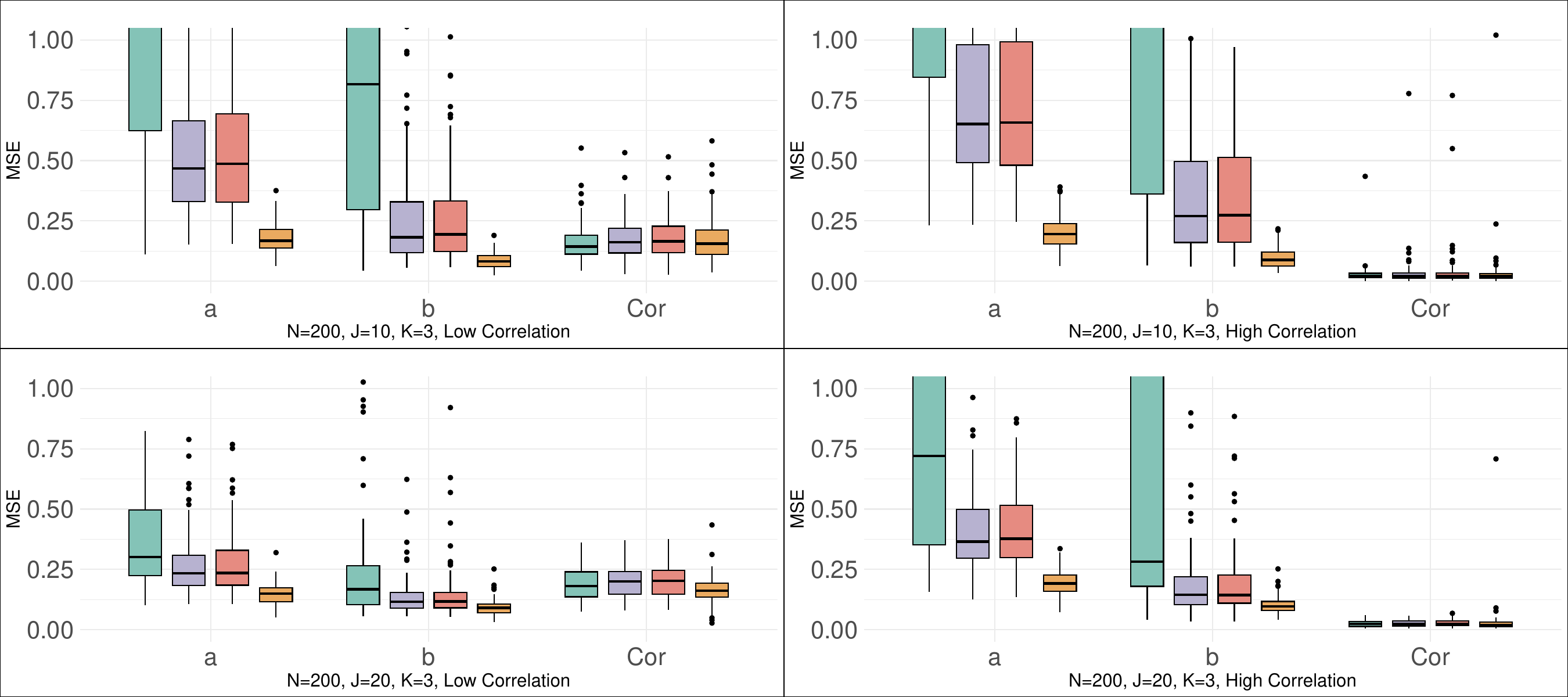}}
\subfigure{\includegraphics[width=6.8in,height=3in]{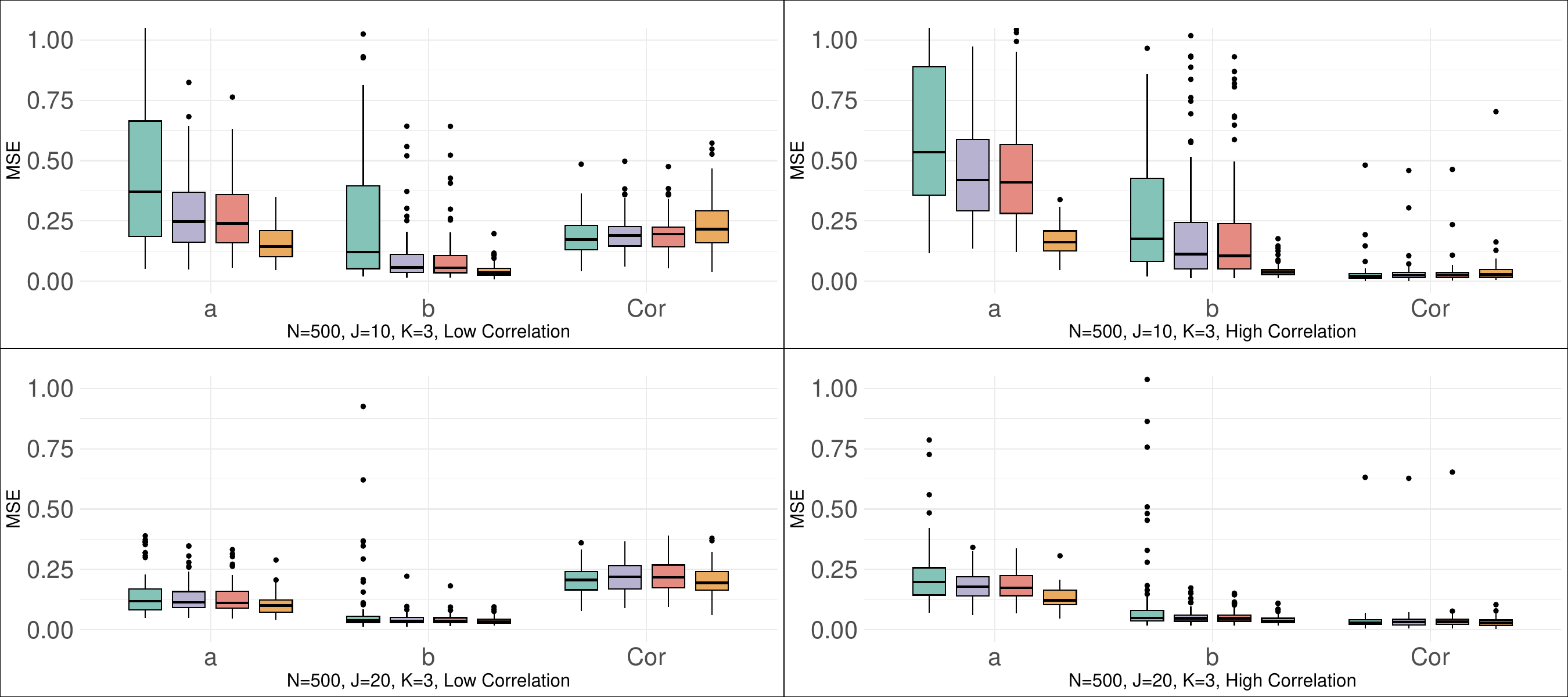}}
\subfigure{\includegraphics[width=5in,height=0.4in]{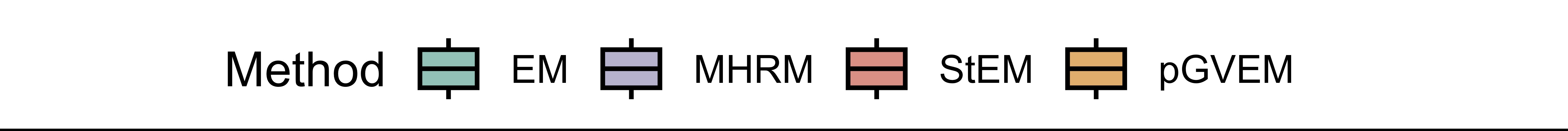}}
\end{center}
\caption{\rvs{Mean Squared Error of the estimation for the \addrvs{MGPCM} of 3 categories from exploratory factor analysis with small scale loadings using different methods.}\label{fig:mse_k3}}
\end{figure}

\begin{figure}
\begin{center}
\subfigure{\includegraphics[width=6.8in,height=3.2in]{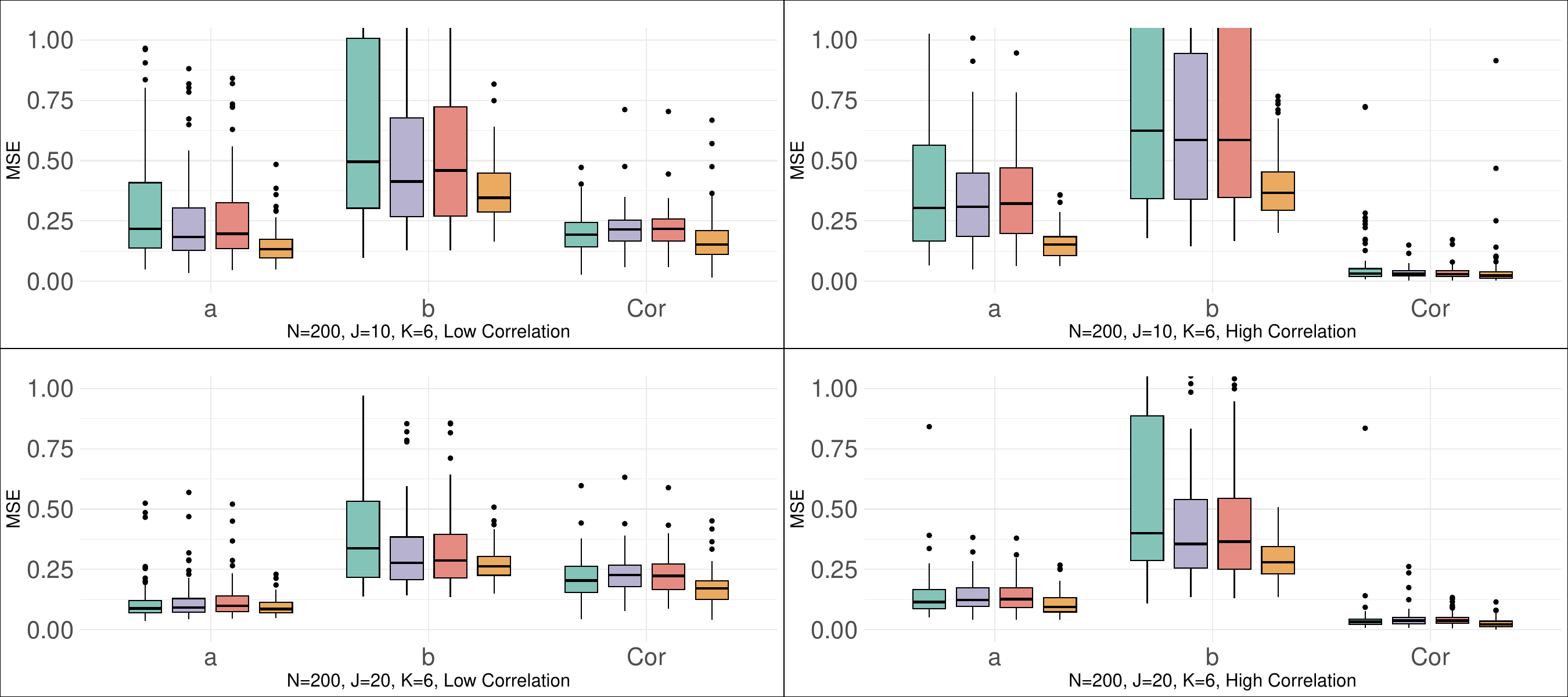}}
\subfigure{\includegraphics[width=6.8in,height=3.2in]{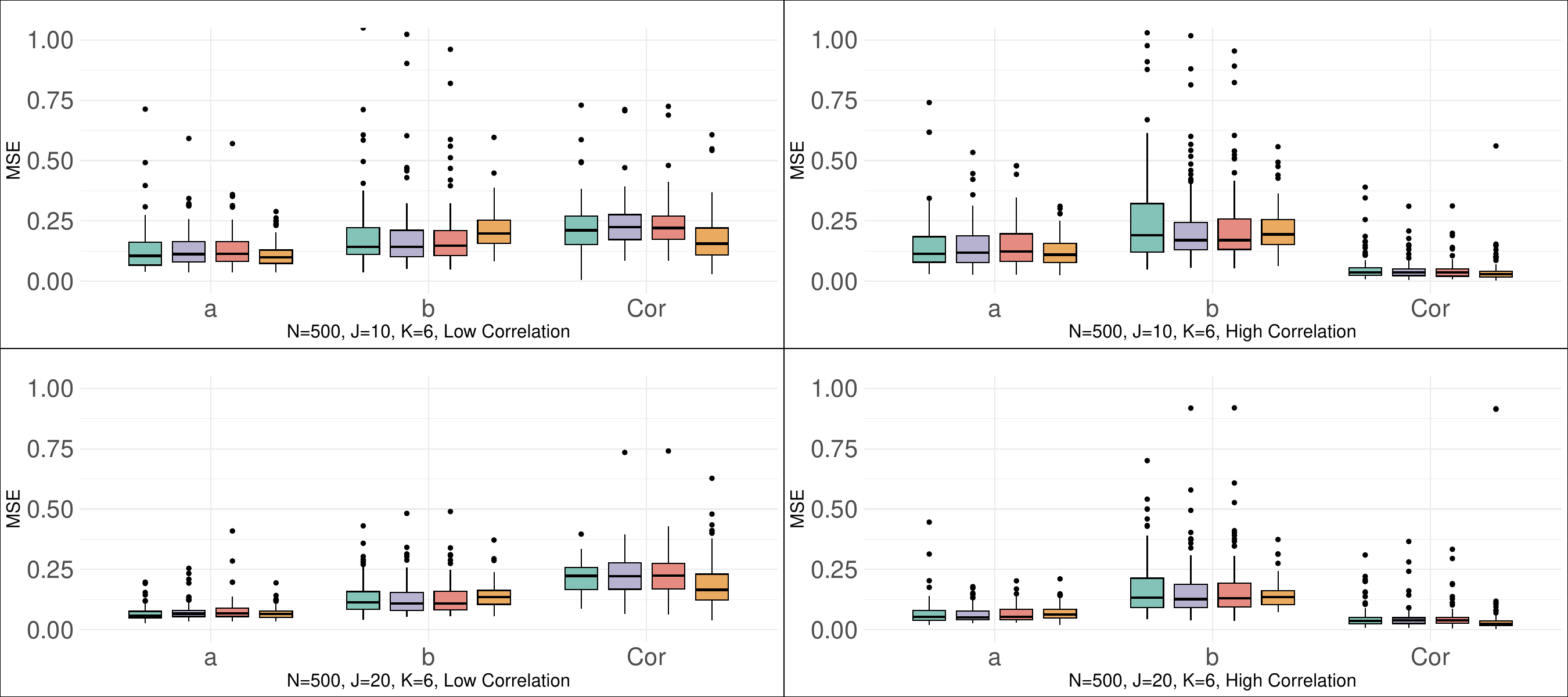}}
\subfigure{\includegraphics[width=4in,height=0.3in]{method.jpeg}}
\end{center}
\caption{\rvs{Mean Squared Error of the estimation for \addrvs{MGPCM} of 6 categories from exploratory factor analysis with small scale loadings using different methods.}\label{fig:mse_k5}}
\end{figure}

\begin{figure}
\begin{center}
\subfigure{\includegraphics[width=6.8in,height=3.2in]{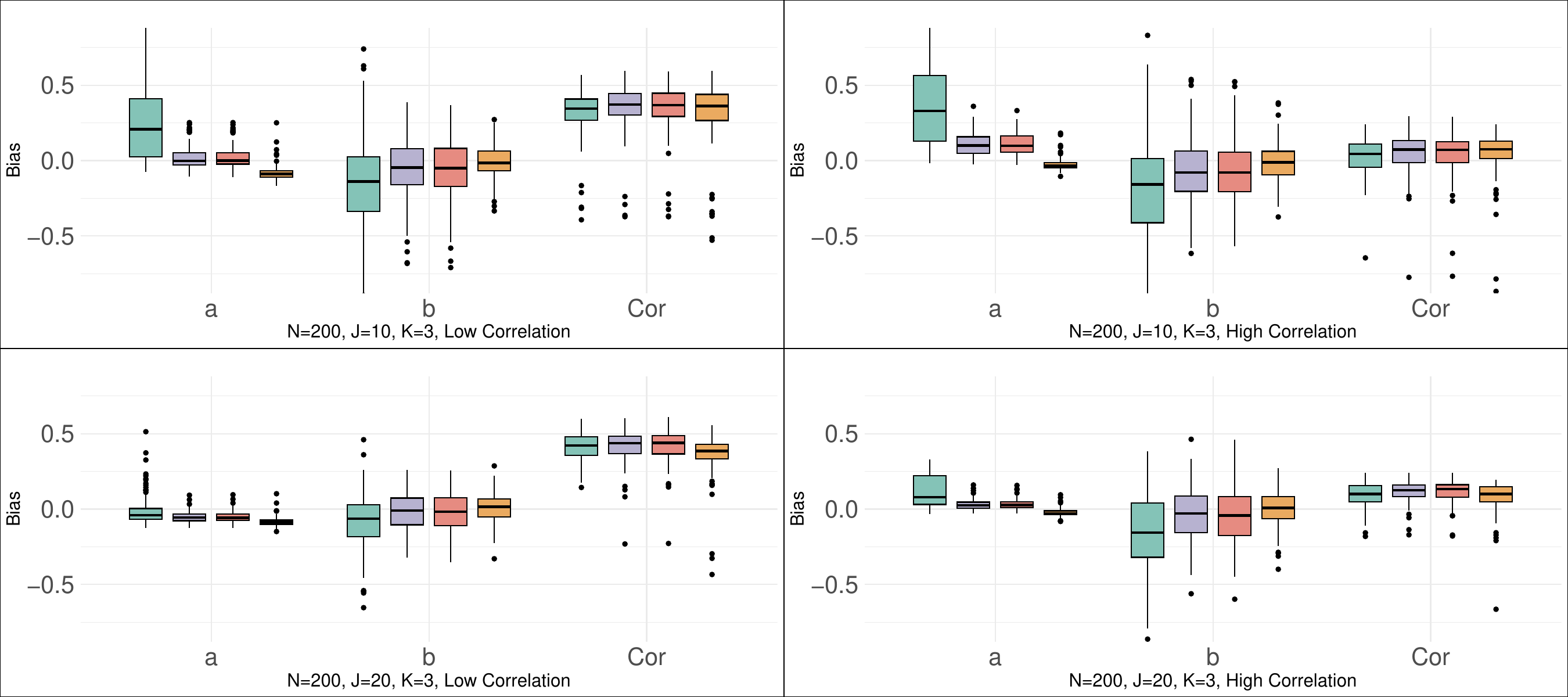}}
\subfigure{\includegraphics[width=6.8in,height=3.2in]{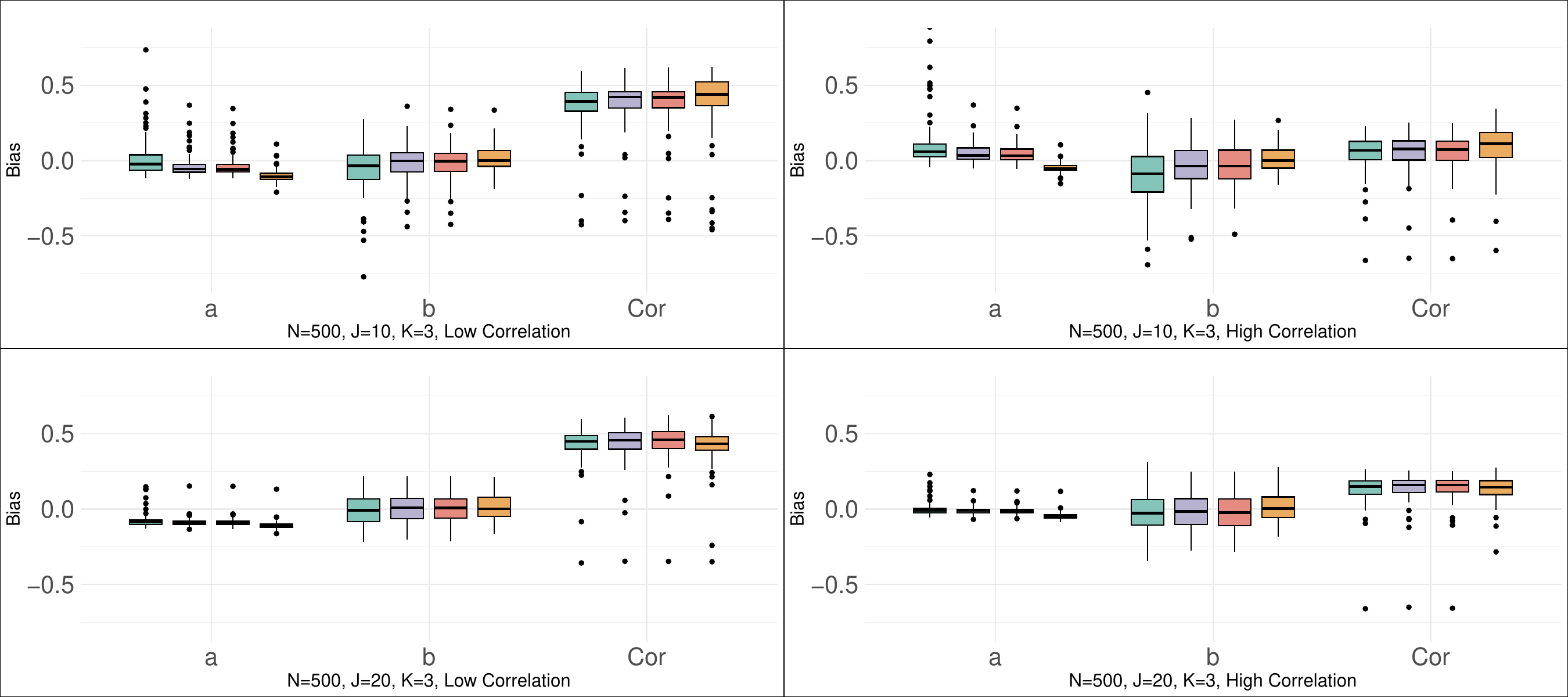}}
\subfigure{\includegraphics[width=4in,height=0.3in]{method.jpeg}}
\end{center}
\caption{\rvs{Bias of the estimation for the \addrvs{MGPCM} of 3 categories from exploratory factor analysis with small scale loadings using different methods.}\label{fig:bias_k3}}
\end{figure}

\begin{figure}
\begin{center}
\subfigure{\includegraphics[width=6.8in,height=3.2in]{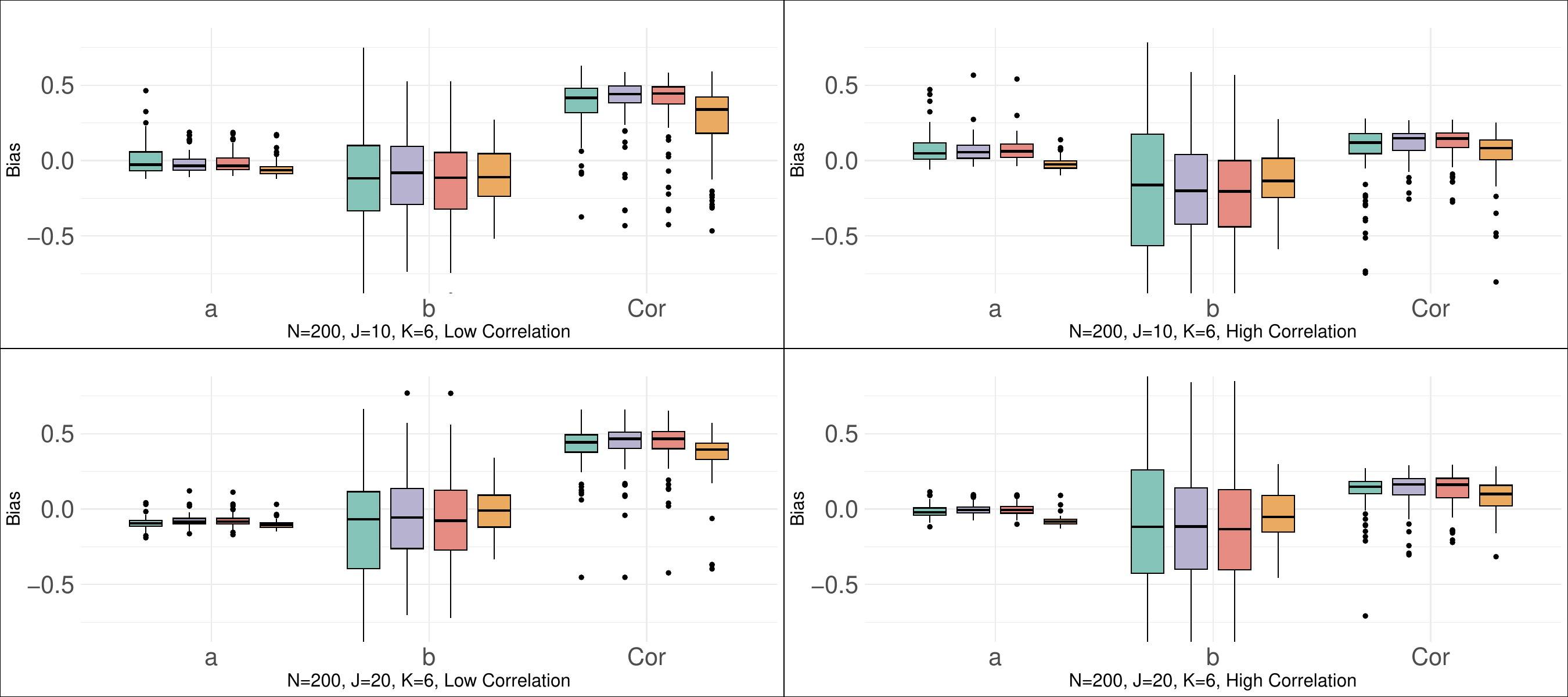}}
\subfigure{\includegraphics[width=6.8in,height=3.2in]{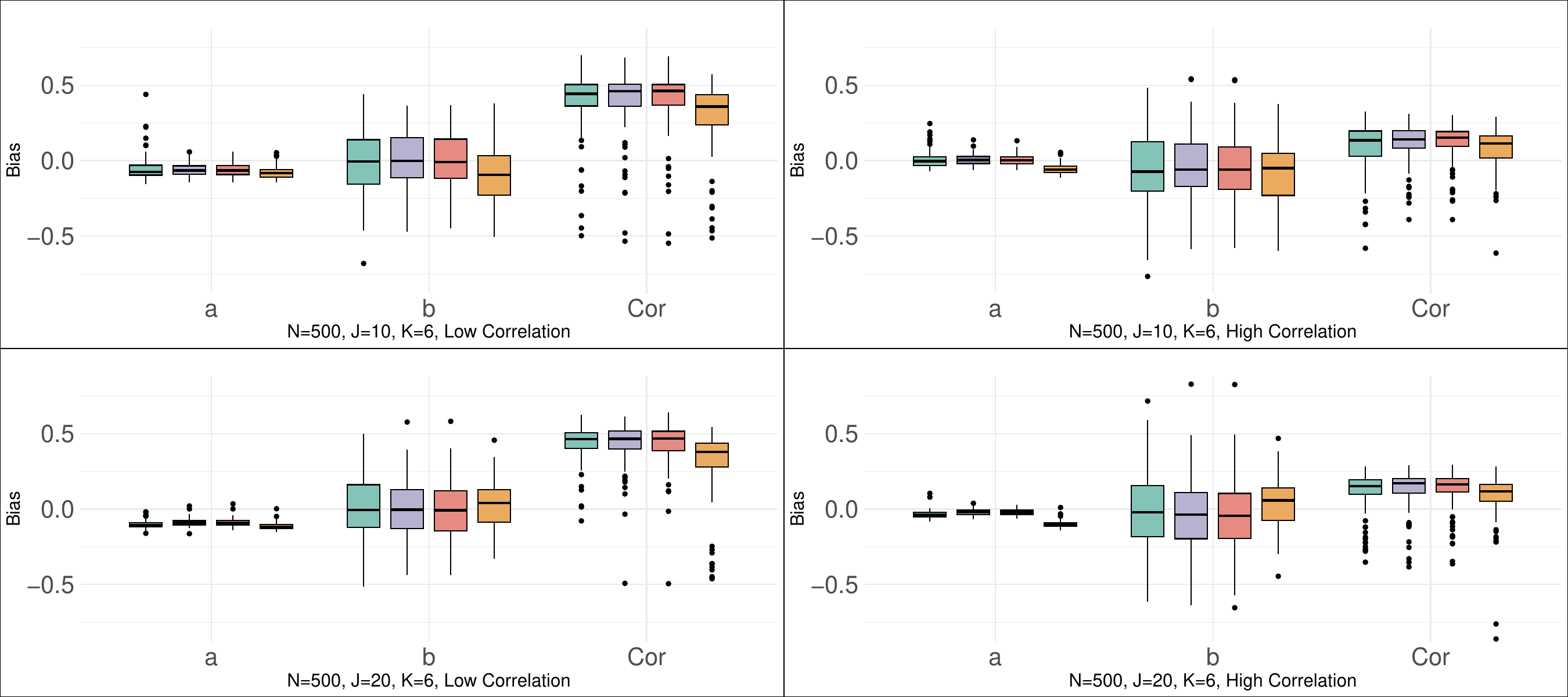}}
\subfigure{\includegraphics[width=4in,height=0.3in]{method.jpeg}}
\end{center}
\caption{\rvs{Bias of the estimation for the \addrvs{MGPCM} of 6 categories from exploratory factor analysis with small scale loadings using different methods.}\label{fig:bias_k5}}
\end{figure}

\begin{figure}
\begin{center}
\subfigure{\includegraphics[width=6.8in,height=3in]{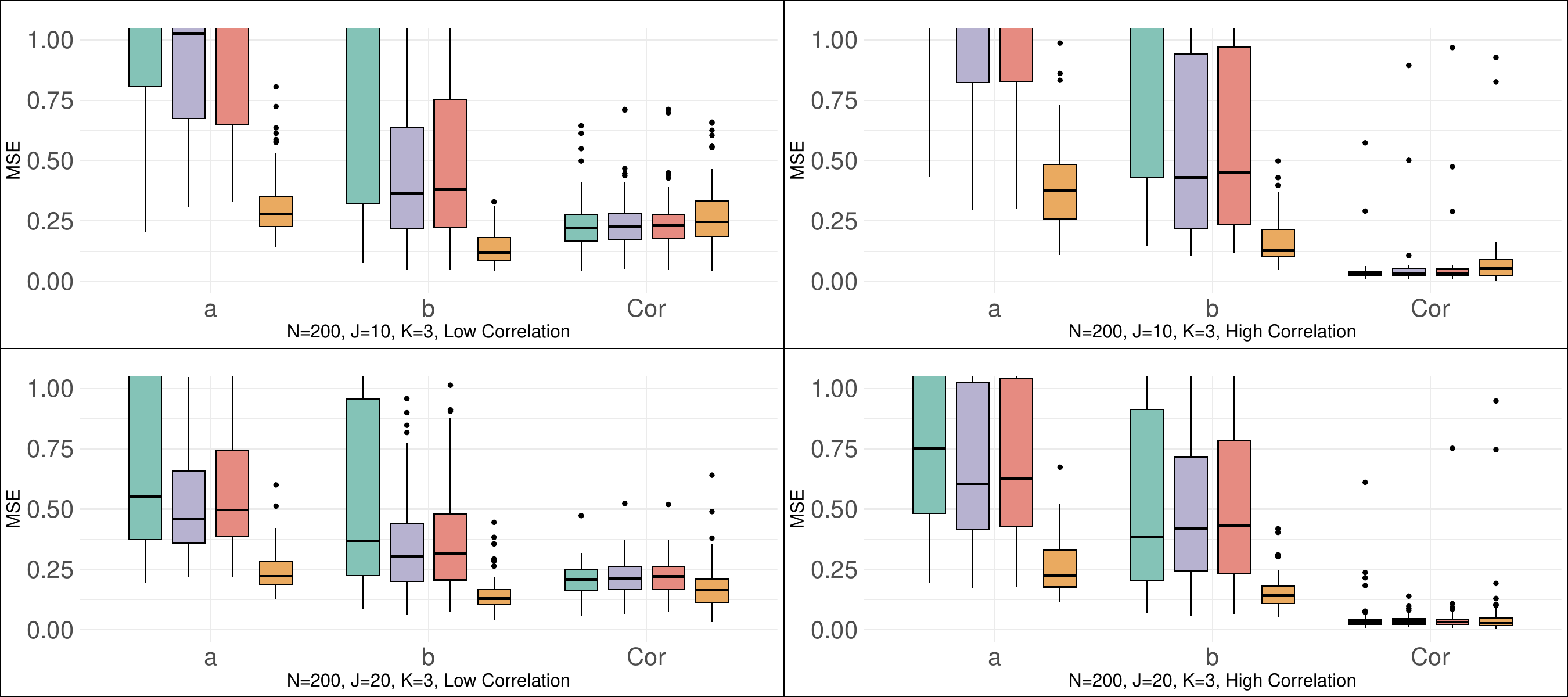}}
\subfigure{\includegraphics[width=6.8in,height=3in]{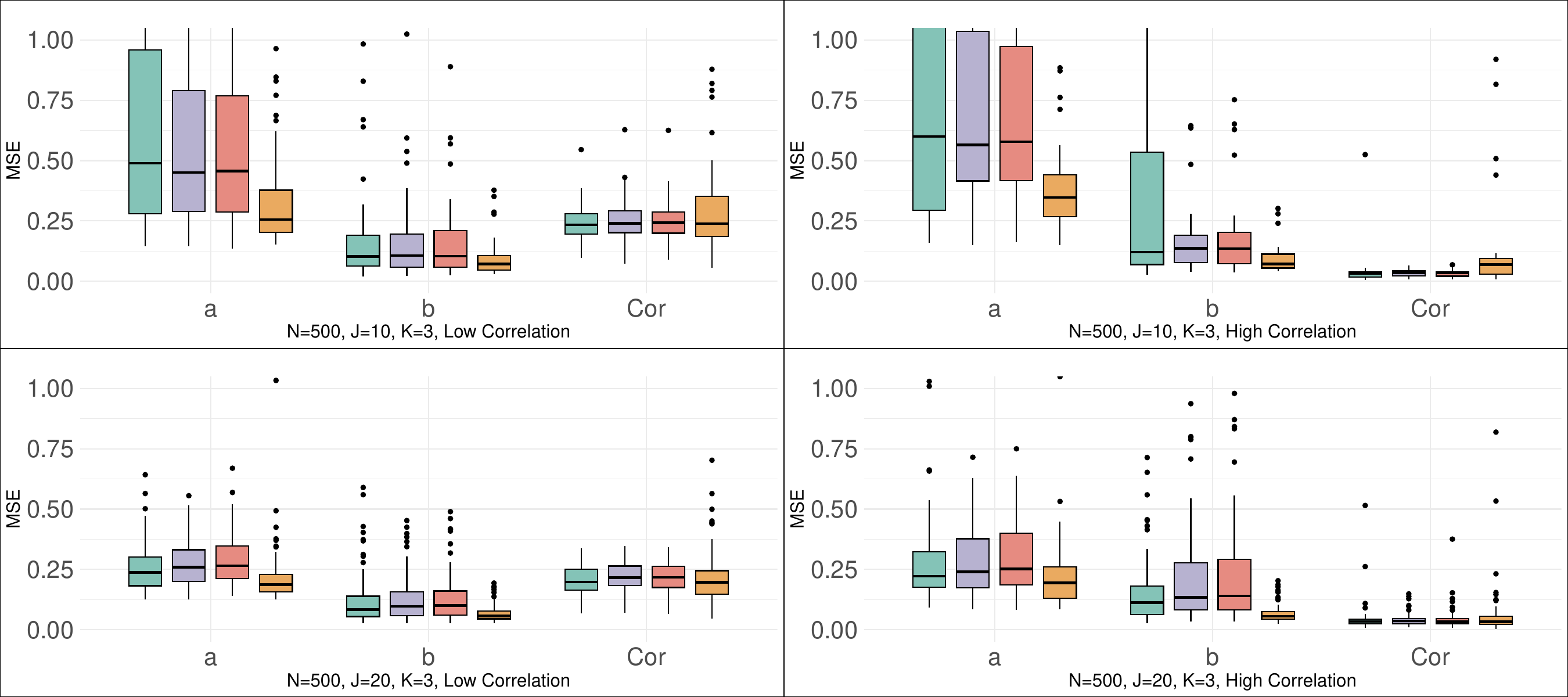}}
\subfigure{\includegraphics[width=5in,height=0.4in]{method.jpeg}}
\end{center}
\caption{\rvs{Mean Squared Error of the estimation for the \addrvs{MGPCM} of 3 categories from exploratory factor analysis with large scale loadings using different methods.}\label{fig:mse_k3p}}
\end{figure}

\begin{figure}
\begin{center}
\subfigure{\includegraphics[width=6.8in,height=3.2in]{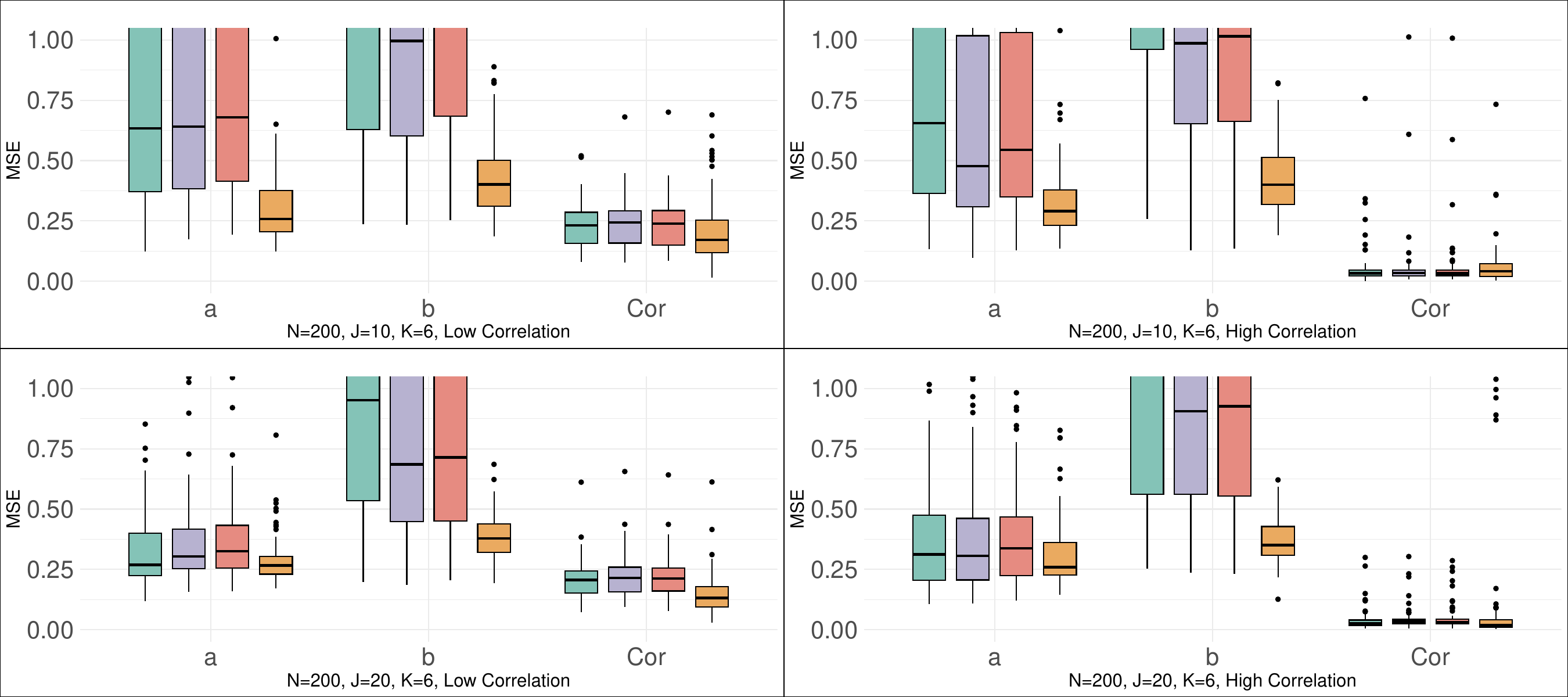}}
\subfigure{\includegraphics[width=6.8in,height=3.2in]{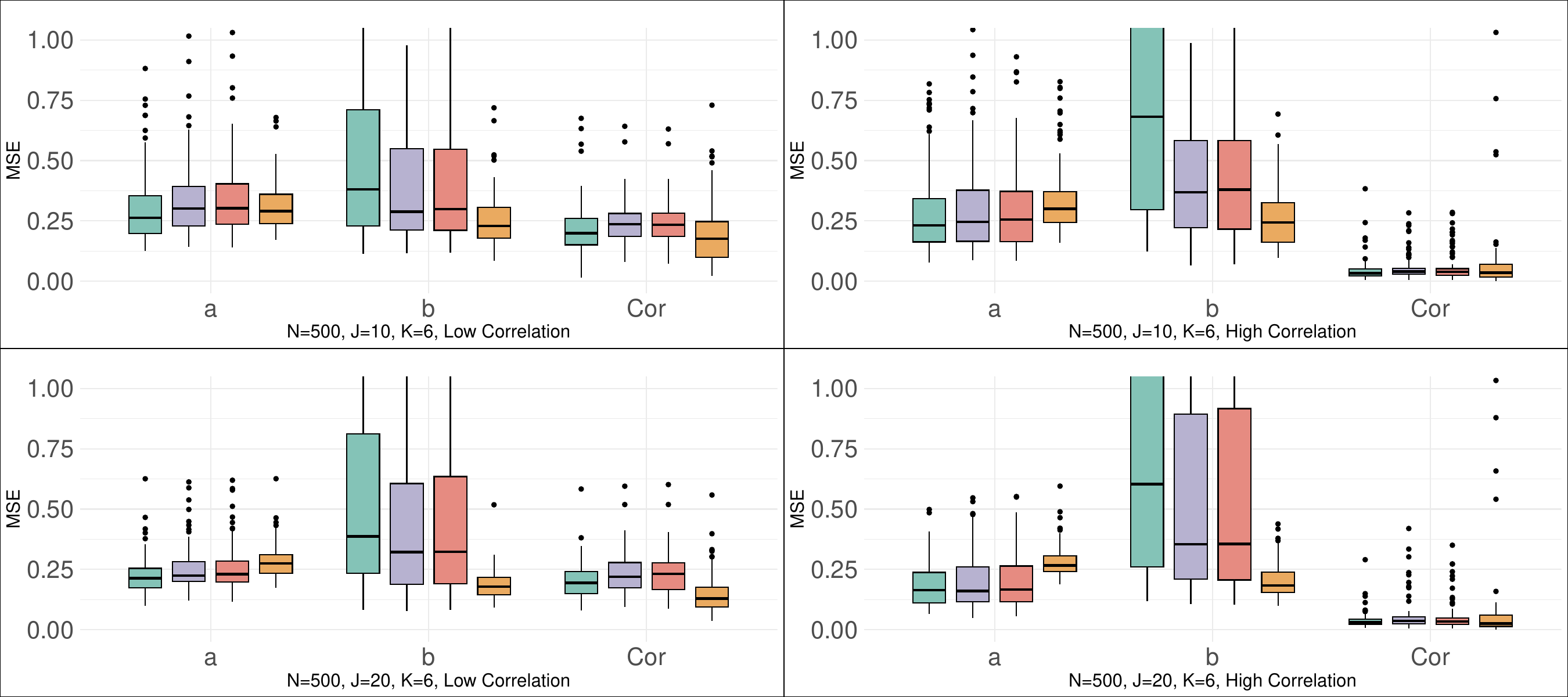}}
\subfigure{\includegraphics[width=4in,height=0.3in]{method.jpeg}}
\end{center}
\caption{\rvs{Mean Squared Error of the estimation for \addrvs{MGPCM} of 6 categories from exploratory factor analysis with large scale loadings using different methods.}\label{fig:mse_k5p}}
\end{figure}

\begin{figure}
\begin{center}
\subfigure{\includegraphics[width=6.8in,height=3.2in]{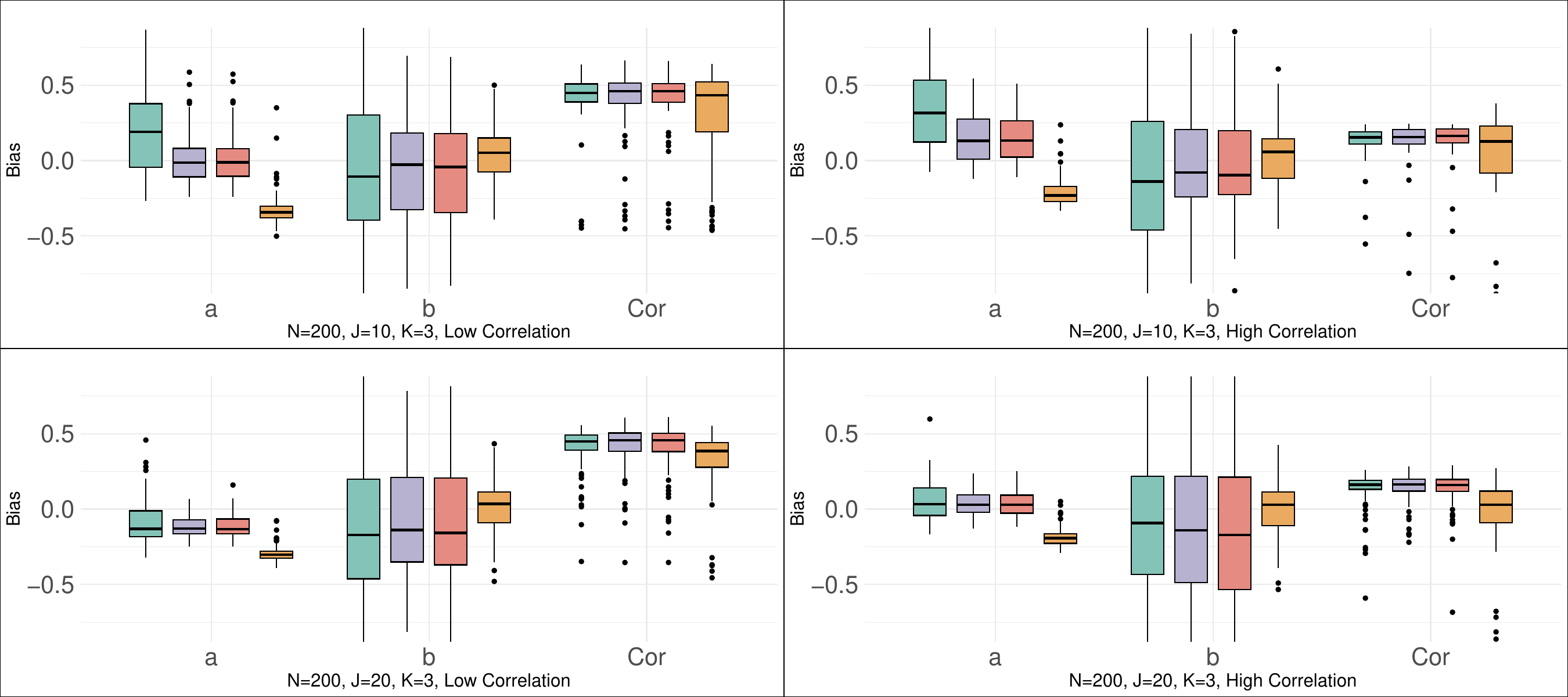}}
\subfigure{\includegraphics[width=6.8in,height=3.2in]{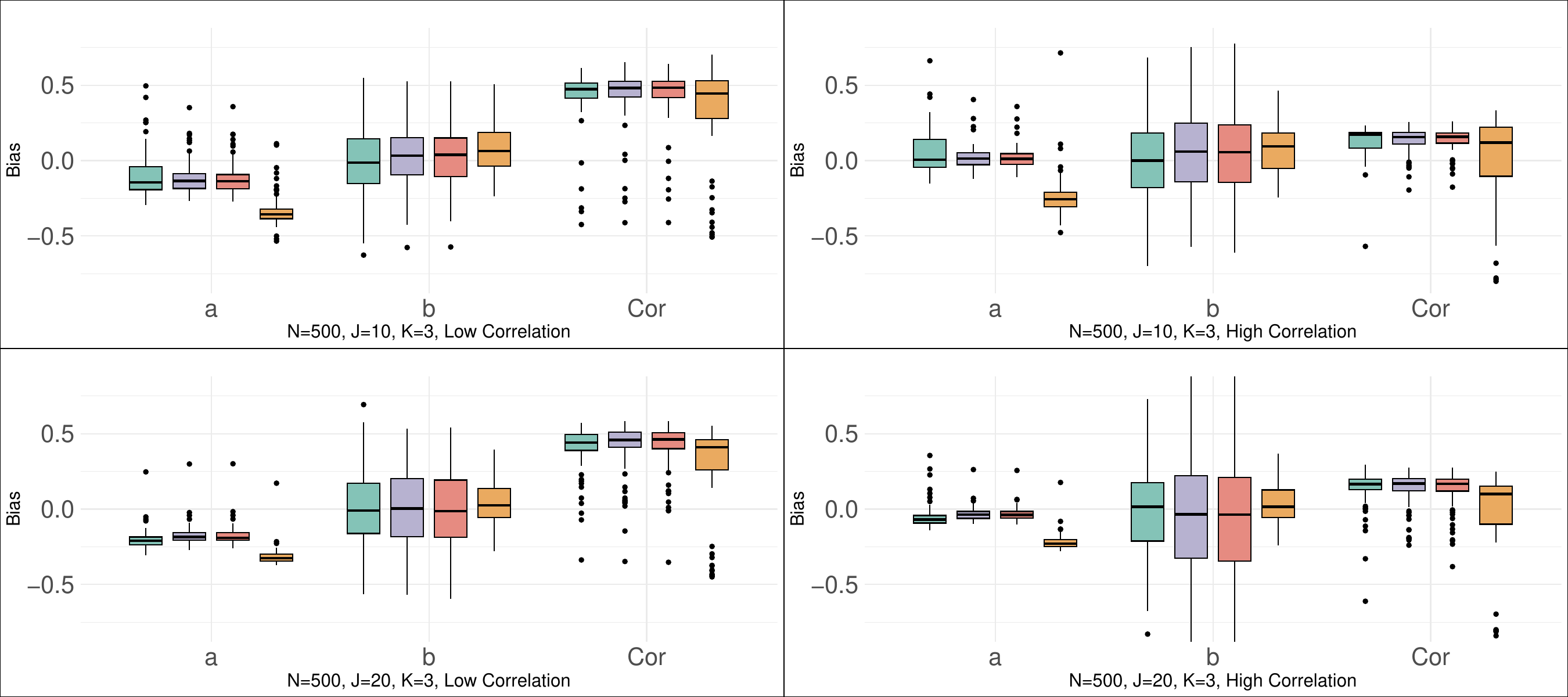}}
\subfigure{\includegraphics[width=4in,height=0.3in]{method.jpeg}}
\end{center}
\caption{\rvs{Bias of the estimation for the \addrvs{MGPCM} of 3 categories from exploratory factor analysis with large scale loadings using different methods.}\label{fig:bias_k3p}}
\end{figure}

\begin{figure}
\begin{center}
\subfigure{\includegraphics[width=6.8in,height=3.2in]{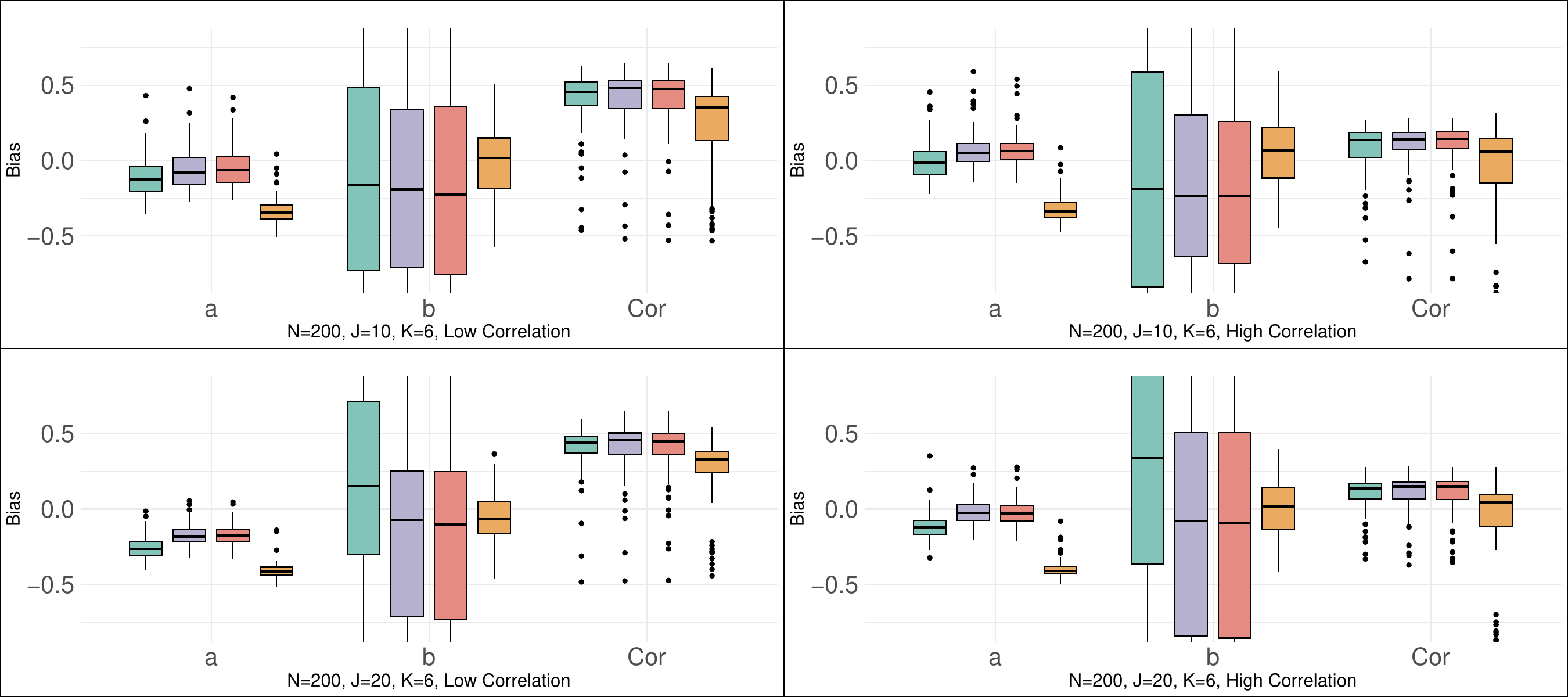}}
\subfigure{\includegraphics[width=6.8in,height=3.2in]{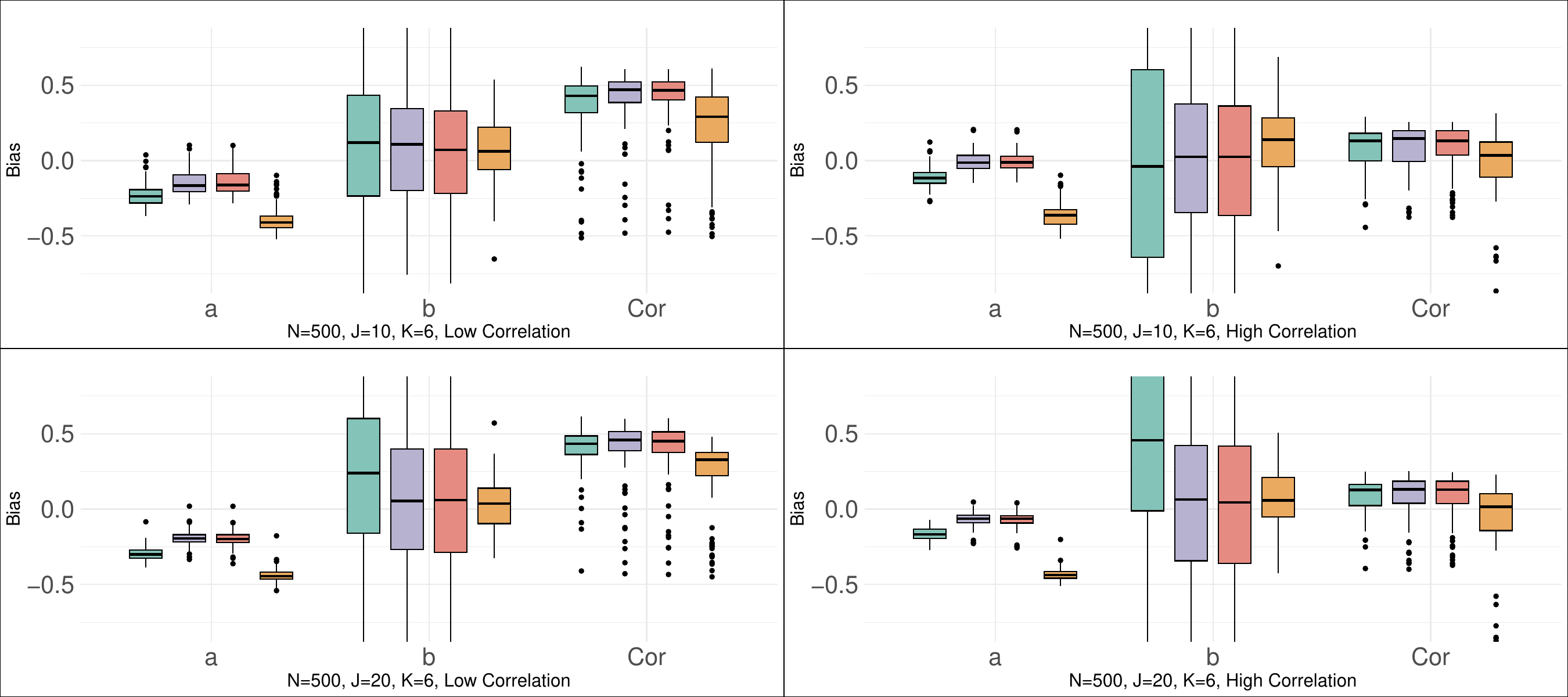}}
\subfigure{\includegraphics[width=4in,height=0.3in]{method.jpeg}}
\end{center}
\caption{\rvs{Bias of the estimation for the \addrvs{MGPCM} of 6 categories from exploratory factor analysis with large scale loadings using different methods.}\label{fig:bias_k5p}}
\end{figure}

Under the setting of small scale loadings (i.e., $a\sim Unif(0.5, 1)$), Figures \ref{fig:mse_k3} and \ref{fig:mse_k5} show the MSE of the four estimation methods for 3-category and 6-category cases, respectively. Each box represents the distribution of errors from 100 replications. We truncated the scale of the $y$-axis of the plot to make it easier to compare the estimation precision across different scenarios. We present the full version of the boxplots in Appendix B.
\rvs{Overall, our method provides more stable and accurate estimates of the MGPCM, as seen from both figures. 
The observed results indicate a reduced variability in estimation errors for the model parameters when comparing pGVEM to the other methods.  
Both MHRM and StEM exhibit better stability and accuracy compared to the standard EM algorithm. 
Notably, pGVEM demonstrates good stability, particularly evident when the sample size $N$ is 200. 
Additionally, it is noteworthy that as the number of categories increases, the estimation of the threshold parameters becomes more challenging, 
%the precision of estimating difficulty parameters diminishes — 
an anticipated result given that the model becomes more complicated for multi-category cases.
%, which then requires a larger sample size for satisfactory accuracy.
Interestingly, despite a decrease in accuracy,
the proposed pGVEM method exhibits a comparatively modest increase in variability in many cases compared with alternative methods, indicating the capability of the pGVEM method to handle more complex scenarios. 
%the increased variability is comparatively modest in many situations when contrasted with alternative methods.
%This observation underscores the capability of our proposed pGVEM method to effectively handle more complex scenarios. 
Furthermore, we present the bias of the estimation using the four different methods in Figures \ref{fig:bias_k3} and \ref{fig:bias_k5} for the considered cases. In general, the bias observed in pGVEM estimation tends to be more moderate across various cases, particularly with regard to the threshold parameter. 
%However, in terms of the discrimination parameters, pGVEM exhibits a tendency to underestimate the scale. This phenomenon can be interpreted in light of our estimation framework, which relies on the relative scale of  $k\bm a_j^\prime\bm\theta_i-b_{jk}$ across different categories. When this difference is relatively small, pGVEM tends to provide more conservative estimations of the loading, attributing a greater proportion of the variance in the item response function to the difficulty parameter.

%In the following we show the case with large scale in the loading, that is the discriminination parameters $a_{jr}$ is simulated from $Unif(1,2)$ for all $j=1,\cdots,J,r=1,\cdots,D$. 
Under the setting of large scale loadings (i.e., $a\sim Unif(1,2)$), the MSE and bias results are presented in Figures \ref{fig:mse_k3p}-\ref{fig:bias_k5p}. We can see from the simulation results that in this setting the error of estimation gets larger in most cases. The reason is that high level of loading increases the frequency of extreme scores, thus making the estimation more challenging. Yet pGVEM still outperforms the other methods with the regime of small sample size or low correlation. Furthermore, in terms of the recovery of the threshold parameter, pGVEM provides estimates with less bias and error. It is noteworthy that in this scenario, it occurs more frequently that for some item, there is no record of certain category from any individual, leading to us discarding such cases from the analysis. This suggests that when dealing with multiple categories of responses, the discrimination parameters may tend to be small for model interpretability.}

%aj sim Uni(1,2)

\begin{figure}
\begin{center}
\includegraphics[width=6in,height=3in]{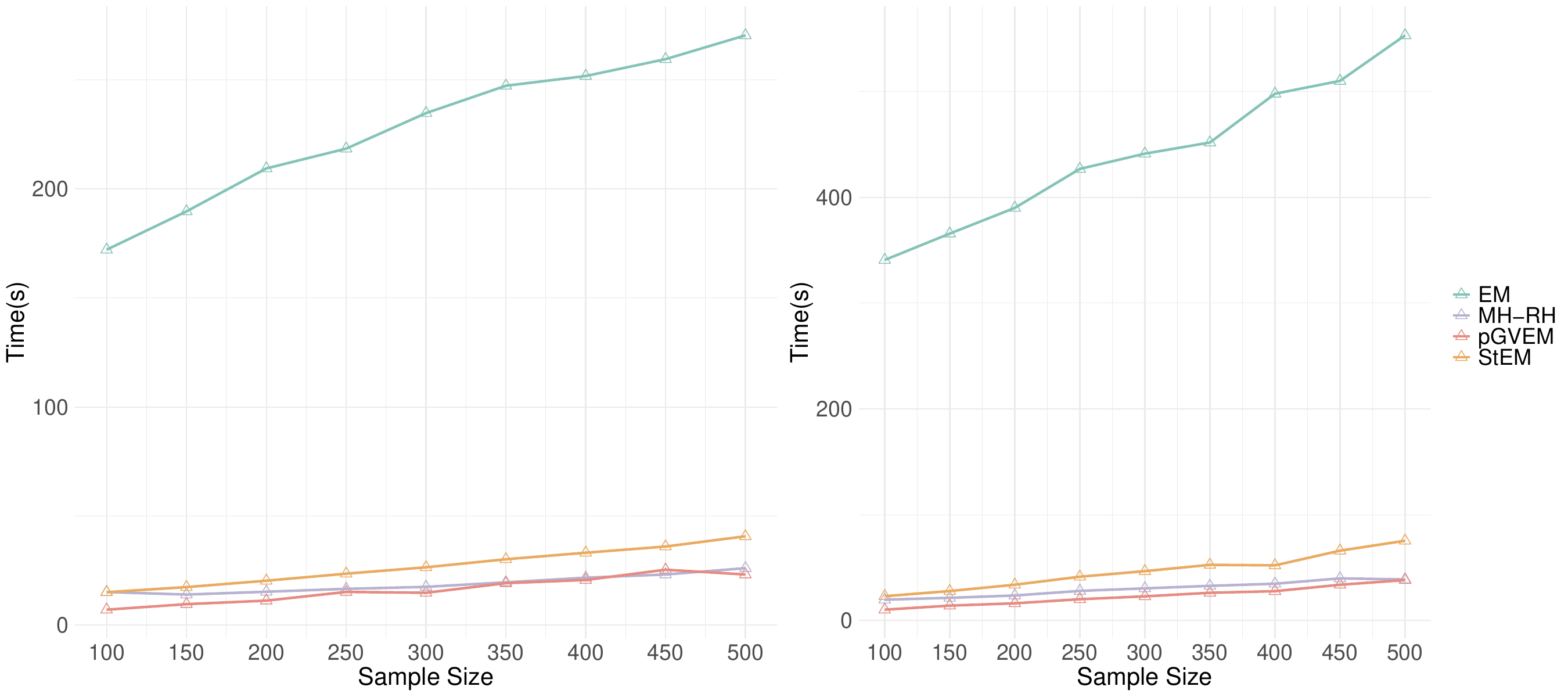}
\end{center}
\caption{\rvs{Computational Time Comparison for Four Methods with $K=6$: Left Side depicts $J=10$, while the Right Side corresponds to $J=20$.}\label{fig:time}}
\end{figure}

\rvs{In Figure \ref{fig:time} we present the averaged computation time of the four methods under different settings of sample size $N$ and test length $J$. The results exhibit similarity across the cases, and for brevity, we displayed the case where the discrimination parameters are simulated from $Unif(0.5,1)$ and correlation coefficient from $Unif(0.1,0.3)$. We take a total number of 6 categories. It is obvious that, compared with pGVEM, the traditional EM algorithm is inefficient especially when the sample size is large. Also, pGVEM is slightly faster than the stochastic EM algorithm and achieves similar computation efficiency compared with MH-RM algorithm in the displayed case. The computational efficiency of our pGVEM algorithm makes it possible to provide fast estimation on large datasets. }

The results of our study demonstrate the superiority of the pGVEM algorithm over the traditional EM algorithm along with MH-RM and StEM in terms of parameter recovery and computational efficiency. Specifically, the pGVEM algorithm achieves comparable, and in some cases, superior parameter recovery compared to the other algorithms. Moreover, pGVEM generates fewer estimation outliers, particularly in situations where the sample size and test length are high.
%Furthermore, pGVEM is much faster computationally than the EM algorithm , as demonstrated in Figure 3, which shows the average computation time of 20 replications with varying test length and sample size. The results indicate that pGVEM's computation time is only a fraction of that of EM, and its computation time only slightly increases when test length and sample size increase. In contrast, doubling test length almost doubles the computation time with the traditional EM algorithm.

As a side check, we compared the pGVEM algorithm with the GVEM algorithm proposed by \cite{cho2021gaussian} for the special case of M2PL. The comparison \rvs{was made under identical setting in this section, except for that we take $k=2$}. The results, presented in Appendix B, indicate that our algorithm performs similarly to \cite{cho2021gaussian} with binary data.
Overall, our findings demonstrate that the pGVEM algorithm is a robust and efficient method for parameter recovery for the  \addrvs{MGPCM}. It can also be applied to other models, including M2PL, with similar success. The results suggest that the pGVEM algorithm may be a valuable tool for researchers seeking to analyze complex data structures efficiently and accurately.

\subsection{Study \MakeUppercase{\romannumeral2}}

\begin{figure}
\begin{center}
\subfigure[Bias of Standard Error]{\includegraphics[width=4.8in,height=1.75in]{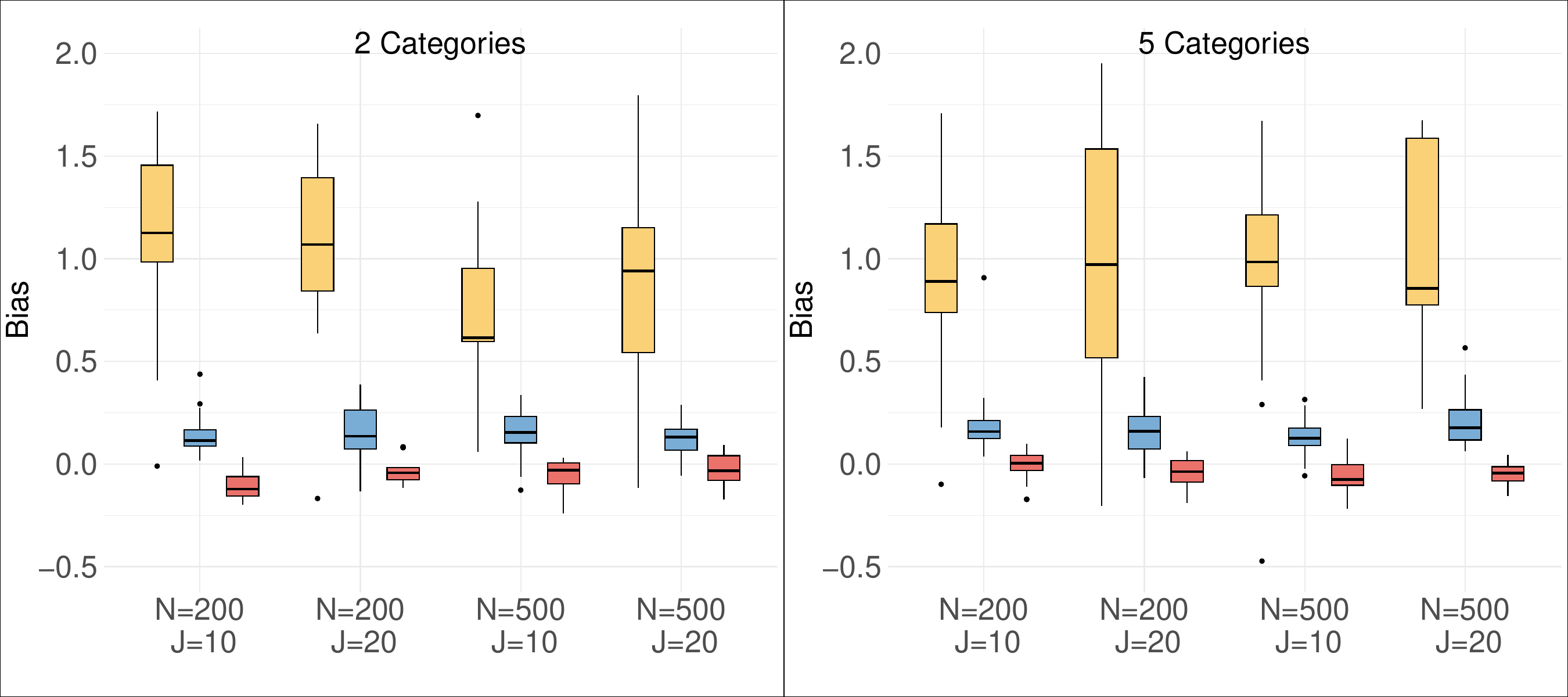}}
\subfigure[Relative Bias of Standard Error]{\includegraphics[width=4.8in,height=1.75in]{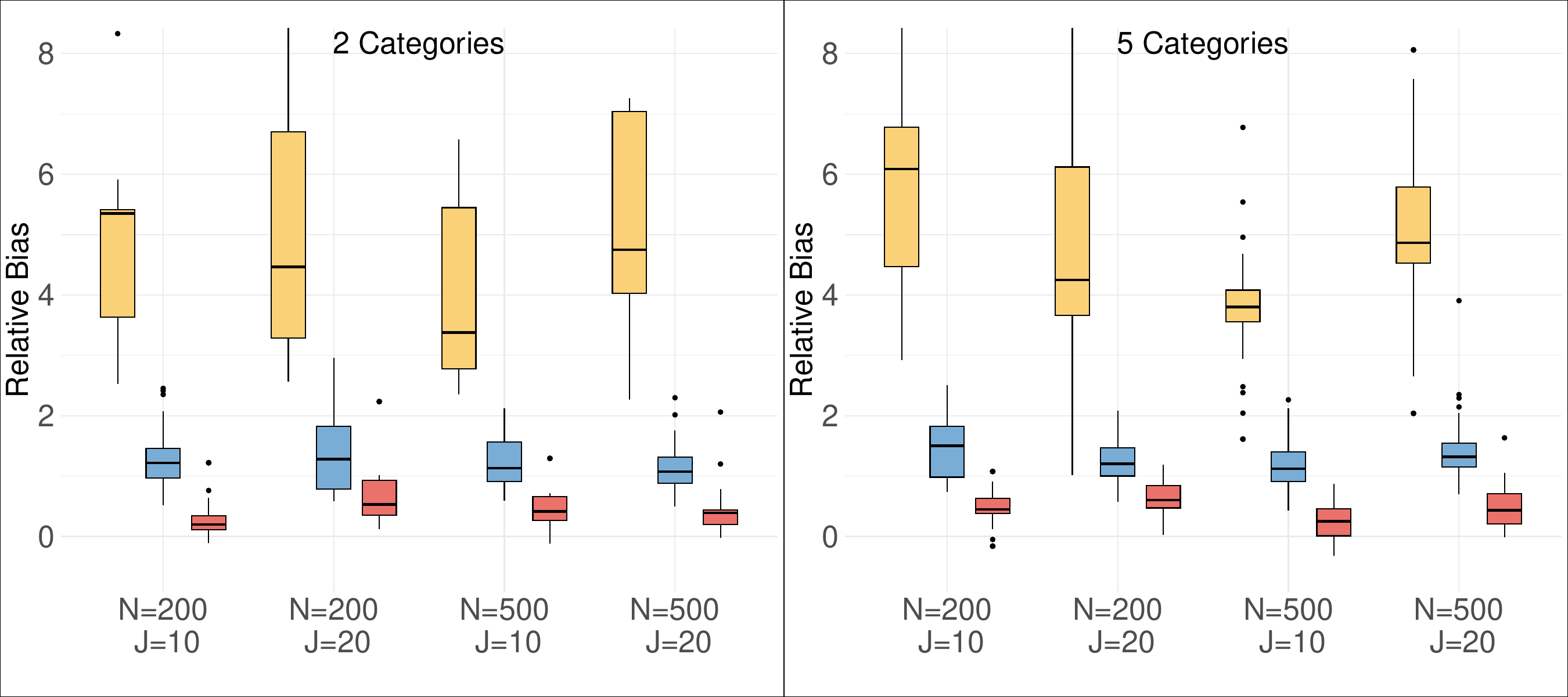}}
\subfigure[Estimated Standard Error]{\includegraphics[width=4.8in,height=1.75in]{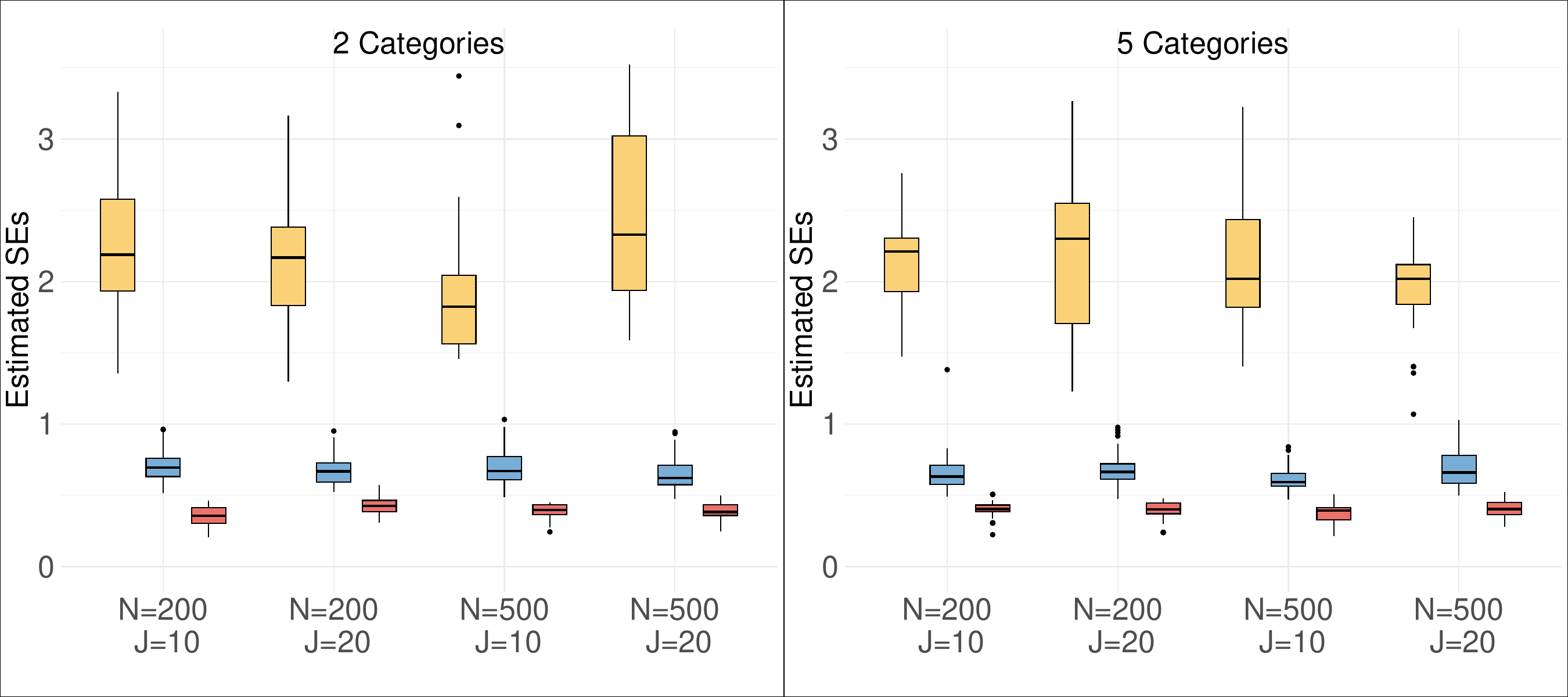}}
\subfigure{\includegraphics[width=3in,height=0.25in]{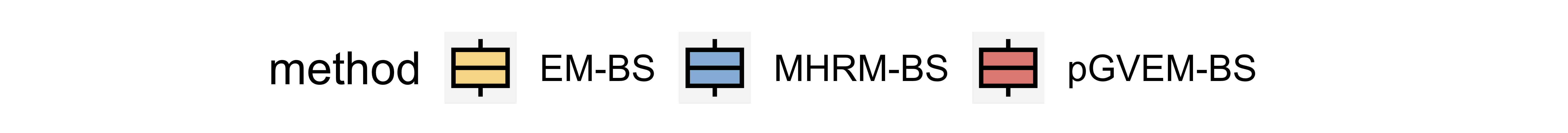}}
\end{center}
\caption{\rvs{Standard error assessment of estimation for the \addrvs{MGPCM} from exploratory factor analysis using different methods with $D=3$ in low factor correlation setting.}\label{fig:se2}}
\end{figure}

\begin{figure}
\begin{center}
\subfigure[Bias of Standard Error]{\includegraphics[width=4.8in,height=1.75in]{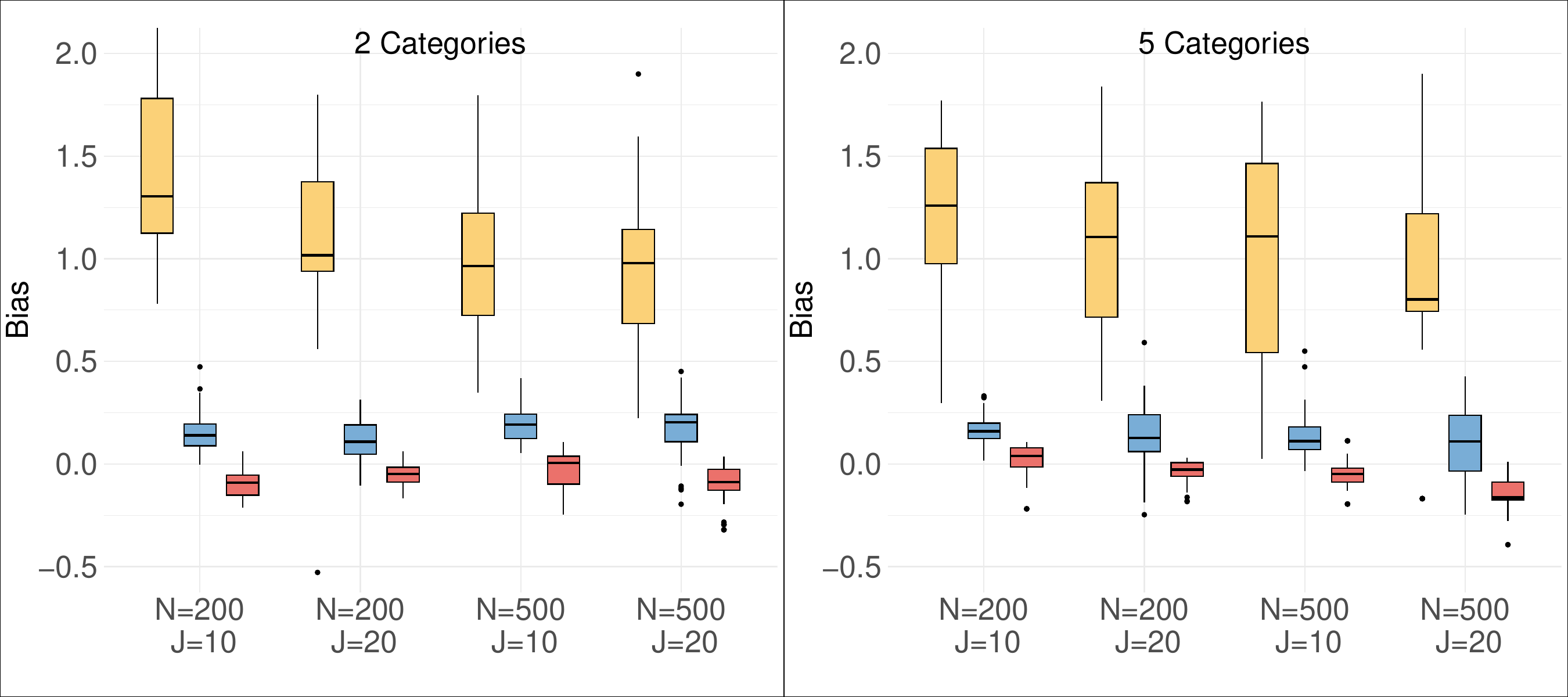}}
\subfigure[Relative Bias of Standard Error]{\includegraphics[width=4.8in,height=1.75in]{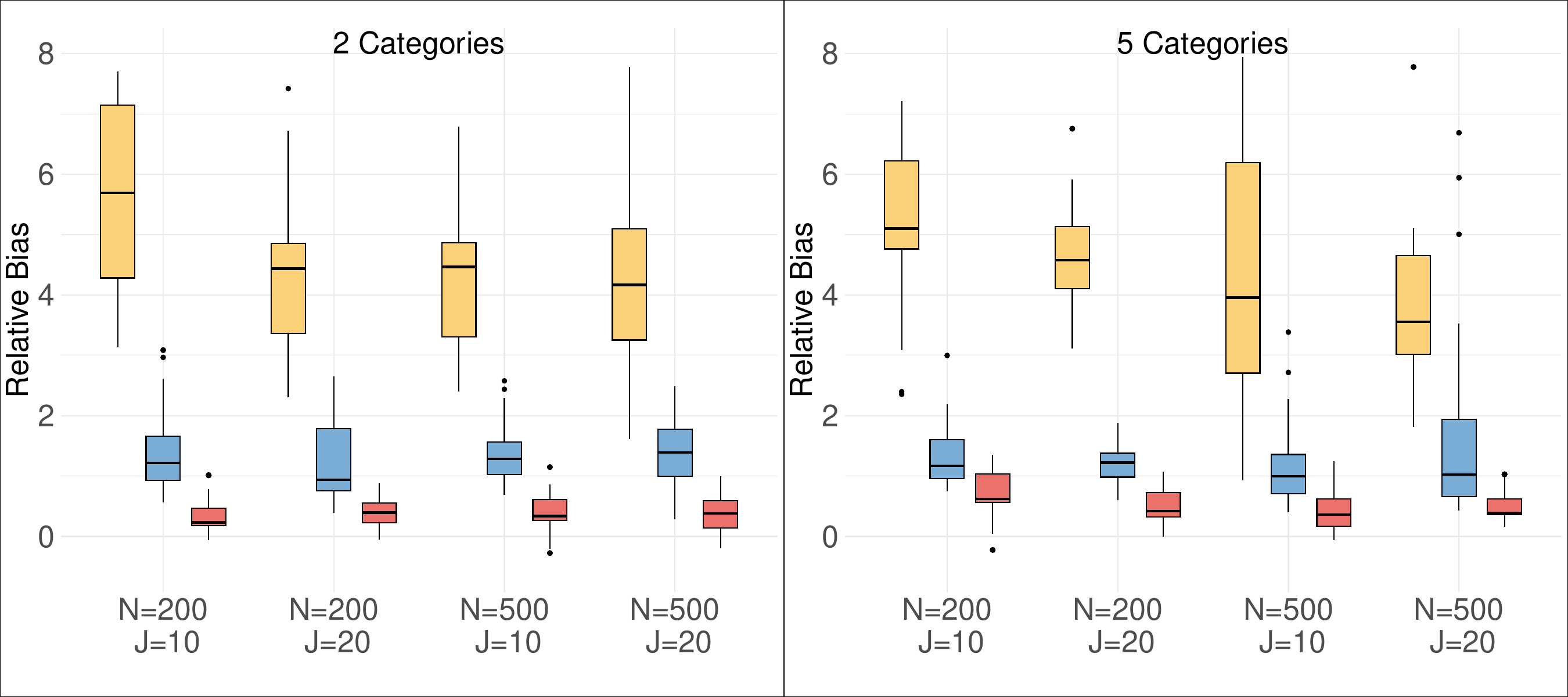}}
\subfigure[Estimated Standard Error]{\includegraphics[width=4.8in,height=1.75in]{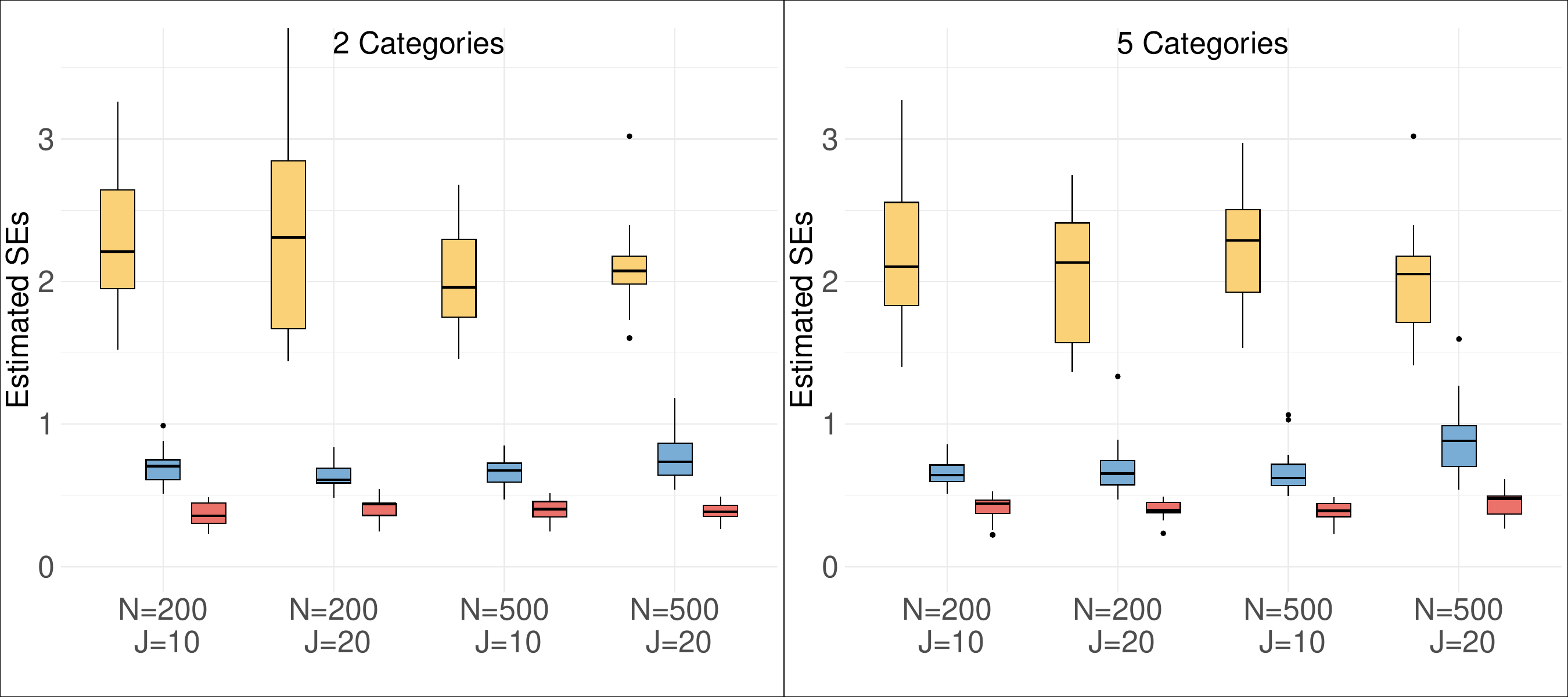}}
\subfigure{\includegraphics[width=3in,height=0.25in]{d3h_se_method.jpeg}}
\end{center}
\caption{\rvs{SE assessment of estimation for the \addrvs{MGPCM} from exploratory factor analysis using different methods with $D=3$ in high factor correlation setting. }\label{fig:se3}}
\end{figure}

\rvs{We conducted simulation study II to assess the performance of the proposed bootstrap standard error (SE) estimation procedure. We explored the \addrvs{bootstrap} estimation in the multidimensional case with multiple categories.
In the multidimensional case, as clarified in Section \ref{ssc_e}, we found that the traditional methods (EM and MH-RM) exhibited instability and produced infeasible results across numerous settings. Consequently,  in this section, we focus on the bootstrap-based SE estimates under the EM algorithm, MH-RM, and the proposed pGVEM algorithm.

The comparisons were conducted under the simulation setting similar to Simulation Study \MakeUppercase{\romannumeral1} and the manipulated factors include sample size, test length,
factor correlations, and the number of categories.
The empirical standard deviations of the estimated item parameters as 
\begin{equation*}
    SE_{v}=\frac{1}{R-1}\sum_{r=1}^R(\hat v^{(r)}-v)^2
\end{equation*}across replications per condition are considered as the approximations of true SEs for each method. Here $v$ stands for item parameters to represent $a_{jd}$ or $b_{jk}$, and $\hat v^{(r)}$ is the estimated parameter in the $r$th replicate.
To assess the performance of the propsed method ,we computed SE estimations along with their Bias and relative bias as follow:
\begin{align*}
    Average \;SE=&\frac{1}{J(D+K)}\sum_{j=1}^J\big[\sum_{r=1}^D\widehat{SE}_{a_{jr}}+\sum_{k=1}^{K-1}\widehat{SE}_{b_{jk}}\big]\\
    Bias=&\frac{1}{J(D+K)}\sum_{j=1}^J\big[\sum_{r=1}^D\widehat{SE}_{a_{jr}}-SE_{a_{jr}}+\sum_{k=1}^{K-1}\widehat{SE}_{b_{jk}}-SE_{b_{jk}}\big]\\
    Relative\;Bias=&\frac{1}{J(D+K)}\sum_{j=1}^J\big[\sum_{r=1}^D(\widehat{SE}_{a_{jr}}-SE_{a_{jr}})/SE_{a_{jr}}+\sum_{k=1}^{K-1} (\addrvs{\widehat{SE}_{b_{jk}}}-SE_{b_{jk}})/SE_{b_{jk}}\big],
\end{align*}
 providing a comprehensive assessment of the reliability and accuracy of the proposed bootstrap methods. Here we present the SEs of \addrvs{the discrimination and threshold parameters} pooled together with the aim of showing the effectiveness of our method compared with its alternatives. The results for \addrvs{the estimated SEs for the two type of parameters} are similar to the pooled ones.
 The variability of discrimination and threshold parameters has been well illustrated in the previous study.
 
 The results are shown in Figure \ref{fig:se2} for the low correlation setting and \addrvs{Fig} \ref{fig:se3} for the high correlation setting. In the multidimensional case, where most of the methods fail to estimate the SEs numerically or from Fisher's information matrix, the bootstrap method still generates stable results. In comparison to bootstrap methods from alternative estimations, the pGVEM-based bootstrap exhibits a lower bias. It is observed that when the sample size is 200, the pGVEM-based bootstrap may slightly underestimate standard errors. Nevertheless, its overall performance remains relatively strong across diverse settings.}

\subsection{Study \MakeUppercase{\romannumeral3}}
\rvs{In this section we conduct a simulation study to examine the performance of the proposed AIC and BIC in determining the latent dimension. Considering the complexity of the model setting of MGPCM, we explore the accuracy of factor identification in sample size $N=500,800$ and test length $J=20,40$. We consider both low correlation settings (simulated from $Unif(0.1,0.3)$) and high correlation settings (simulated from $Unif(0.5,0.7)$) for dimensions $D=2,3,4$. In each configuration, a total number of 100 independent samples were generated and we recorded the number of cases where the number of factors were correctly identified. Discrimination parameters are generated from $Unif(1,2)$.
%, with variability prevailing across all dimensions.

Tables \ref{tab:ic1} and \ref{tab:ic2} present the correct estimation rates for the number of dimensions of 3 and 6 categories, respectively. The results indicate that an increase in sample size generally contributes to higher correct estimation rates. %Still, this may be attributed to the model's complexity, particularly within a limited test scale, where deterministic identification or estimation becomes challenging. 
We can also observe that a lower correlation generally leads to higher correct estimation rates, especially when there are fewer categories and shorter test lengths, as the latent structure may be more challenging to identify in such settings. Overall, we observe that AIC performs slightly better in the case of $K=3$ and BIC has an advantage in the regime of $K=6$. In conclusion, the proposed criterion is efficient in general, and with a larger sample size, the model is more likely to be correctly identified. Our findings are also consistent with existing studies under the MIRT model \citep{cho2021gaussian}.}
 
\begin{table}
    \centering
    \begin{tabular}{c|cc|cc|cc|cc|cc|cc|}
    & \multicolumn{6}{c|}{Low Correlation} & \multicolumn{6}{c|}{High Correlation} \\
    Correctness & \multicolumn{2}{c|}{D=2} & \multicolumn{2}{c|}{D=3} & \multicolumn{2}{c|}{D=4} & \multicolumn{2}{c|}{D=2} & \multicolumn{2}{c|}{D=3} & \multicolumn{2}{c|}{D=4} \\
    & AIC & BIC & AIC & BIC & AIC & BIC & AIC & BIC & AIC & BIC & AIC & BIC \\ \hline
    N=500, J=20 & 99 &  97&  53&  42&  7&  4&  66&  52& 3 & 2 & 0 &0  \\
    N=500, J=40 &  100&  100& 96 & 92 & 68 & 56 & 100 & 100 & 79 & 39 & 25 & 12 \\
    N=800, J=20 &  99& 98 & 60 & 54 & 26 & 13 & 64 & 51 & 23 & 12 & 17 &  3\\
    N=800, J=40 & 100 & 100 & 99 &97  &80  &70  & 100 &100  & 83 & 67 & 44 & 32 \\
\end{tabular}
    \caption{Correct number of trials in determining the latent dimension, K=3.}
    \label{tab:ic1}
\end{table}
\begin{table}
    \centering
    \begin{tabular}{c|cc|cc|cc|cc|cc|cc|}
        & \multicolumn{6}{c|}{Low Correlation} & \multicolumn{6}{c|}{High Correlation} \\
        Correctness& \multicolumn{2}{c|}{D=2 }&\multicolumn{2}{c|}{ D=3 }&\multicolumn{2}{c|}{ D=4 }&\multicolumn{2}{c|}{ D=2 }&\multicolumn{2}{c|}{ D=3 }& \multicolumn{2}{c|}{ D=4} \\
        &AIC&BIC&AIC&BIC&AIC&BIC&AIC&BIC&AIC&BIC&AIC&BIC\\\hline
        N=500, J=20 & 9 & 29 & 72 &  87& 89 &87  & 37 & 75 & 85 & 85 & 39 & 32 \\
        N=500, J=40 & 6&38&25&75&61&76& 10 & 18 & 40 & 96 & 86 & 90 \\
        N=800, J=20 & 28 &48  & 90 & 94 & 91 &85  & 77 & 91 & 92 &  88& 41 & 33 \\
        N=800, J=40 & 34&62&46&71&58&87 & 31 & 43 & 79 & 98 &93  & 92 \\
    \end{tabular}
    \caption{Correct number of trials in determining the latent dimension, K=6.}
    \label{tab:ic2}
\end{table}

\section{Real Data Analysis}
\label{sec:rda}
\subsection{Trend in International Mathematics and Science Study Dataset}
In this section, we demonstrate the application of the pGVEM algorithm by analyzing a dataset from the Trend in International Mathematics and Science Study (TIMSS) \citep{mullis2017timss,martin2019timss,fishbein2018timss,martin2020methods}. TIMSS provides reliable and timely trend data on the mathematics and science achievement of U.S. students compared to that of students in other countries. The assessment consists of a large pool of mathematics and science questions, which are divided into different blocks using the item matrix sampling design to relieve response burden. Specifically, 14 booklets were assembled, and each student was required to complete one of them. Different booklets may include the same items for linking. TIMSS 2019 divided the test items into 28 blocks for each grade, with 13 mathematics items and 15 science items. In the data of students in grade eight, each block contained 12 to 18 items.
For our analysis, we selected a mathematics and a science block that appeared in booklet 5 of grade 8. In the mathematics block there were 2 polytomous items and in the science there were 3. 
%There were also missing data in the database, and 
Of the 1,252 students who responded to these booklets,  918 students' responses were completely recorded. In this study, we only estimated the parameters using the data from these 918 students. Appendix C provides details of the item code and corresponding test content. The IRT parameters provided in the TIMSS assessment document \citep{martin2019timss} were used as the true parameters, to which our estimated parameters were compared.

It should be noted that in operational analysis of TIMSS, uni-dimensional IRT models were used for math and science domains separately. When analyzing the items from math and science domain separately, the modified information criterion in Equation \eqref{modaic} and \eqref{modbic} computed under the EFA framework both attain the smallest value when the latent dimension is 1 (i.e., $D=1$), which implies that both domains are essentially uni-dimensional under the generalized partial credit model setting. However, according to the information criterion provided in the `mirt' package (i.e., using its default EM algorithm), the latent dimension cannot be clearly decided.  In the following we display the parameters estimated by EM with fixed quadrature and pGVEM. The results are as follows.

\begin{figure}[htbp]
\begin{center}
\includegraphics[width=4in]{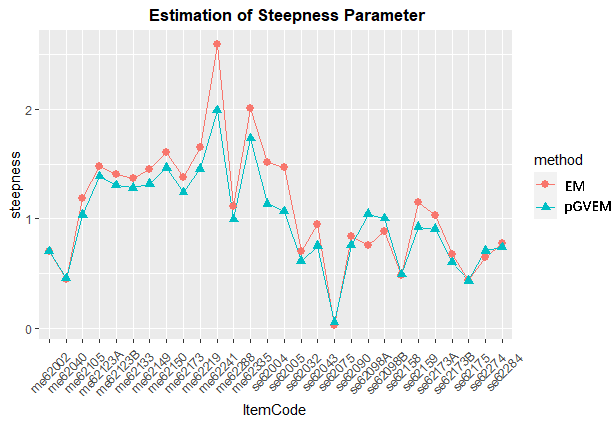}
\end{center}
\caption{Comparison of Discrimination Parameters Estimated from `mirt' and pGVEM\label{fig:forth}}
\end{figure}
\begin{figure}[htbp]
\begin{center}
\includegraphics[width=4in]{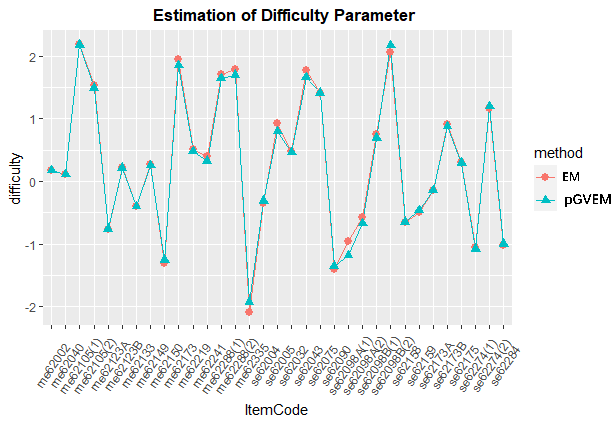}
\end{center}
\caption{Comparison of Threshold Parameters Estimated from `mirt' and pGVEM \label{fig:fifth}}
\end{figure}

To show the results visually, we plot the estimated parameters, by EM and pGVEM, in Figures \ref{fig:forth} and \ref{fig:fifth}. Based on our analysis, it can be inferred that the $b$-parameters obtained from both methods exhibit a significant level of proximity. Similarly, the $a$-parameters demonstrate close values. However, it is worth noting that the estimates derived from the EM algorithm tend to be slightly larger for the majority of the items.
%We can see from the results that the estimated parameters can well approximate those reported in the assessment file. However the estimation is not completely accurate because we leave out missing data and all items are estimated under MGPCM framework without assuming guessing parameters for the dichotomous data. 

To reveal the relation of latent ability across different domains, we estimate all 28 items jointly. First, compute the modified information criterion in Equation \eqref{modaic} and 
 \eqref{modbic} for the setting of joint estimating the data from mathematics and science block, with results presented in Table 1 and 2. From the results the optimal number of latent dimensions is 2. This result is reasonable as math proficiency and science proficiency should emerge as two separate factors. Yet the information criterion output from `mirt' package again did not provide sensible results. By assuming the latent dimension is 2, we present the results estimated by the two methods after promax rotation in Figure \ref{sec:conc}. The analysis of estimated parameters reveals a clear two-factor loading structure as shown in Figure \ref{sec:conc}. Notably, the second dimension emerges as primarily the students' proficiency in solving mathematical problems, whereas the first dimension tends to measure students' science proficiency. This finding suggests that the two dimensions capture distinct but interconnected aspects of overall cognitive ability. Moreover, we can infer that certain types of questions are particularly effective in assessing students' latent abilities in either mathematics or science. These questions demonstrate a higher degree of typicality in evaluating students' competence within their respective domains. This insight underscores the importance of carefully selecting assessment items that align with the targeted cognitive skills, as they offer a more accurate reflection of students' underlying abilities.

\begin{table}[htbp]\centering
\label{tab:first}
\caption{Information Criterion from `mirt'. Note: Math, Sci, Joint denote respectively information criterion for math items, science items and the collection of all items jointly}
\begin{tabular}{|c|c|c|c|c|c|c|}
     \hline
     Dimension&Math(AIC)&Math(BIC)&Sci(AIC)&Sci(BIC)&Joint(AIC)&Joint(BIC)\\\hline 1& \textbf{12923.50}&19087.96&\textbf{13048.86}&\textbf{19256.74}&\textbf{31296.65}&\textbf{31735.79}\\\hline
     2&12939.39&\textbf{19022.30}&13117.81&19263.41&31311.77&31765.61\\\hline
     3&12940.75&19027.24&13167.39&19335.86&31341.26&31861.50\\\hline
     4&12963.25&19039.05&13233.29&19410.39&31441.63&31966.93
     \\\hline
\end{tabular}
\end{table}

\begin{table}[htbp]\centering
\label{tab:second}
\caption{Information Criterion from pGVEM Algorithm. Note: Math, Sci, Joint denote respectively information criterion for math items, science items and the collection of all items jointly}
\begin{tabular}{|c|c|c|c|c|c|c|}
     \hline
     Dimension&Math(AIC)&Math(BIC)&Sci(AIC)&Sci(BIC)&Joint(AIC)&Joint(BIC)\\\hline 1&\textbf{13795.40}&\textbf{18533.71}&\textbf{13935.24}&\textbf{18697.66}&31873.35&32172.33\\\hline
     2&14744.12&18809.71&14961.12&19060.46&\textbf{31806.70}&\textbf{32162.33}\\\hline
     3&15656.59&19630.67&15960.38&19977.87&32392.96&33000.55\\\hline
     4&16588.21&20575.46&16988.45&21028.75&33187.73&33964.10
     \\\hline
\end{tabular}
\end{table}

\begin{figure}[htbp]
\begin{center}
\subfigure[The First Component of Estimated Parameters]{\includegraphics[width=3.5in]{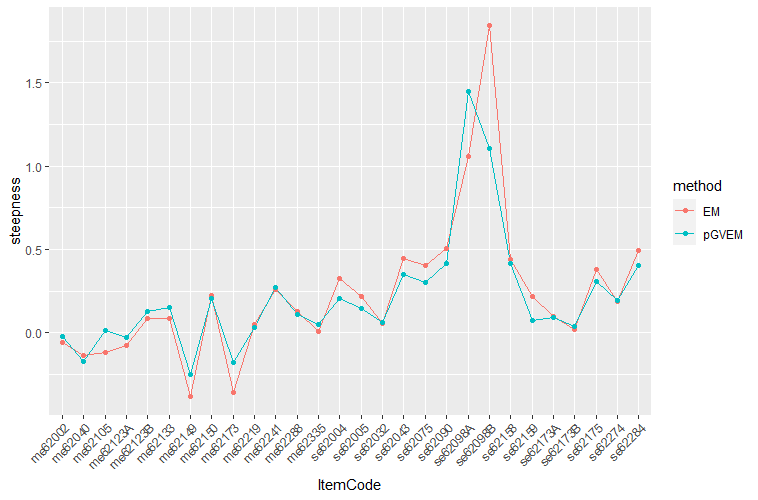}}\subfigure[The Second Component of Estimated Parameters]{\includegraphics[width=3.5in]{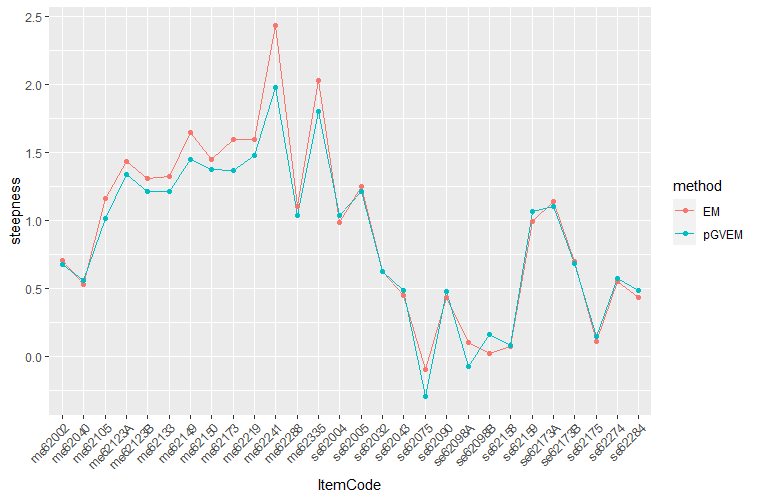}}

\end{center}
\caption{Jointly Estimated Discrimination Parameters from `mirt' and pGVEM} \label{fig:seventh_joint}
\end{figure}

\subsection{Big-Five Factor Personality Dataset}
\begin{figure}[htbp]
\begin{center}
\includegraphics[width=7in,height=4.4in]{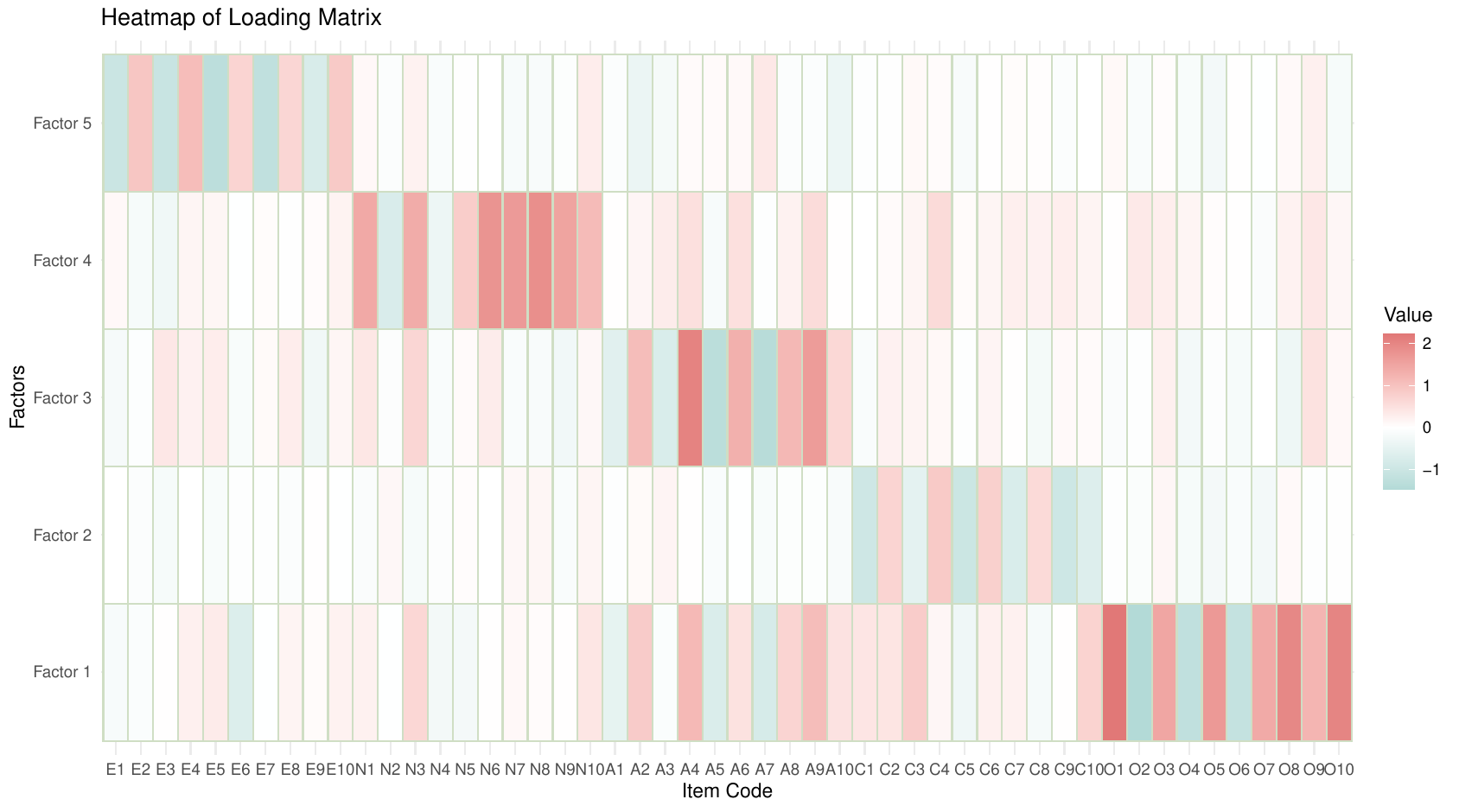}
\end{center}\caption{\rvs{Heatmap of Loading matrix from BiG-Five Personality Study}} \label{fig:hmp_big5}
\end{figure}

\rvs{We further demonstrate the application of the pGVEM algorithm by analyzing a dataset from the Big Five personality assessment.  The Five-Factor Model (FFM) stands as the most widely recognized and extensively used model in psychology for understanding and measuring personality \citep{goldberg1992development}. This model provides a framework to capture the richness and complexity of individual differences, assuming five domains that encompass a comprehensive spectrum of personality traits. The five dimensions include Openness, Conscientiousness, Extraversion, Agreeableness, and Neuroticism (often referred to by the acronym OCEAN) \citep{costa2008revised}.

To measure the latent traits, various assessment tools have been developed to assess an individual's standing on the Big Five dimensions \citep{wiggins1997personality,mccrae1996evaluating,goldberg1999broad}. In our study, we employ the Big-Five Factor Markers derived from the International Personality Item Pool (IPIP), a widely recognized instrument developed by \cite{goldberg1992development}. The dataset is publicly available at \url{http://openpsychometrics.org/tests/IPIP-BFFM/}.
This dataset consists of responses from a substantial sample of 19,718 individuals, each evaluated on the fifty-item Big-Five Factor Markers. The IPIP consists of fifty items, each requiring respondents to rate the extent to which they perceive each statement as true about themselves on a five-point scale. This scale ranges from 1 (Disagree) to 3 (Neutral) to 5 (Agree), providing a nuanced and graded assessment across the five categories.

Under the setting of exploratory factor analysis, we fit the MGPCM using the proposed pGVEM algorithm. The estimated factor loadings are represented in Figure \ref{fig:hmp_big5}. The heatmap reveals that each factor exhibits a distinct and salient association with the grouped items. Also, we can see that not all latent factors have a positive influence on the overall rating -- a phenomenon commonly observed in personality tests \citep{mccrae1996evaluating,goldberg1999broad}. A significant correlation between Factor 3 and Factor 5 can be observed, as they both significantly impact responses to the A-items.
To further illustrate these relationships, we present the estimated correlation matrix of the factors in Table \ref{tab:cor_big5}.
The largest correlation is between Factor 1 and Factor 3 with a value of 0.529, which is consistent with our observation from Figure  \ref{fig:hmp_big5}.
%. This table provides a detailed insight into the correlation of the five domains.
%In summary, our study utilizes the established Big-Five Factor Markers from the IPIP and conducted an exploratory factor analysis under the MGPCM framework. 
Overall, the estimated factor structure aligns with existing literature, demonstrating the usefulness of the proposed method as a computationally efficient tool to analyze large scale assessment data.    
%Our findings yield interpretable estimations that enhance the assessment of personality traits through the calibrated items. These results contribute valuable insights into the complex personality structure, fostering a more comprehensive understanding of individual differences.
}
\begin{table}[ht]
\centering
\begin{tabular}{|c|c|c|c|c|c|}
\hline
&F1&F2&F3&F4&F5\\\hline
F1&1.000 &\textbf{0.237}  &\textbf{0.529}  &\textbf{ 0.459}  & 0.030  \\\hline
F2& \textbf{0.237} &1.000 &\textbf{0.294} &-0.098 &  0.031 \\\hline
F3& \textbf{0.529} &\textbf{0.294} &1.000 & \textbf{0.305}  & 0.096 \\\hline
F4&\textbf{ 0.459}  &-0.098 &  \textbf{0.305} &1.000 & \textbf{-0.213}\\\hline
F5&  0.030 &  0.031 &  0.096& \textbf{-0.213}&1.000 \\
\hline
\end{tabular}\caption{\rvs{Estimated correlation matrix of the 5 latent factors for the Big-Five dataset}}\label{tab:cor_big5}
\end{table}

\section{Conclusion}
\label{sec:conc}

In this paper, we proposed a new variational GVEM algorithm (namely, pGVEM) for the \addrvs{multidimensional generalized partial credit Model}. The \addrvs{GPCM} is one of the most widely used models in cognitive assessment and educational measurement for items that are scored polytomously, such as assigning partial scores for intermediate correct steps.  There are \addrvs{limited number of methods} for an efficient estimation of MGPCM, and it is shown that our pGVEM algorithm outperforms many existing approaches in that it is relatively stable and much faster without resulting in much aberrant estimates or convergence issues. When the sample size and test length are both large, our proposed variational lower bound seems to approximate the target marginal likelihood closely. The computation efficiency is achieved by replacing the intractable high-dimensional integral with a variational lower bound that contributes to faster EM-type updates involving only small-scale linear equations. The simulation study in Section 4 provides simulation evidence to support pGVEM in producing accurate parameter estimates quickly, as compared to the traditional EM implementation of \addrvs{marginal maximum likelihood estimator}s. The real data analysis demonstrates that our estimation scheme is capable of extracting proper information about the latent variables. 

\rvs{Variational inference has emerged as a prominent and efficient methodology in the field of psychometrics, particularly due to its ability to handle large-scale datasets with both accuracy and computational efficiency. In addition to IRT models with continuous latent variables, \cite{yamaguchi2020variational} recently proposed a variational Bayesian (VB) inference algorithm for the saturated Cognitive Diagnosis Models, which represents a notable advancement in scalable and computationally efficient Bayesian estimation for discrete latent variable models. \cite{oka2023scalable} developed a scalable estimation algorithm for the DINA Q-matrix, which employs an iteration scheme utilizing stochastic optimization and variational inference. 
Our method, on the other hand, extends the literature on continuous latent variable estimation \citep{cho2021gaussian,cho2022regularized} by considering multiple response categories. It is encouraging to further explore the relationship of various variational approximation methods, which may lead to a more robust and flexible estimation framework for many psychometric models.}

There are also some other potential directions to extend the current work. First, as with many nonconvex optimization problems, 
%our algorithm may rely on the choice of initial values., i.e., when the initial values differ a lot from the true parameters, the algorithm may reach a local instead of global optima. 
the efficacy of our algorithm may be influenced by the chosen initial values. Instances where initial values deviate substantially from the actual parameters may lead the algorithm to converge to local optima rather than the global one.  Given the propensity for variational approximations to yield multiple local optimal values, investigating the initialization strategy for our estimation process warrants further exploration.
%Second, there are still possibilities that the GPCM model with random-effect latent traits is not identifiable. 
%Second, it would also be interesting to study the identifiability of the  GPCM model, where a proof for the model identification shall guarantee that there exist a unique global optimal yet detection for it may require exploring the structure of the likelihood function, in which way we can develop robust algorithm for the MMLE. 
%Another topic worth studying is to show that the estimation is consistent when the sample size $N\to\infty$. The consistency and asymptotic properties can be very hard to explore for MMLE as it involves analysis of $D$-dimensional integrals. But building such properties can give us useful insight into the model structures.
Second, \rvs{the current method of determining the number of latent dimensions may not work for certain cases, especially when the dimension is very high or there exists a high correlation between latent factors. Therefore, a more accurate and robust model selection method that will work in these challenging scenarios is needed.}

\bibliography{ref}
\bibliographystyle{apalike}
\newpage

\begin{center}
{\LARGE\bf
Supplementary Material for ``Variational Estimation for Multidimensional Generalized Partial Credit Model"}
\end{center}

This supplement contains additional material for the article ``Variational Estimation for Multidimensional Generalized Partial Credit Model''. In Appendix A, we provide the detailed derivation of the pGVEM algorithm. Additional numerical results for Section \ref{sec:simu} are presented in Appenix B. In Appendix C, we provide the item information for the datasets in Section \ref{sec:rda}, and in Appendix D we display the detailed estimation results of the TIMSS dataset.

\begin{center}
\section*{ Appendix A: Derivation of pGVEM for Generalized Partial Credit Model}
\end{center}
In this appendix we give the detailed derivation for the pGVEM update Equation (\ref{updatemean}), (\ref{updatecov}) and (\ref{updatedisc})-(\ref{updatecovtheta})
%Equation(\ref{updatemean}Equation(\ref{updatecov}Equation(\ref{updatedisc}Equation(\ref{updatethres}Equation(\ref{updatevarpara}Equation(\ref{updatecovtheta}
for the Generalized Partial Credit Model.

For the zero step, all item parameters are randomly generated within their domain, denoted by $\bm a_j^{(0)}, b_{jk}^{(0)}$ and $\xi_{ijk}^{(0)}$ for $i=1,\cdots,N$, $j=1,\cdot,J$ and $k=1,\cdot,K_j-1$. Recall $b_{j0}$ are fixed as 0 for all $j$. For the $t^{th}$ step of minimizing the KL divergence $KL\{q_i^{(t)}(\bm\theta_i)||P(\bm\theta_i|Y_i,M_p^{(t-1)})\}=KL\{q_i^{(t)}(\bm\theta_i)|P(\bm\theta_i,Y_i|M_p^{(t-1)})\}$ with density $q_i^{(t)}(\bm\theta_i)$ selected from a Gaussian distribution, we can derive its mean and variance extracted from Equation (\ref{gvem2}):
\begin{align*}\big[\bm\Sigma_i^{(t)}\big]^{-1}=&\,\bm\Sigma_\theta^{-1}+2\sum_{j=1}^J\sum_{k=0}^{K_j-1}\eta\big(\xi_{ijk}^{(t-1)}\big)(k-k_{ij})^2\bm{a}_j^{(t-1)}{\bm{a}_j^{(t-1)}}^\prime;\\\bm{\mu}_i^{(t)}=&\,\bm\Sigma_i^{(t)}\times \Big\{\sum_{j=1}^J\sum_{k=0}^{K_j-1}(k-k_{ij})\big[2\eta\big(\xi_{ijk}^{(t-1)}\big)(b_{jk}^{(t-1)}-b_{jk_{ij}}^{(t-1)})-0.5\big]{\bm{a}_j^{(t-1)}}^\prime\Big\}.\end{align*}
Then we can compute the lower bound
\begin{align*}\underline{E}^{(t)}(M_p,\bm\xi)=&\sum_{i=1}^N\sum_{j=1}^J\Big\{\log 2+\sum_{i=1}^N\sum_{j=1}^J\sum_{k=0}^{K_j-1}\Big[ 
-\eta(\xi_{ijk})(k-k_{ij})^2\bm a_{j}^\prime(\bm\Sigma_i^{(t)}+(\bm\mu_i^{(t)})(\bm\mu_i^{(t)})^\prime)\bm a_{j}\nonumber\\
&+(k-k_{ij})\big(2\eta(\xi_{ijk})(b_{jk}-b_{jk_{ij}})-0.5\big)\bm a_j^\prime\bm\mu_i^{(t)}-\eta(\xi_{ijk})(b_{jk}-b_{jk_{ij}})^2+\frac12(b_{jk}-b_{jk_{ij}})\nonumber\\&+\eta(\xi_{ijk}){\xi_{ijk}}^2+\frac12\xi_{ijk}-\log(1+e^{\xi_{ijk}})
\Big] \Big\} \\&-\frac N2\log|\bm\Sigma_{\theta}^{(t)}|-\sum_{i=1}^N\frac12Tr\Big\{(\bm\Sigma_{\theta}^{(t)})^{-1}\big[\bm\Sigma_i^{(t)}+(\bm\mu_i^{(t)})\big(\bm\mu_i^{(t)}\big)^\prime \big]\Big\}.\end{align*}
The cross term makes a simultaneous maximization rather difficult, therefore we instead adopt a Gauss-Seidel approach. First by
\begin{align*}\frac{\partial \underline{E}^{(t)}(M_p,\bm\xi)}{\partial \bm a_j}=&\sum_{i=1}^N\sum_{k=0}^{K_j-1}\Big\{-2\eta\big(\xi_{ijk}^{(t-1)}\big)(k-k_{ij})^2\big[\bm\Sigma_i^{(t)}+(\bm\mu_i)^{(t)}\big(\bm\mu_i^{(t)}\big)^\prime\big]\bm a_j\\&+(k-k_{ij})\big[2\eta\big(\xi_{ijk}^{(t-1)}\big)(b_{jk}^{(t-1)}-b_{jk_{ij}^{(t-1)}})-0.5\big]\bm\mu_i^{(t)}\Big\},\end{align*}
therefore by plugging in the value of $b_{jk}$ and $\xi_{ijk}$ from previous iteration, we have
\begin{align*}\bm a_j^{(t)}=\frac12&\Big[\sum_{i=1}^N\sum_{k=0}^{K_j-1}\eta\big(\xi_{ijk}^{(t-1)}\big)(k-k_{ij})^2\big(\bm\Sigma_i^{(t)}+(\bm\mu_i^{(t)})\big(\bm\mu_i^{(t)}\big)^\prime \big)\Big]^{-1}\\&\Big[\sum_{i=1}^N\sum_{k=0}^{K_j-1}(k-k_{ij})\big(2\eta\big(\xi_{ijk}^{(t-1)}\big)(b_{jk}^{(t-1)}-b_{jk_{ij}}^{(t-1)})-0.5\big)\bm\mu_i^{(t)}\Big].\end{align*}
For the threshold parameters, the derivatives are given as
\begin{align*}\frac{\partial \underline{E}^{(t)}(M_p,\bm\xi)}{\partial b_{jk}}=&\sum_{i=1}^N\Big[-2\eta\big(\xi_{ijk}^{(t-1)}\big)b_{jk}+2\eta\big(\xi_{ijk}^{(t-1)}\big)(k-k_{ij}){\bm a_j^{(t)}}^\prime\bm\mu_i^{(t)}+2\eta\big(\xi_{ijk}^{(t-1)}\big)b_{jk_{ij}}^{(t-1)}+\frac12\Big]\\&+\sum_{i=1}^N\sum_{v=0}^{K_j-1}\Big[I_{(k=k_{ij})}(-2\eta(\xi_{ijv}^{(t-1)})-2\eta(\xi_{ijv}^{(t-1)})(v-k_{ij}){\bm a_j^{(t)}}^\prime\bm\mu_i^{(t)}+2\eta\big(\xi_{ijk}^{(t-1)}\big)b_{jv}^{(t-1)}-\frac12\Big],\end{align*}
and similarly we have update for $b_{jk}^{(t)}$:

$$b_{jk}^{(t)}=\frac{\sum_{i=1}^N\big[\mbox B_1^{(t)}(i,j,k)I_{(k\neq k_{ij})}+I_{(k=k_{ij})}\sum_{v=0,v\neq k}^{K_j-1}\mbox B_2^{(t)}(i,j,v,k)\big]}{2\sum_{i=1}^N\big(\eta\big(\xi_{ijk}^{(t-1)}\big)I_{(k\neq k_{ij})}+I_{(k=k_{ij})}\sum_{v=0,v\neq k}^{K_j-1}\eta\big(\xi_{ijv}^{(t-1)}\big)\big)}$$
with
\begin{align*}
\mbox B_1^{(t)}(i,j,k)=2\eta\big(\xi_{ijk}^{(t-1)}\big)(k-k_{ij}){\bm a_j^{(t)}}^\prime\bm\mu_i^{(t)}+0.5+2\eta\big(\xi_{ijk}^{(t-1)}\big);\\\mbox B_2^{(t)}(i,j,v,k)=-2\eta\big(\xi_{ijk}^{(t-1)}\big)(v-k){\bm a_j^{(t)}}^\prime\bm\mu_i^{(t)}-0.5+2\eta\big(\xi_{ijv}^{(t-1)}\big).
\end{align*}
For the variational parameter, notice that the function
$$g(x)=\eta(x)(x^2-c)+\frac12x-\log(1+e^x)$$
is maximized for $x^2=\max\{c,0\}$, so the update for $\xi_{ijk}$ is
$${\xi_{ijk}^{(t)}}^2=\Big[(k-k_{ij}){\bm a_j^{(t)}}^\prime\bm\mu_i^{(t)}-\big(b_{jk}^{(t)}-b_{jk_{ij}}^{(t)}\big)\Big]^2+(k-k_{ij})^2{\bm a_j^{(t)}}^\prime \bm\Sigma_i^{(t)}\bm a_j^{(t)}.$$

The derivative with respect to $\bm\Sigma_\theta^{-1}$ is
$$\frac{\partial \underline{E}^{(t)}(M_p,\bm\xi)}{\partial \bm\Sigma_\theta^{-1}}=\frac N2\bm\Sigma_\theta-\frac12\sum_{i=1}^N\Big[\bm\Sigma_i^{(t)}+(\bm\mu_i^{(t)})\big(\bm\mu_i^{(t)}\big)^\prime\Big]$$
Therefore the update when covariance is given by
\begin{equation*}
    \bm\Sigma_\theta=\frac 1N\sum_{i=1}^N\Big[\bm\Sigma_i^{(t)}+(\bm\mu_i^{(t)})\big(\bm\mu_i^{(t)}\big)^\prime\Big].
\end{equation*}
Note again that the update for $\bm\Sigma_\theta$ is only performed in confirmatory factor analysis and fixed as identity for exploratory factor analysis.\\

%\newpage
\begin{center}
{\section*{Appendix B: Additional Numerical Results}}
\end{center}

\subsection*{Recovery of the latent ability}
\begin{figure}
\begin{center}
\subfigure[Low correlation]{\includegraphics[width=6.8in,height=2.8in]{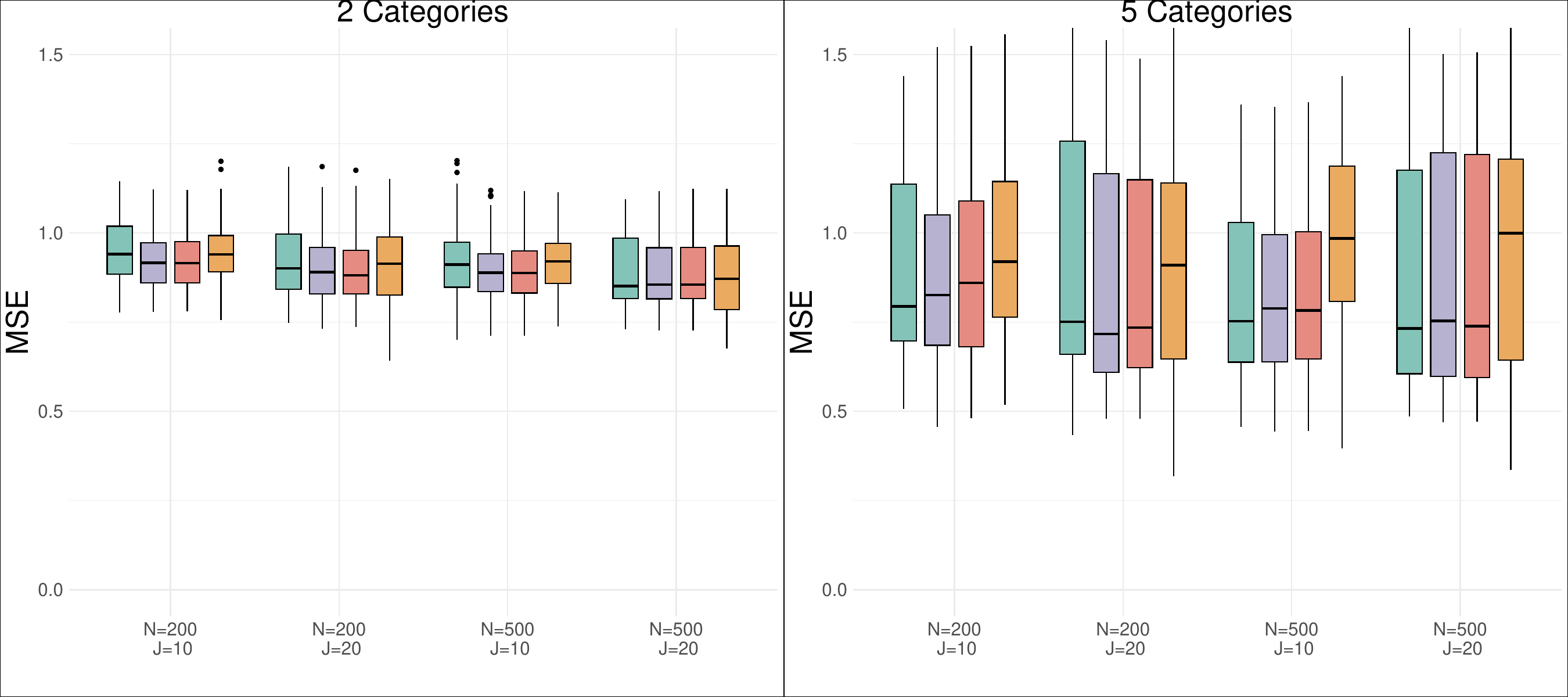}}
\subfigure[High correlation]{\includegraphics[width=6.8in,height=2.8in]{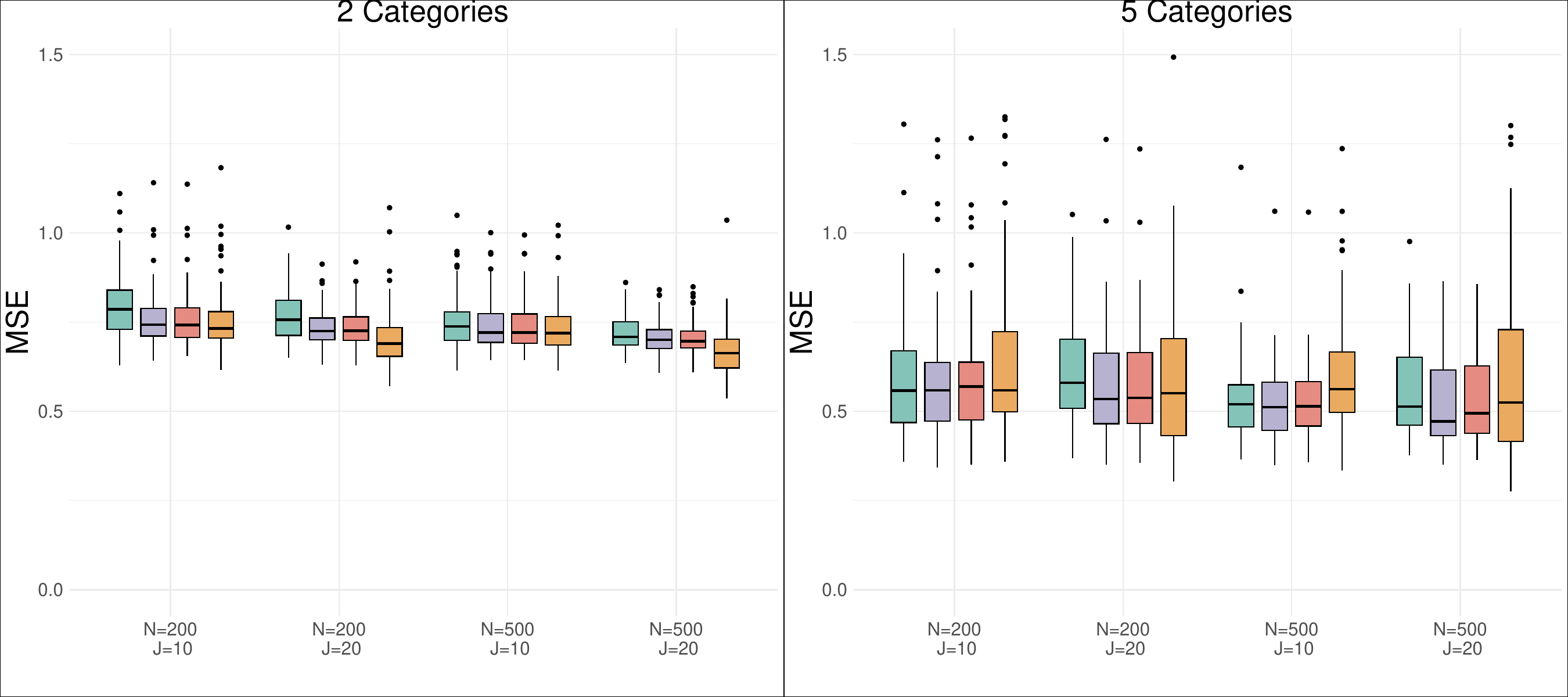}}
\subfigure{\includegraphics[width=4in,height=0.3in]{method.jpeg}}
\end{center}
\caption{\rvs{Mean Squared Error of the $\bm\theta$ estimation for \addrvs{MGPCM} from exploratory factor analysis using different methods.}\label{fig:theta_mse}}
\end{figure}

\begin{figure}
\begin{center}
\subfigure[Low correlation]{\includegraphics[width=6.8in,height=2.8in]{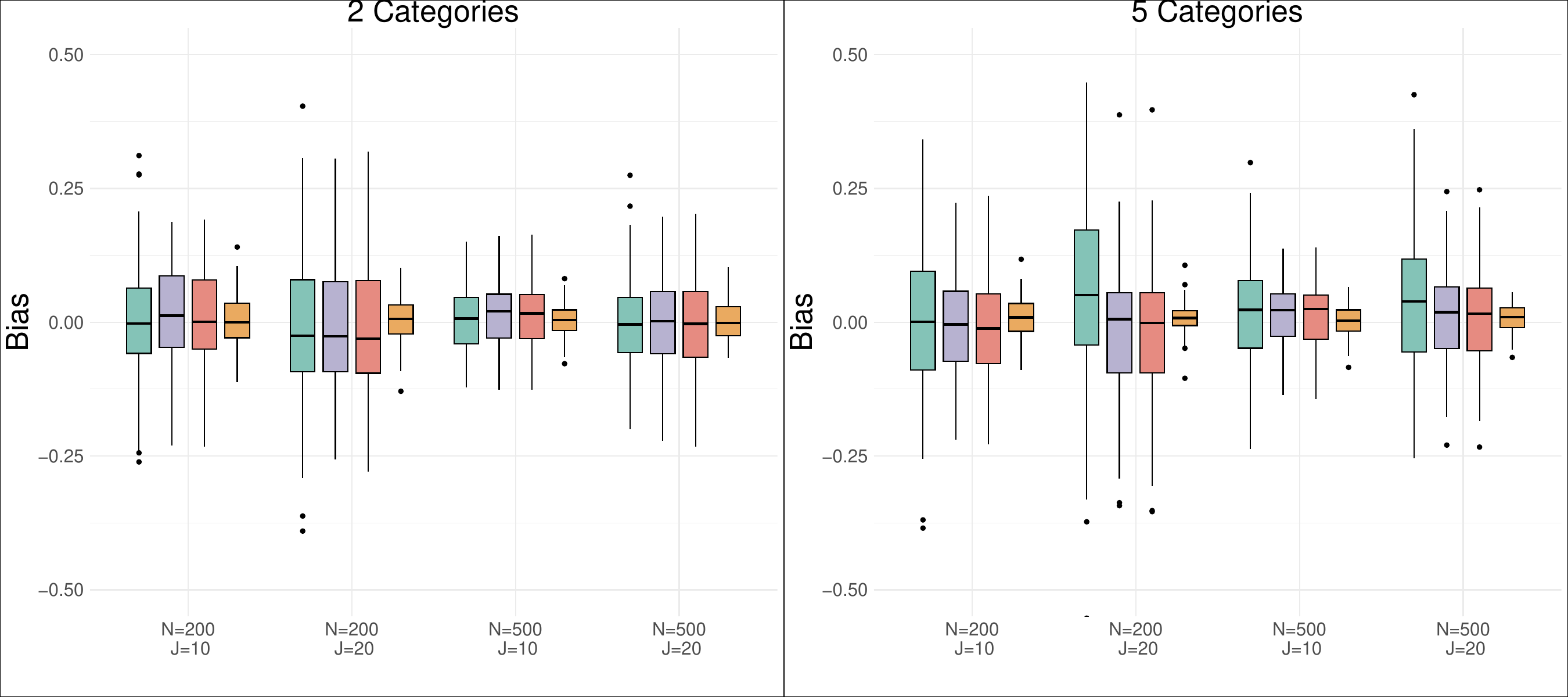}}
\subfigure[High correlation]{\includegraphics[width=6.8in,height=2.8in]{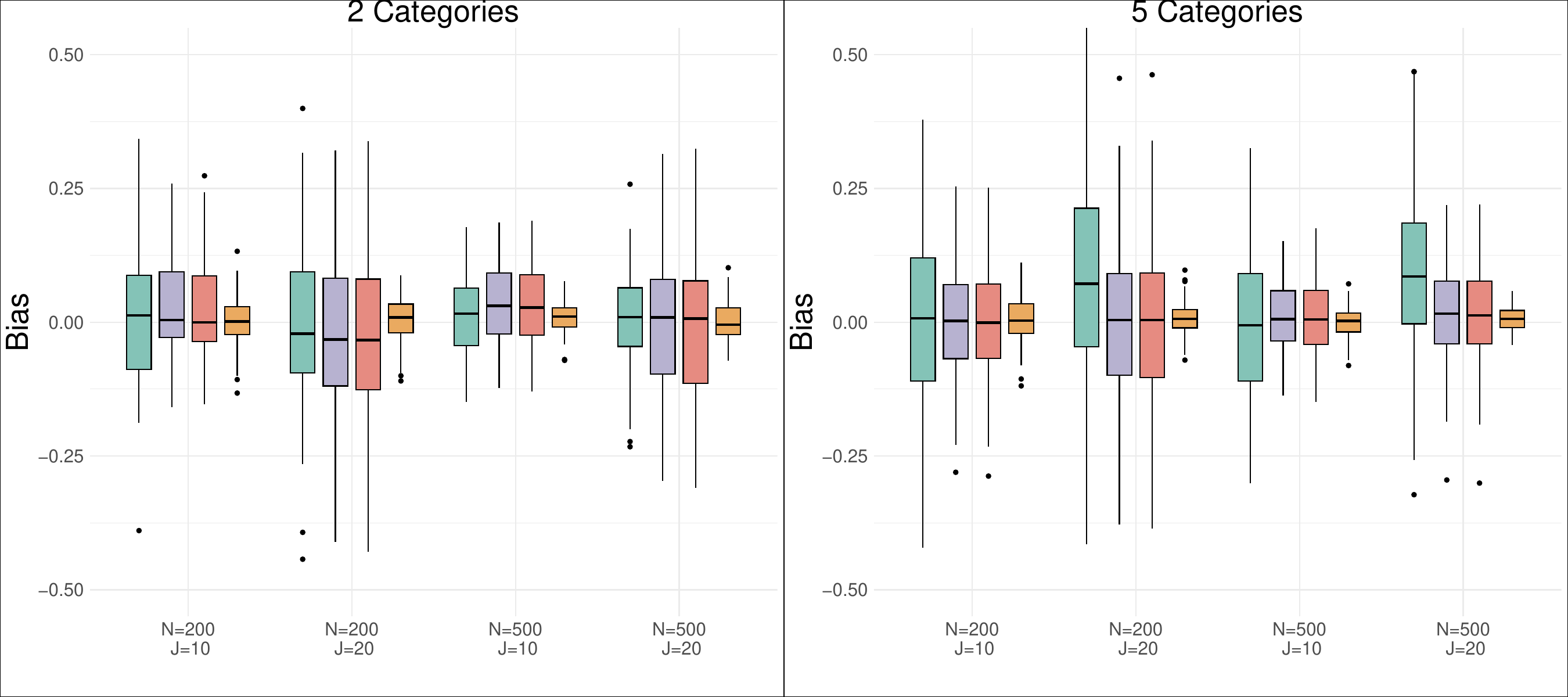}}
\subfigure{\includegraphics[width=4in,height=0.3in]{method.jpeg}}
\end{center}
\caption{\rvs{Bias of the $\bm\theta$ estimation for \addrvs{MGPCM} from exploratory factor analysis using different methods.}\label{fig:theta_bias}}
\end{figure}

\rvs{In this subsection, we present the result of recovery of the latent ability. In pGVEM framework, we can only obtain the estimated correlation of the latent factors along with the variational distribution of the factor for each individuals. Here we used the mean of each distribution $\hat q_i^{(t)}(\bm \theta_i)$ (i.e., see Equation \eqref{updatemean}) as point estimates of the latent ability to compute the estimated item response functions, i.e., the probability of presence of each partial credit score. Because $\hat q_i^{(t)}(\bm \theta_i)$ is the minimizer of the KL divergence $KL\{q_i^{(t)}(\bm \theta_i)||\Pr(\bm Y_i|M_p,\bm \theta_i)\}$ selected from the Gaussian family, \textit{it is reasonable to use the mean value of $\hat q_i^{(t)}(\bm\theta_i)$ as an approximation to the latent ability}. And for the methods implemented in `mirt' package, we use the built-in function to obtain the estimations. The results are presented in Figure \ref{fig:theta_mse} and \ref{fig:theta_bias}.}

\subsection*{Comparison with GVEM}
In this subsection we compare the results between GVEM and pGVEM. These two algorithms are similar in their derivation but differ in the detailed implementation. We present the comparison here to demonstrate that our proposed pGVEM is still efficient in estimation of M2PL model with GVEM as benchmark. The following Figures \ref{fig:seventh} and \ref{fig:eighth} provide comparison of GVEM and pGVEM for 2-category MGPCM (i.e., MGPCM with binary responses). \rvs{Here the discrimination parameters are simulated from $Unif(1,2)$. For the convergence criterion of GVEM, we stop the iteration when the average change of all parameters falls below a threshold:
\begin{equation*}
    \frac{1}{J\times K+J}\sum_{j=1}^J\big(\|\hat{\bm a}_j-\bm a_j\|^2+(\hat b_j-b_j)^2\big)<1.0\time10^{-5}
\end{equation*}}
 
\begin{figure}
\begin{center}
\subfigure[MSE of discrimination parameter]{\includegraphics[width=6in,height=2in]{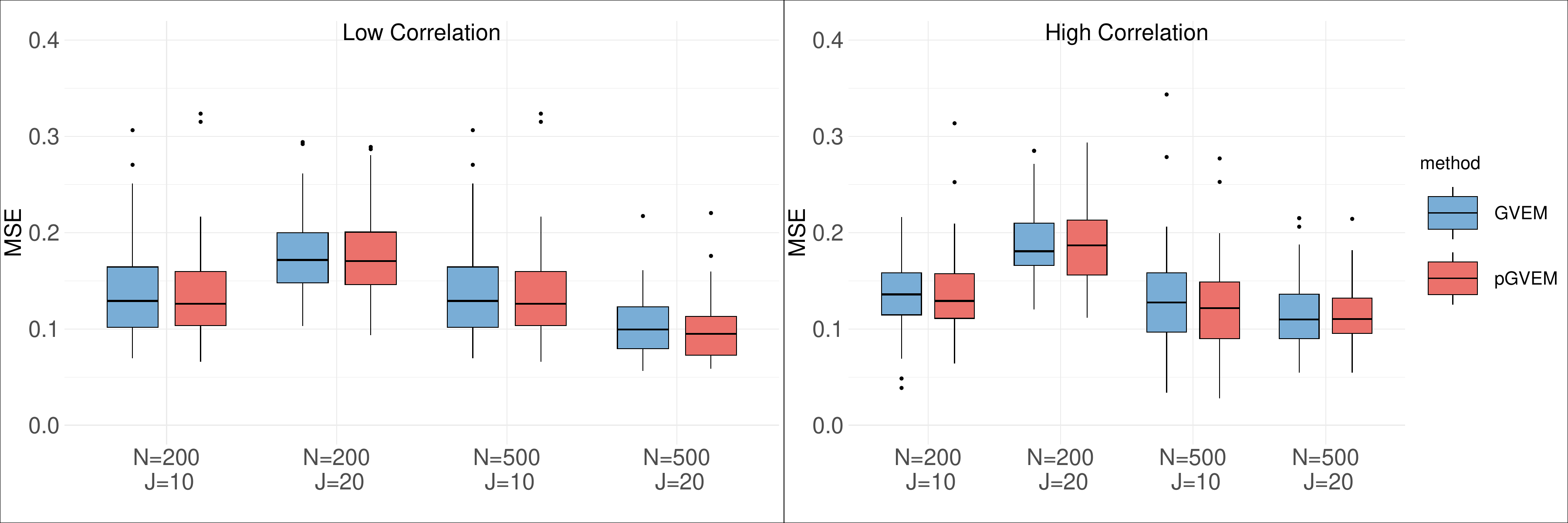}}
\subfigure[MSE of difficulty parameter]{\includegraphics[width=6in,height=2in]{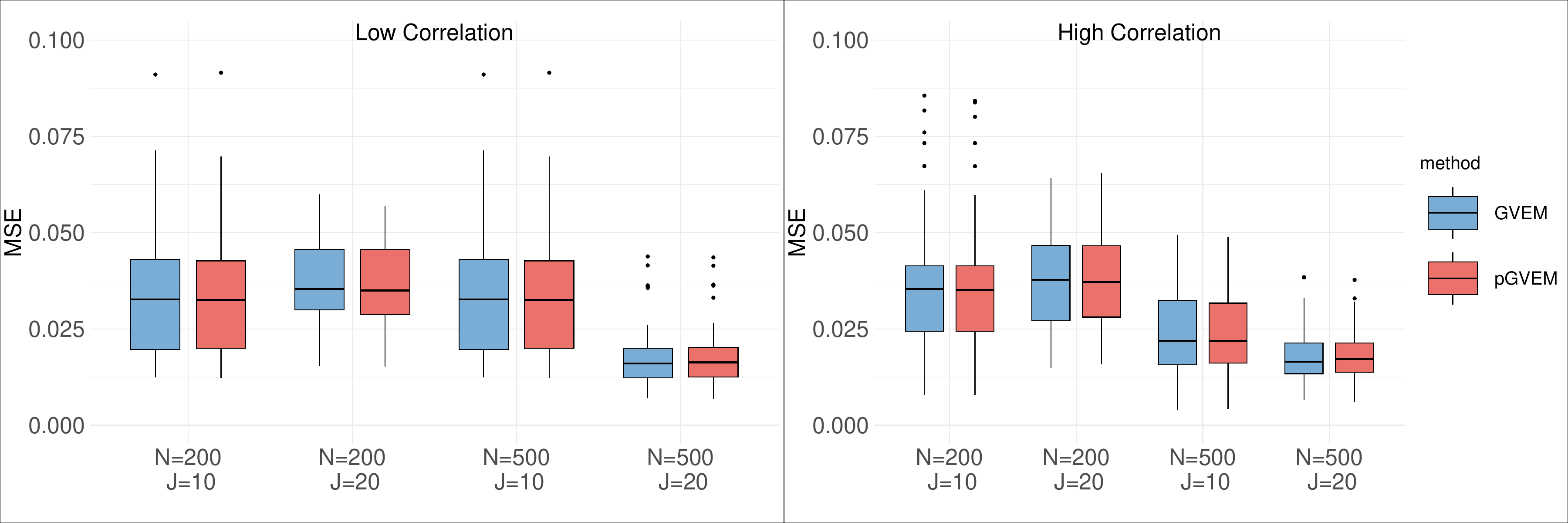}}
\subfigure[MSE of correlation]{\includegraphics[width=6in,height=2in]{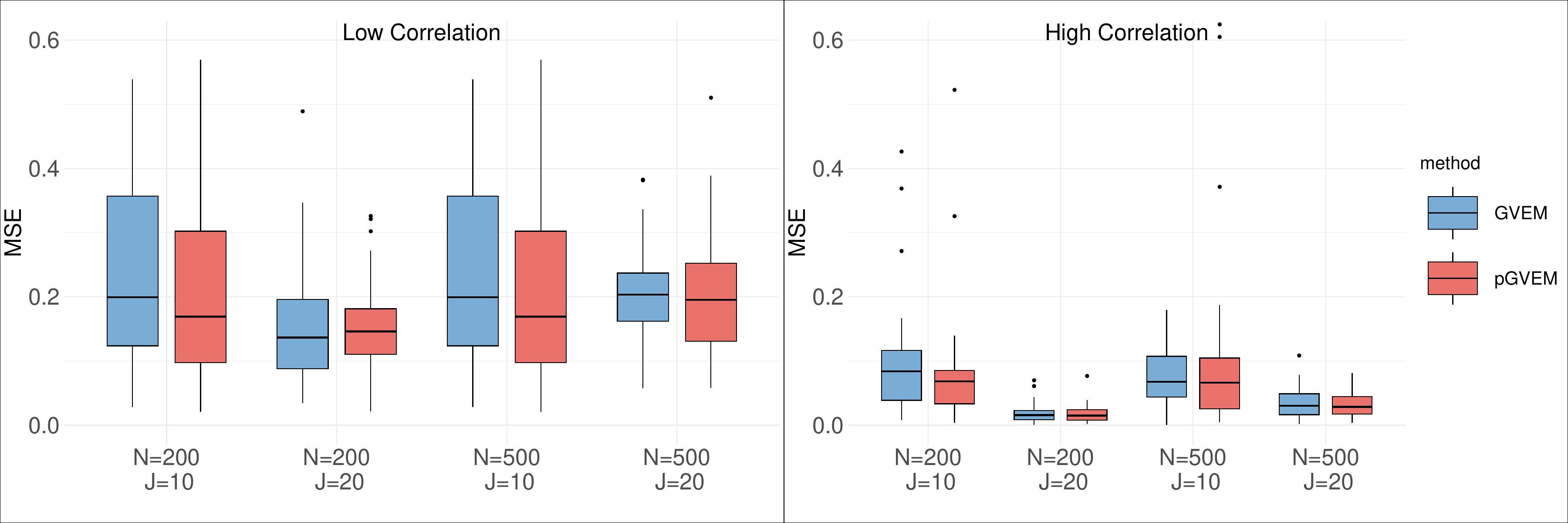}}
\end{center}
\caption{\rvs{Mean Squared Error of estimation for the M2PL Model from exploratory factor analysis using GVEM and pGVEM method.}\label{fig:seventh}}
\end{figure}

\begin{figure}
\begin{center}
\subfigure[Bias of discrimination parameter]{\includegraphics[width=6in,height=2in]{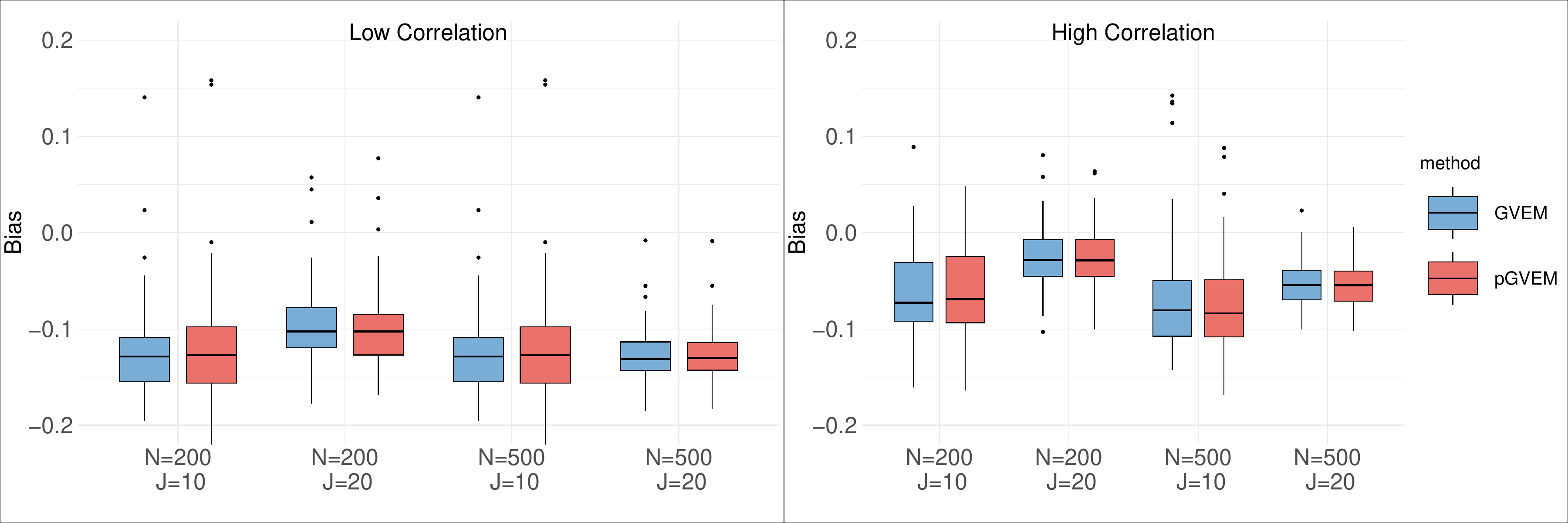}}
\subfigure[Bias of difficulty parameter]{\includegraphics[width=6in,height=2in]{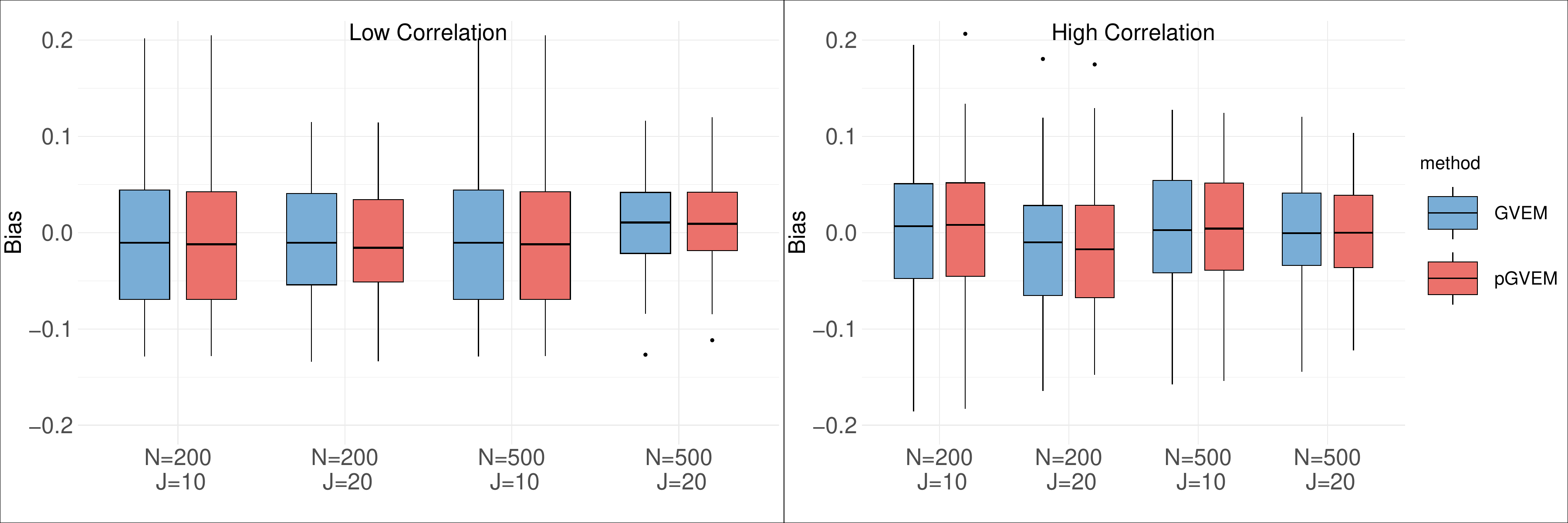}}
\subfigure[Bias of correlation]{\includegraphics[width=6in,height=2in]{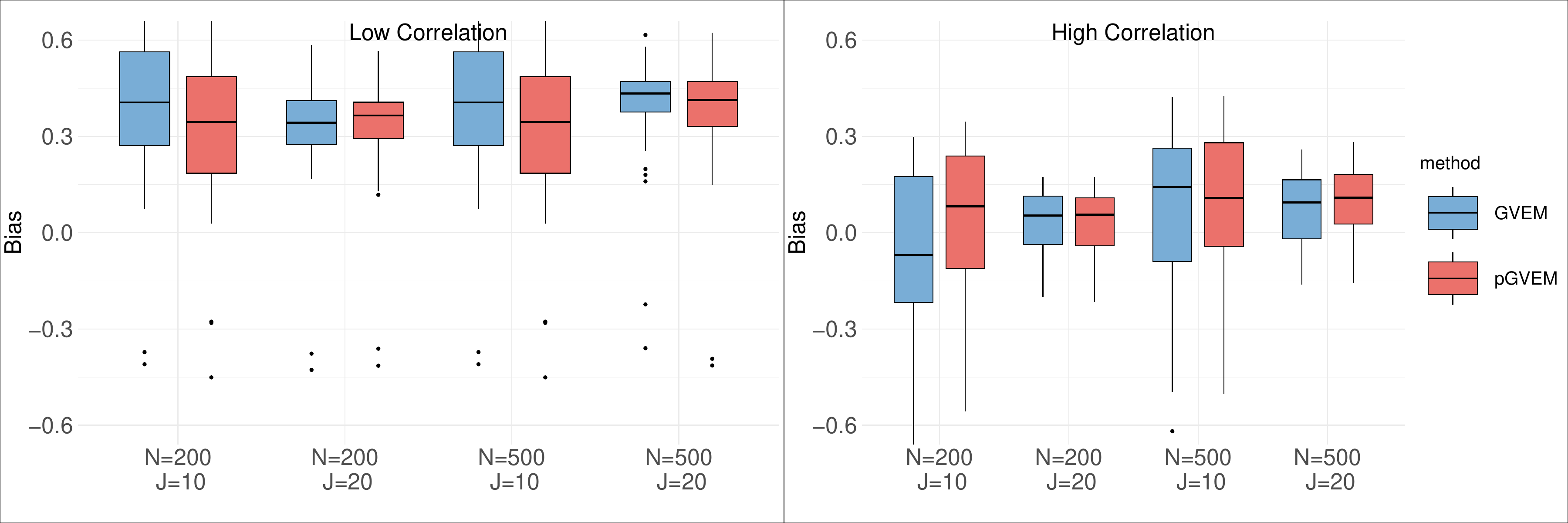}}
\end{center}
\caption{\rvs{Bias of estimation for the M2PL Model from exploratory factor analysis using GVEM and pGVEM method.}\label{fig:eighth}}
\end{figure}

\subsection*{Full Version of Figure in Simulation Study \MakeUppercase{\romannumeral1}}
\begin{figure}
\begin{center}
\subfigure{\includegraphics[width=6.8in,height=3.2in]{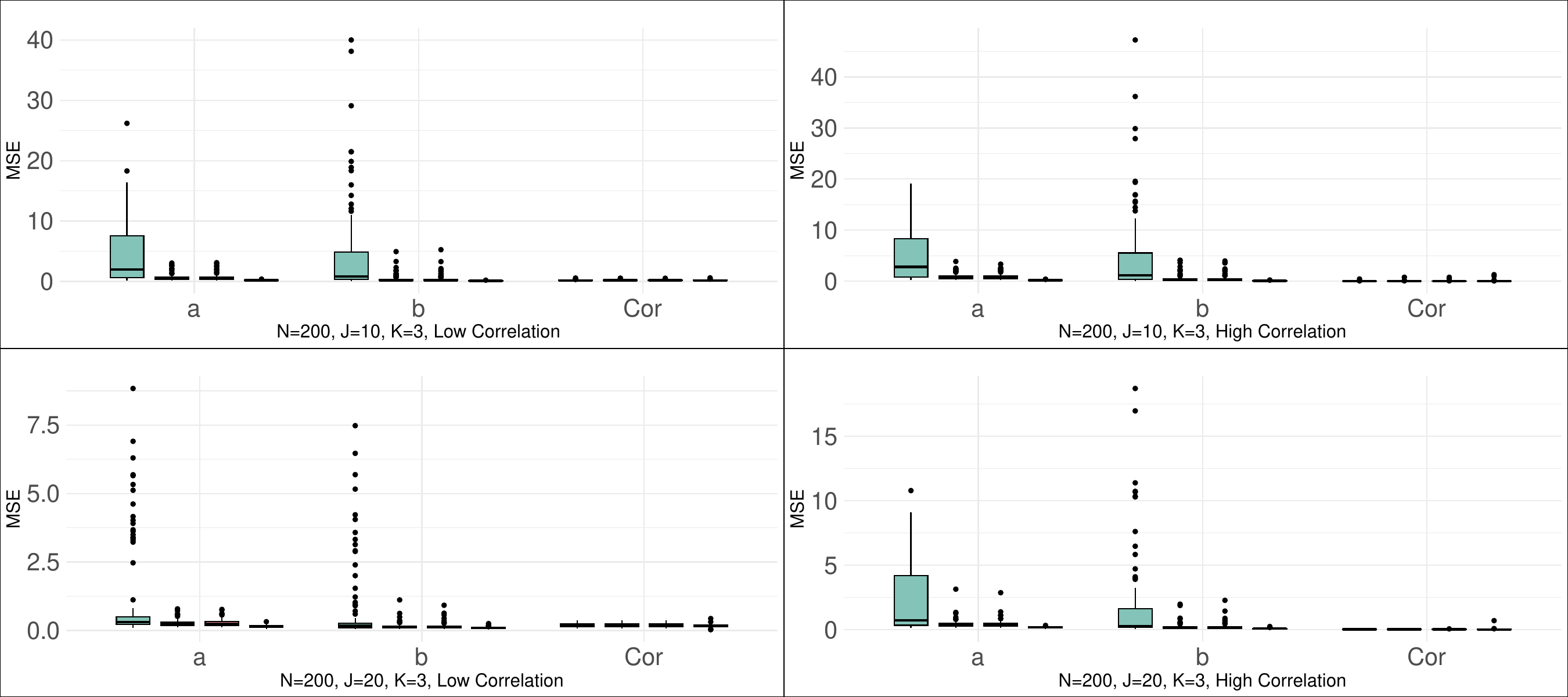}}
\subfigure{\includegraphics[width=6.8in,height=3.2in]{full_I200K2mse.pdf}}
\subfigure{\includegraphics[width=4in,height=0.3in]{method.jpeg}}
\end{center}
\caption{\rvs{Full version of Figure \ref{fig:mse_k3} and \ref{fig:mse_k5} for $N=200$.}\label{fig:full_mse_k}}
\end{figure}

\begin{figure}
\begin{center}
\subfigure{\includegraphics[width=6.8in,height=3in]{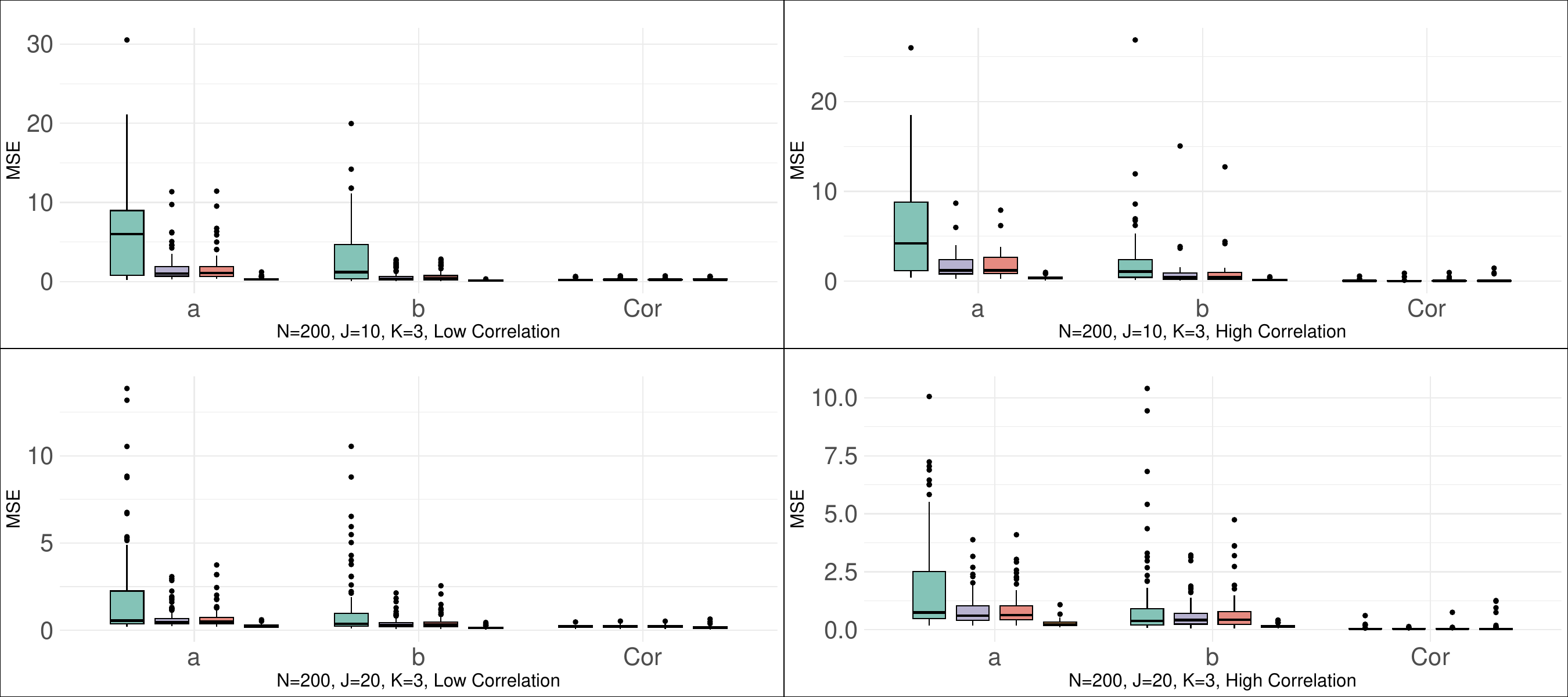}}
\subfigure{\includegraphics[width=6.8in,height=3in]{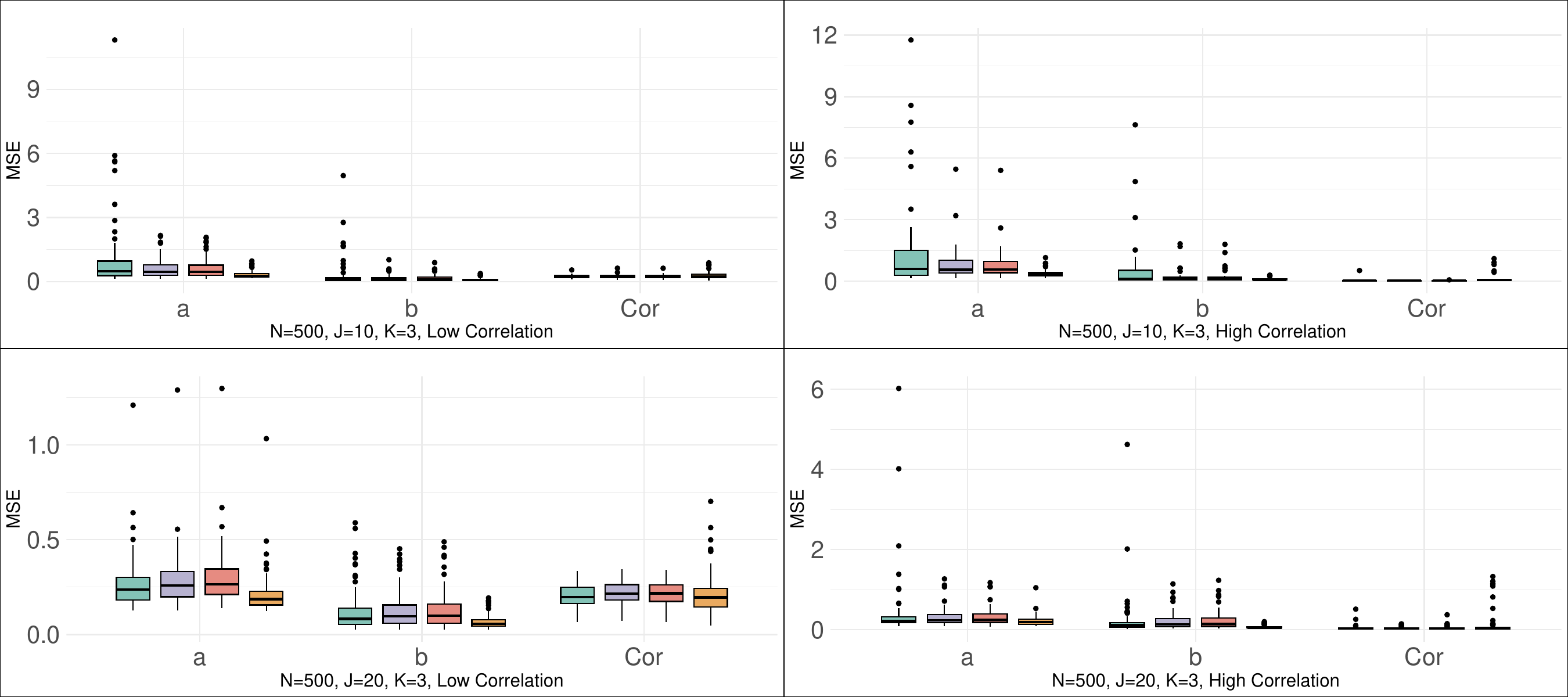}}
\subfigure{\includegraphics[width=5in,height=0.4in]{method.jpeg}}
\end{center}
\caption{\rvs{Full version of Figure \ref{fig:mse_k3p}.}\label{fig:full_mse_k3p}}
\end{figure}

\begin{figure}
\begin{center}
\subfigure{\includegraphics[width=6.8in,height=3.2in]{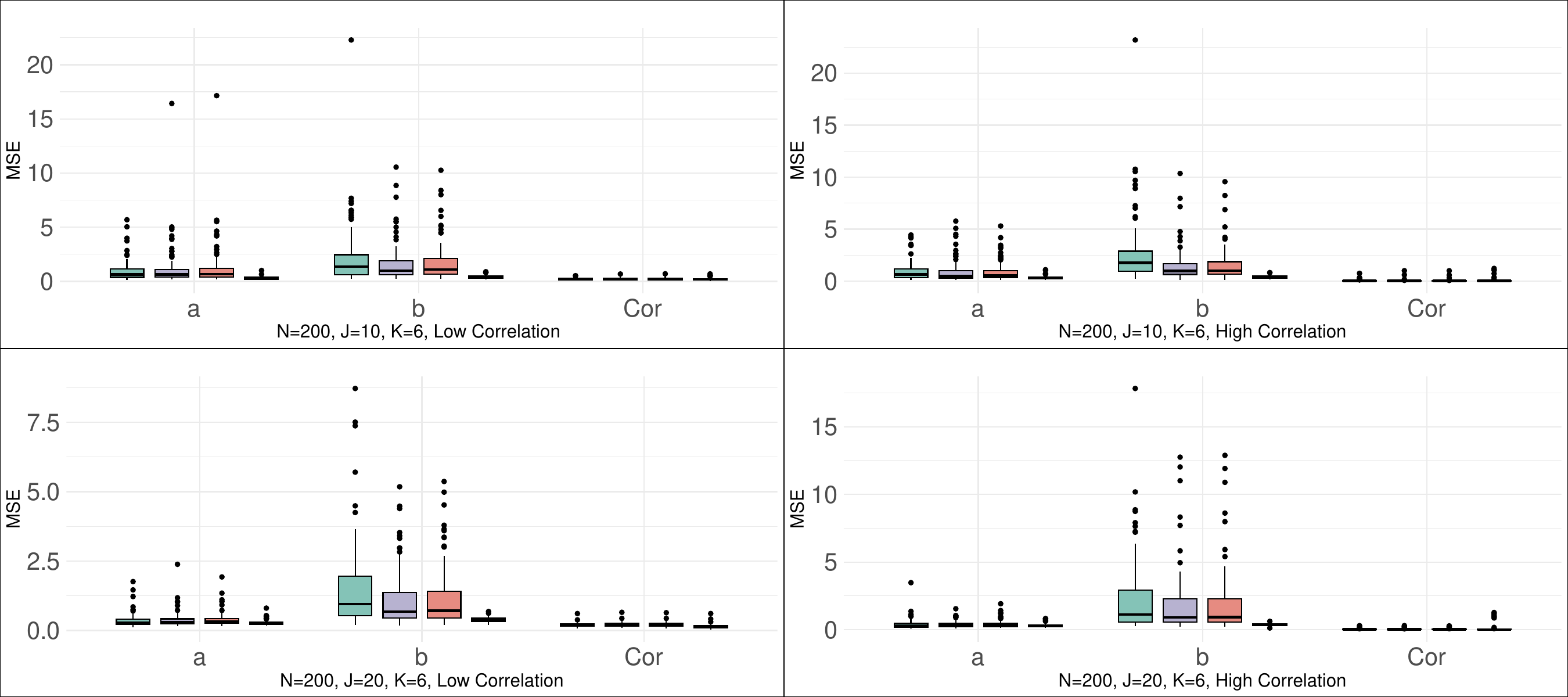}}
\subfigure{\includegraphics[width=6.8in,height=3.2in]{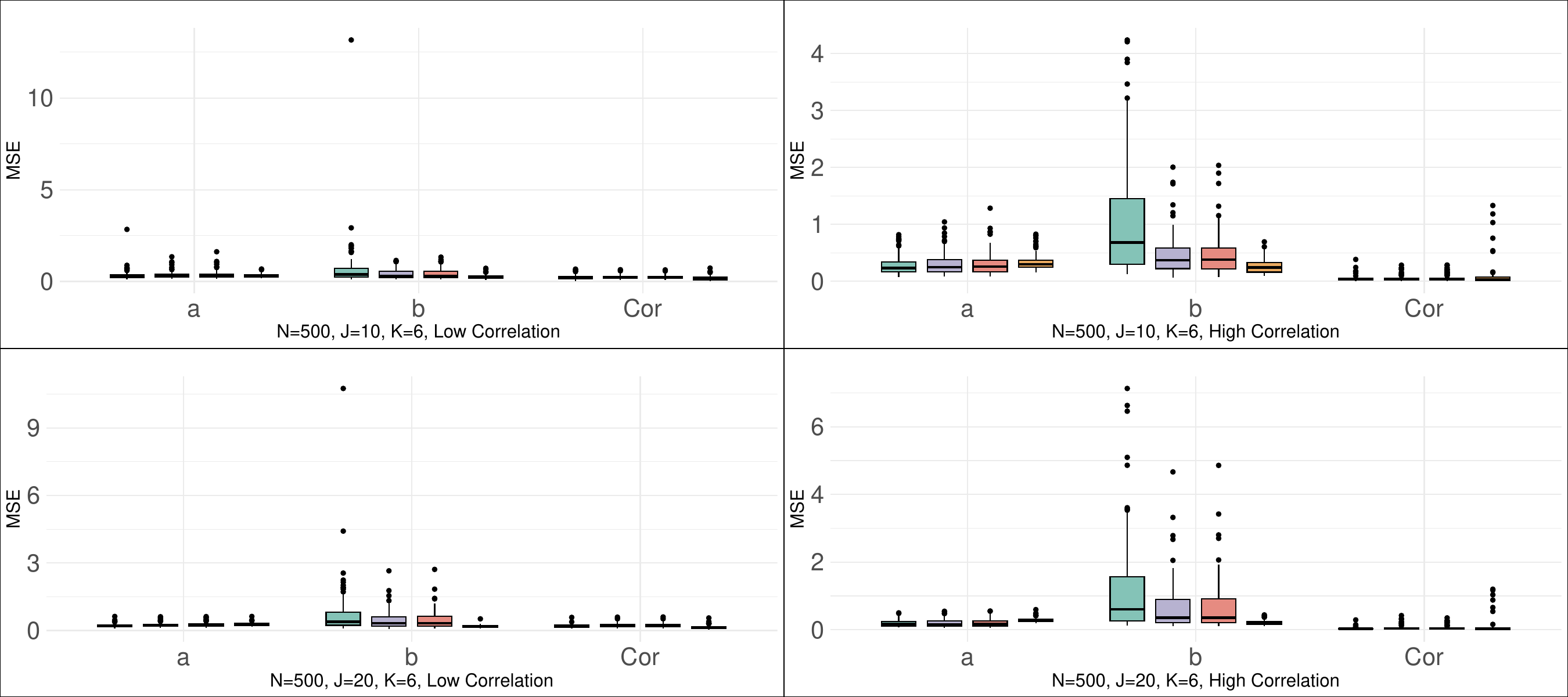}}
\subfigure{\includegraphics[width=4in,height=0.3in]{method.jpeg}}
\end{center}
\caption{\rvs{Full version of Figure \ref{fig:mse_k5p}.}\label{fig:full_mse_k5p}}
\end{figure}

\begin{figure}
\begin{center}
\subfigure{\includegraphics[width=6.8in,height=3.2in]{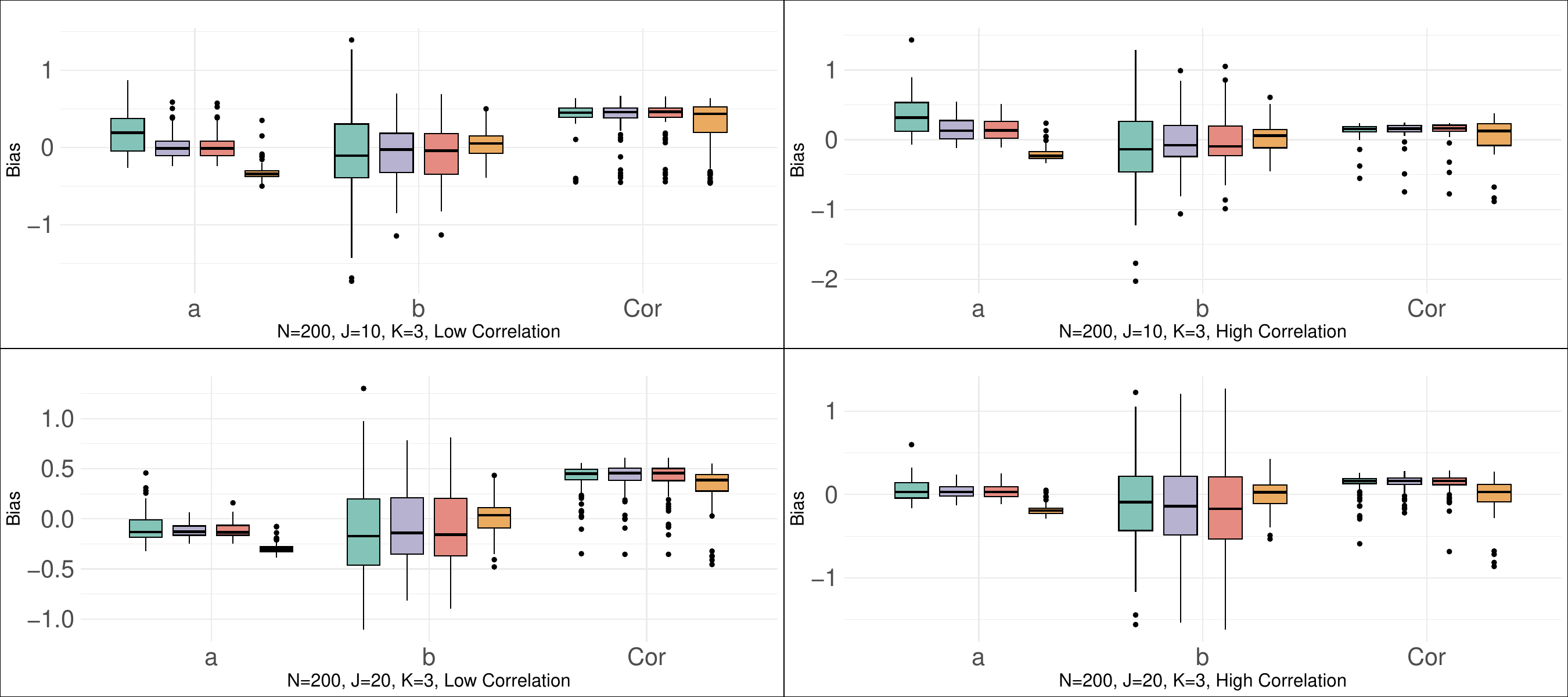}}
\subfigure{\includegraphics[width=6.8in,height=3.2in]{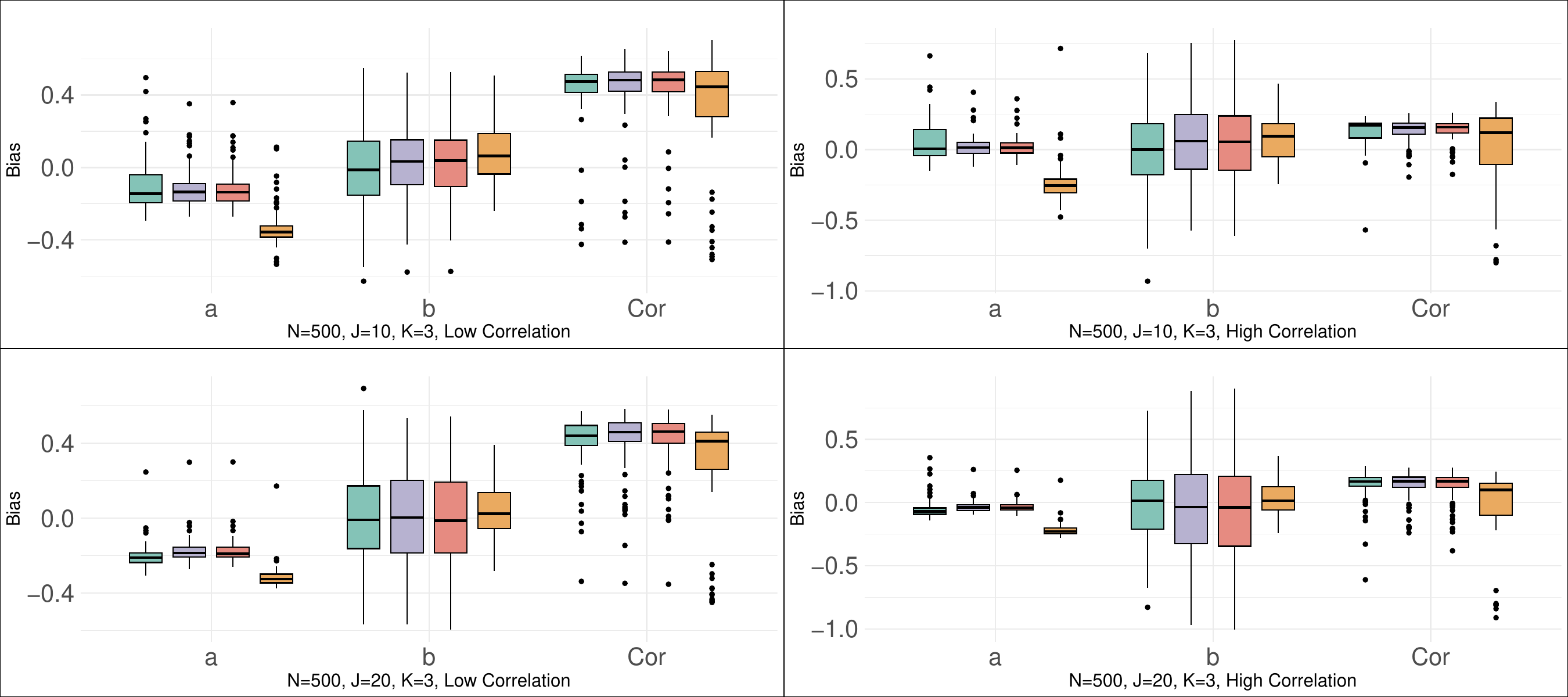}}
\subfigure{\includegraphics[width=4in,height=0.3in]{method.jpeg}}
\end{center}
\caption{\rvs{Full version of Figure \ref{fig:bias_k3p}.}\label{fig:full_bias_k3p}}
\end{figure}

\begin{figure}
\begin{center}
\subfigure{\includegraphics[width=6.8in,height=3.2in]{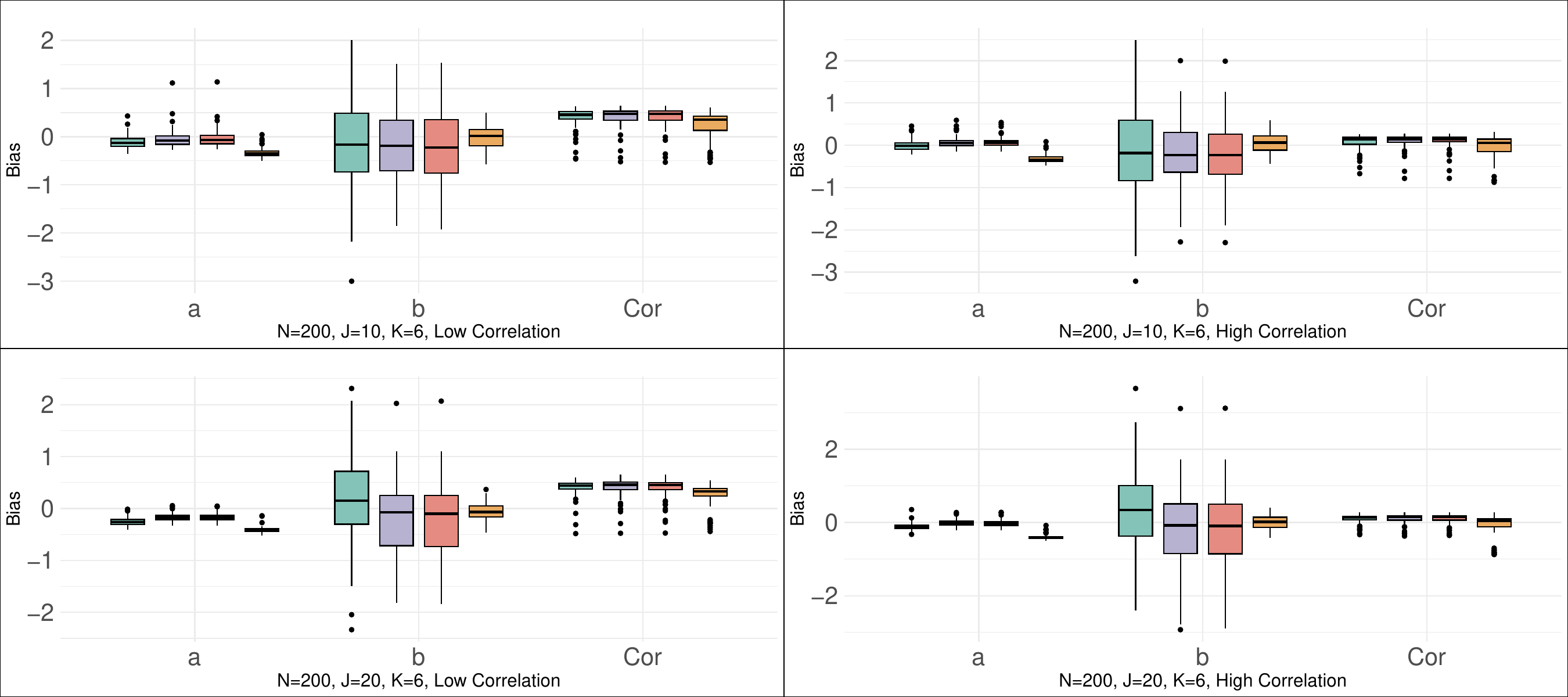}}
\subfigure{\includegraphics[width=6.8in,height=3.2in]{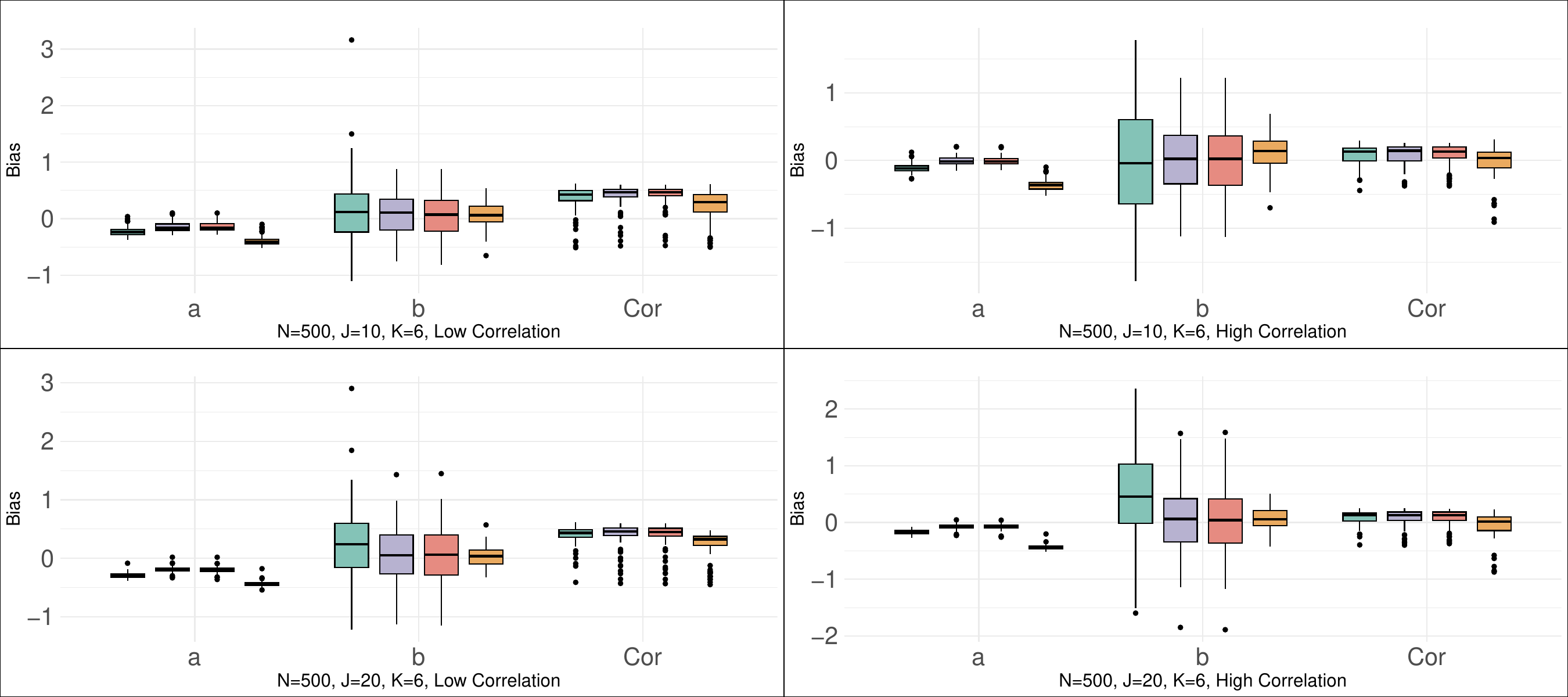}}
\subfigure{\includegraphics[width=4in,height=0.3in]{method.jpeg}}
\end{center}
\caption{\rvs{Full version of Figure \ref{fig:bias_k5p}.}\label{fig:full_bias_k5p}}
\end{figure}

This section is the full version of Figures presented in Simulation Study \MakeUppercase{\romannumeral1}. Specifically, we present the MSE for the case of $N=200$ in Figure \ref{fig:mse_k3} and \ref{fig:mse_k5} in Figure \ref{fig:full_mse_k} and the results of MSE and bias in the setting of large loading in Figure \ref{fig:full_mse_k3p}-\ref{fig:full_bias_k5p}

\newpage
\begin{center}
\section*{ Appendix C: Test Items of the Datasets}
\end{center}
The following tables include the item codes in the TIMSS dataset and the questions. In the bracket is category information for the items. For details, see \cite{martin2019timss}.
\begin{center}
\begin{tabular}{ |p{2cm}|p{15cm}| }
 \hline
 \multicolumn{2}{|c|}{mathematics test items and their code in TIMSS database}\\
 \hline
 Code&Questions\\
\hline
me62150&DIFFERENCE BETWEEN LOW TEMPERATURE IN CITY X AND Y (1)\\
me62335&SELECT EQUIVALENT RATIO TO 3:2 (B)\\
me62219&KATY ENLARGES A PHOTO - NEW HEIGHT (A)\\
me62002&FILL IN BOXES TO MAKE THE SMALLEST PRODUCT (1)\\
me62149&IDENTIFY EXPRESSION TO CALCULATE ROBIN'S EARNINGS (D)\\
me62241&ROY'S PHONE BUSINESS - EQUATION FOR Y (1)\\
me62105&AREA OF RECTANGLE WITH SIDES X AND 2X + 1 (2)\\
me52040&ESTIMATE AREA OF IRREGULAR SHAPE ON 1 CM GRID (C)\\
me62288&FIND VERTICES OF TRAPEZOIDS M AND N (DERIVED) (2)\\
me62173&FIND ANGLE X ON A FOLDED PIECE OF PAPER (1)\\
me62133&BLACK AND WHITE MARBLES IN A BAG WITH REPLACEMENT (D)\\
me62123a&RELAY RACE - MEAN TIME OF RUNNERS (C)\\
me62123b&RELAY RACE - MEAN TIME WHEN 2 RUNNERS IMPROVE (B)\\
\hline
\end{tabular}
\end{center}
\begin{center}
\begin{tabular}{ |p{2cm}|p{15cm}| }
 \hline
 \multicolumn{2}{|c|}{science test items and their code in TIMSS database}\\
 \hline
 Code&Questions\\
\hline
se62090&WATER CYCLE IN FOREST ECOSYSTEM (C)\\
se62274&RAW MATERIALS FOR PHOTOSYNTHESIS (2)\\
se62284&HAIR COLOR OF YOUNG RABBITS (B)\\
se62098B&PLANT AND ANIMAL CELLS DIFFERENT (2)\\
se62098A&PLANT AND ANIMAL CELLS SIMILAR (2)\\
se62032&HOT METAL BALL ON BALANCE (C)\\
se62043&ELECTROMAGNET AND PAPER CLIPS (1)\\
se62158&GRAPHS OF MUSICAL NOTES (B)\\
se62159&BOX PULLED BY THREE FORCES (D)\\
se62004&BLOCK POUNDED INTO FLAT SHEET (B)\\
se62075&HUGO'S CHEMICAL REACTION (D)\\
se62004&BLOCK POUNDED INTO FLAT SHEET (B)\\
se62175&POWER PLANT GEOGRAPHIC FACTOR (1)\\
se62173A&TEMPERATURE AND GEOGRAPHY (DERIVED) (1)\\
se62173B&TEMPERATURE AND GEOGRAPHY: OCTOBER (B)\\
\hline
\end{tabular}
\end{center}

\newpage

The following tables include the item codes in the Big-Five dataset and the corresponding questions.
\begin{center}
\begin{tabular}{ |p{2cm}|p{15cm}| }
 \hline
 \multicolumn{2}{|c|}{Items of Extraversion and their code in Big-Five dataset} \\
 \hline
 Code & Questions \\
 \hline
 E1 & I am the life of the party. \\
 E2 & I don't talk a lot. \\
 E3 & I feel comfortable around people. \\
 E4 & I keep in the background. \\
 E5 & I start conversations. \\
 E6 & I have little to say. \\
 E7 & I talk to a lot of different people at parties. \\
 E8 & I don't like to draw attention to myself. \\
 E9 & I don't mind being the center of attention. \\
 E10 & I am quiet around strangers. \\
   \hline
\end{tabular}
\end{center}
 \begin{center}
\begin{tabular}{ |p{2cm}|p{15cm}| }
 \hline
 \multicolumn{2}{|c|}{Items of Neuroticism and their code in Big-Five dataset} \\
 \hline
 Code & Questions \\
 \hline
 N1 & I get stressed out easily. \\
 N2 & I am relaxed most of the time. \\
 N3 & I worry about things. \\
 N4 & I seldom feel blue. \\
 N5 & I am easily disturbed. \\
 N6 & I get upset easily. \\
 N7 & I change my mood a lot. \\
 N8 & I have frequent mood swings. \\
 N9 & I get irritated easily. \\
 N10 & I often feel blue. \\
    \hline
\end{tabular}
\end{center}
 \begin{center}
\begin{tabular}{ |p{2cm}|p{15cm}| }
 \hline
 \multicolumn{2}{|c|}{Items of Agreeableness and their code in Big-Five dataset} \\
 \hline
 Code & Questions \\
 \hline
 A1 & I feel little concern for others. \\
 A2 & I am interested in people. \\
 A3 & I insult people. \\
 A4 & I sympathize with others' feelings. \\
 A5 & I am not interested in other people's problems. \\
 A6 & I have a soft heart. \\
 A7 & I am not really interested in others. \\
 A8 & I take time out for others. \\
 A9 & I feel others' emotions. \\
 A10 & I make people feel at ease. \\
    \hline
\end{tabular}
\end{center}
 \begin{center}
\begin{tabular}{ |p{2cm}|p{15cm}| }
 \hline
 \multicolumn{2}{|c|}{Items of Conscientiousness and their code in Big-Five dataset} \\
 \hline
 Code & Questions \\
 \hline
 C1 & I am always prepared. \\
 C2 & I leave my belongings around. \\
 C3 & I pay attention to details. \\
 C4 & I make a mess of things. \\
 C5 & I get chores done right away. \\
 C6 & I often forget to put things back in their proper place. \\
 C7 & I like order. \\
 C8 & I shirk my duties. \\
 C9 & I follow a schedule. \\
 C10 & I am exacting in my work. \\
    \hline
\end{tabular}
\end{center}
 \begin{center}
\begin{tabular}{ |p{2cm}|p{15cm}| }
 \hline
 \multicolumn{2}{|c|}{Items of Openness and their code in Big-Five dataset} \\
 \hline
 Code & Questions \\
 \hline
 O1 & I have a rich vocabulary. \\
 O2 & I have difficulty understanding abstract ideas. \\
 O3 & I have a vivid imagination. \\
 O4 & I am not interested in abstract ideas. \\
 O5 & I have excellent ideas. \\
 O6 & I do not have a good imagination. \\
 O7 & I am quick to understand things. \\
 O8 & I use difficult words. \\
 O9 & I spend time reflecting on things. \\
 O10 & I am full of ideas. \\
 \hline
\end{tabular}
\end{center}

\begin{center}
\section*{ Appendix D: Detailed Report on Parameter Estimation of the TIMSS Datset}
\end{center}

\begin{table}
\renewcommand\arraystretch{0.8}\begin{center}
\begin{tabular}{ p{2cm} p{2cm} p{2cm} p{2.5cm} p{2cm} p{2cm} p{2cm}|}
% \multicolumn{6}{|c|}{IRT parameters from eTIMSS Adjusted Model Calibration provided by TIMSS}\\
 \toprule
 Item&Slope $a_j$&Location $b_j$&Guessing $c_j$& Step 1 $D_{j1}$&Step 2 $D_{j2}$\\
\midrule
me62150&1.111&-0.193&&&\\
me62335&1.377&0.004&0.175&&\\
me62219&2.050&0.961&0.218&&\\
me62002&0.447&0.846&&&\\
me62149&1.089&0.617&0.111&&\\
me62241&1.708&0.743&&&\\
me62105&0.757&0.960&&-1.718&1.718\\
me62040&0.769&1.057&0.224&&\\
me62288&0.776&1.250&&-0.880&0.880\\
me62173&1.119&0.922&&&\\
me62133&1.315&0.726&0.214&&\\
me62123A&1.562&0.464&0.306&&\\
me62123B&1.444&0.814&0.138&&\\
se62090&1.011&0.180&0.304&&\\
se62274&0.577&0.879&&1.149&-1.149\\
se62284&0.375&0.478&0.172&&\\
se62098A&0.639&0.500&&-0.050&0.050\\
se62098B&0.798&1.337&&-0.091&0.091\\
se62032&1.742&1.504&0.287&&\\
se62043&0.907&0.981&&&\\
se62158&0.679&0.678&0.299&&\\
se62159&0.983&0.400&0.204&&\\
se62005&1.250&0.666&&&\\
se62075&0.990&0.770&0.314&&\\
se62004&1.806&0.885&0.173&&\\
se62175&0.739&0.674&&&\\
se62173A&0.647&0.253&&&\\
se62173B&0.808&1.862&0.393&&\\
\bottomrule
\end{tabular}
\caption{IRT parameters from eTIMSS Adjusted Model Calibration provided in official TIMSS file of IRT item parameter downloaded from TIMSS 2019 International Database}
\end{center}
\label{TIMSS-trueparameters}\end{table}

\begin{table}\renewcommand\arraystretch{0.8}\begin{center}
\begin{tabular}{ p{2cm}p{2cm}p{2.5cm}p{2.5cm}}
 \toprule
 Item&Slope $a_j$&Location $b_{j1}$&Location $b_{j1}$\\
\midrule
me62150&1.603&-1.306&\\
me62335&2.010&-2.089&\\
me62219&1.647&0.514&\\
me62002&0.705&0.181&\\
me62149&1.453&0.271&\\
me62241&2.587&0.393&\\
me62105&1.183&2.192&1.539\\
me62040&0.448&0.109&\\
me62288&1.108&1.712&1.786\\
me62173&1.379&1.944&\\
me62133&1.370&-0.400&\\
me62123A&1.477&-0.770&\\
me62123B&1.400&0.231&\\

se62090&0.839&-1.409&\\
se62274&0.646&-1.063&1.164\\
se62284&0.777&-1.018&\\
se62098A&0.754&-0.962&-0.575\\
se62098B&0.883&0.748&2.055\\
se62032&0.702&0.476&\\
se62043&0.947&1.767&\\
se62158&0.482&-0.649&\\
se62159&1.150&-0.499&\\
se62005&1.466&0.923&\\
se62075&0.024&1.411&\\
se62004&1.512&-0.352&\\
se62175&0.440&0.298&\\
se62173A&1.036&-0.146&\\
se62173B&0.677&0.906&\\
\bottomrule
\end{tabular}
\caption{IRT parameters Estimated from EM}
\end{center}\end{table}

\begin{table}[htbp]\renewcommand\arraystretch{0.8}\begin{center}
\begin{tabular}{ p{2cm}p{2cm}p{2.5cm}p{2.5cm}}
 \toprule
 Item&Slope $a_j$&Location $b_{j1}$&Location $b_{j1}$\\
\midrule
me62150&1.465&-1.258&\\
me62335&1.735&-1.927&\\
me62219&1.455&0.477&\\
me62002&0.702&0.177&\\
me62149&1.317&0.251&\\
me62241&1.986&0.328&\\
me62105&1.032&2.183&1.493\\
me62040&0.454&0.107&\\
me62288&0.993&1.656&1.699\\
me62173&1.241&1.856&\\
me62133&1.281&-0.398&\\
me62123A&1.385&-0.758&\\
me62123B&1.302&0.215&\\

se62090&0.754&-1.363&\\
se62274&0.706&-1.078&1.193\\
se62284&0.738&-0.997&\\
se62098A&1.041&-1.178&-0.673\\
se62098B&1.002&0.696&2.169\\
se62032&0.613&0.460&\\
se62043&0.749&1.660&\\
se62158&0.493&-0.649&\\
se62159&0.923&-0.462&\\
se62005&1.065&0.797&\\
se62075&0.051&1.411&\\
se62004&1.135&-0.312&\\
se62175&0.433&0.295&\\
se62173A&0.904&-0.142&\\
se62173B&0.599&0.880&\\
\bottomrule
\end{tabular}
\caption{IRT parameters Estimated from pGVEM}\end{center}
\end{table}

\begin{table}[htbp]\renewcommand\arraystretch{0.8}\begin{center}

\begin{tabular}{ p{2cm}p{2cm}p{2cm}p{2.5cm}p{2.5cm}}
 \toprule
 Item&Slope $a_j^{(1)}$&Slope $a_j^{(2)}$&Location $b_{j1}$&Location $b_{j1}$\\
\midrule
me62150&1.449&0.222&-1.307&\\
me62335&2.033&0.005&-2.122&\\
me62219&1.598&0.052&0.510&\\
me62002&0.704&-0.61&0.179&\\
me62149&1.649&-0.385&0.265&\\
me62241&2.432&0.260&0.397&\\
me62105&1.160&-0.117&2.193&1.472\\
me62040&0.530&-0.137&0.108&\\
me62288&1.105&0.127&1.714&1.866\\
me62173&1.595&-0.357&1.977&\\
me62133&1.324&0.087&-0.406&\\
me62123A&1.435&-0.075&-0.753&\\
me62123B&1.312&0.087&0.227&\\

se62090&0.438&0.505&-1.404&\\
se62274&0.551&0.186&-1.073&1.174\\
se62284&0.438&0.497&-1.031&\\
se62098A&0.102&1.058&-1.204&-0.677\\
se62098B&0.022&1.842&0.773&3.153\\
se62032&0.626&0.055&0.468&\\
se62043&0.451&0.444&1.693&\\
se62158&0.071&0.438&-0.649&\\
se62159&0.996&0.216&-0.508&\\
se62005&1.252&0.221&0.890&\\
se62075&-0.093&0.404&1.450&\\
se62004&0.988&0.325&-0.330&\\
se62175&0.109&0.380&0.298&\\
se62173A&1.140&0.100&-0.169&\\
se62173B&0.697&0.019&0.910&\\
\bottomrule
\end{tabular}
\caption{IRT parameters Jointly Estimated from EM}
\end{center}\end{table}

\begin{table}[htbp]\renewcommand\arraystretch{0.8}\begin{center}

\begin{tabular}{ p{2cm}p{2cm}p{2cm}p{2.5cm}p{2.5cm}}
 \toprule
 Item&Slope $a_j^{(1)}$&Slope $a_j^{(2)}$&Location $b_{j1}$&Location $b_{j1}$\\
\midrule
me62150&1.377&0.204&-1.324&\\
me62335&1.802&0.051&-2.047&\\
me62219&1.479&0.034&0.432&\\
me62002&0.680&-0.026&0.154&\\
me62149&1.451&-0.251&0.206&\\
me62241&1.984&0.269&0.273&\\
me62105&1.013&0.012&2.181&1.399\\
me62040&0.557&-0.170&0.093&\\
me62288&1.045&0.107&1.627&1.735\\
me62173&1.371&-0.182&1.827&\\
me62133&1.220&0.153&-0.446&\\
me62123A&1.336&-0.021&-0.783&\\
me62123B&1.213&0.133&0.173&\\

se62090&0.479&0.413&-1.410&\\
se62274&0.580&0.186&-1.097&1.172\\
se62284&0.0491&0.405&-1.046&\\
se62098A&-0.07&1.455&-1.510&-0.881\\
se62098B&0.169&1.104&0.649&2.350\\
se62032&0.628&0.059&0.442&\\
se62043&0.489&0.345&1.639&\\
se62158&0.088&0.414&-0.658&\\
se62159&1.068&0.074&-0.545&\\
se62005&1.220&0.142&0.814&\\
se62075&-0.296&0.303&1.431&\\
se62004&1.047&0.201&-0.372&\\
se62175&0.155&0.305&0.281&\\
se62173A&1.105&0.094&-0.210&\\
se62173B&0.683&0.034&0.879&\\
\bottomrule
\end{tabular}
\caption{IRT parameters Jointly Estimated from GVEM}

\end{center}\end{table}

\end{document}